\documentclass{amsart}

\usepackage{amssymb}

\usepackage{amsfonts}
\usepackage{amsmath}

\usepackage{graphicx}
\usepackage{enumitem}
\usepackage[usenames,dvipsnames]{pstricks}
\usepackage{tikz}
\usepackage{caption}
\usepackage{subcaption}

\newtheorem{theorem}{Theorem}[section]
\newtheorem{lemma}[theorem]{Lemma}
\newtheorem{corollary}[theorem]{Corollary}
\newtheorem{prop}[theorem]{Proposition}

\theoremstyle{definition}

\theoremstyle{remark}

\numberwithin{equation}{section}

\DeclareMathOperator{\re}{Re}
\DeclareMathOperator{\im}{Im}

\usepackage{hyperref}

\begin{document}
\title[Stokes graphs of the Rabi problem]{Stokes graphs of the Rabi problem with real parameters}

    \author{Ren\'e Lang\o en}
    \address{Department of Mathematics, University of Bergen, Norway}
    \email{rene.langoen@uib.no}

    \author{Irina Markina}
    \address{Department of Mathematics, University of Bergen, Norway}
    \email{irina.markina@uib.no}

    \author{Alexander Solynin}
    \address{Department of Mathematics and Statistics, Texas Tech University, Lubbock, TX, USA}
    \email{alex.solynin@ttu.edu}


    \date{}

    \begin{abstract}

The goal of this paper is to study the geometry of the Stokes graphs
associated with the problem, which was introduced by Isidor Rabi in 1937 to model
reactions of atoms to the harmonic electric field with frequency close to the natural frequency of the atoms.
In the standard Garnier form, the Rabi model is a matrix linear differential
equation with three physical parameters, which are: the level
of separation of the fermion mode $\Delta$, the boson-fermion
coupling $g$, and the eigenvalue $E$ of the Hamiltonian relevant to this model.
The qualitative behavior of solutions of this type of problems
is often described in terms of the Stokes graphs of associated
quadratic differential, which in the case of Rabi problem can be represented in the form
$Q_0(z)\, dz^2 =
-\frac{z^4+c_3z^3+c_2z^2+c_1z+c_0}{(z-1)^2(z+1)^2}\, dz^2$ with
the coefficients $c_k$, $k=0,1,2,3$, depending on the parameters
$\Delta$, $g$, and $E$. In this paper, we first give a complete
classification of possible generic topological types of domain
configurations and Stokes graphs of this quadratic differential
assuming that its coefficients $c_k$ are real and the zeros of its
numerator are distinct from its poles. Then we identify the set of
coefficients $(c_3,c_2,c_1,c_0)\in \mathbb{R}^4$, which correspond
to particular choices of the physical parameters $\Delta$, $g$,
and $E$. The structure of Stokes graphs and domain configurations
of quadratic differentials, which appear as asymptotic cases when
the parameters of the Rabi problem tend to infinity, also will be
discussed.

    \end{abstract}

    \maketitle

    \tableofcontents

 \section{Introduction}

    In this work, we study geometry of Stokes graphs and domain
configurations of quadratic differentials associated with the Rabi
problem. These graphs and configurations provide important
information on the qualitative behavior of solutions to this
problem.  The problem was introduced by Isidor Issak
Rabi~\cite{Rabi37} as a model describing how a rapidly varying
weak magnetic field affects an oriented atom possessing nuclear
spin. Thus, the Rabi problem deals with reactions of the atom to
the harmonic electric field with frequency close to the atom's
natural frequency. Despite its simplicity, the quantum Rabi model
is not exactly solvable. The problem, originated in mathematical
physics as a model describing a simple harmonic oscillator or two
level quantum system~\cite{Zhong2013}, has obtained numerous
applications; in particular, in the theory of quantum computing
and other areas of quantum mechanics under different coupling
regimes, see~\cite{Xie17}.

The Rabi model in the standard Garnier form~\cite{Iwasaki1991} is
a system of linear matrix equations on the Riemann sphere, where
the coefficient matrix has two simple poles at $p_1,p_2\in \mathbb
C$ and a pole of order two at $\infty$. It has three physical
parameters: $\Delta\in \mathbb R$ is the level of separation of
the fermion mode, $g\in\mathbb C$ is the boson-fermion coupling,
and $E\in \mathbb C$ is the eigenvalue of the Hamiltonian defined
by the physical problem. The detailed description of the Rabi
problem will be given in Section~2. The governing equation of this
problem is a second order linear ODE, which after linear change of
variables can be written in the form
\begin{equation}\label{eq:1}
    \frac{d^2}{dz^2}y(z)+Q(\Delta, E,g,z)y(z)=0,
\end{equation}
where $Q(\Delta, E,g,z)
    =-(1/4)(z^4+a_3z^3+a_2z^2+a_1z+a_0)z^{-2}(z+4g^2)^{-2}$ is a rational function with coefficients $a_k=a_k(\Delta, E,g)$,
    $k=0,1,2,3$, depending on the parameters of the Rabi problem.

    The general scheme to study the equation of
type~\eqref{eq:1} includes the following 3 steps.
\begin{itemize}
    \item[1.] Construction of the Stokes graph embedded in $\overline{\mathbb{C}}$ and identification of its faces.
    We recall that the Stokes graph of equation (\ref{eq:1}) is the graph consisting of the critical trajectories of the quadratic differential
\begin{equation} \label{eq:2} %
    Q(\Delta, E,g,z)\,dz^2
    =-\frac{1}{4}\frac{z^4+a_3z^3+a_2z^2+a_1z+a_0}{z^2(z+4g^2)^2}\,dz^2, \quad z\in\mathbb C\cup\{\infty\}.
\end{equation}
    \item[2.] Finding the fundamental solution in each domain that is a face of the Stokes graph.
    \item[3.] Identifying the matrices relating the fundamental solutions defined
    in different domains in order to get the ``global'' fundamental solution.
\end{itemize}

In this paper, we focus on the first step of this general scheme
assuming that the parameters $\Delta$, $E$, and $g^2\not=0$ of the
Rabi problem are real numbers. Thus, in our study we assume that
the boson-fermion coupling $g\not=0$ is either real or pure
imaginary nonzero number. These our assumptions imply, in turn,
that the coefficients $a_k$ of the quadratic differential
(\ref{eq:2}) are
 real numbers.

The relations between properties of solutions of certain ODE's and
the structure of critical trajectories of related quadratic
differentials were explored by many authors working with
differential equations. One of the primary references here is the
monograph \cite{Fedor1993} by M.~V.~Fedoryuk. For more recent
results on applications of quadratic differentials to specific
differential equations, we refer to the papers \cite{MMT2016},
\cite{ShapiroSol2017}, and \cite{ChouikhiThabet}.

We structure our paper as follows. In
Section~\ref{sec:Rabi-model}, we describe the Rabi model and
remind useful facts from the theory of linear ODE's. Furthermore,
we use linear change of variables to rewrite the quadratic
differential \eqref{eq:2} in a more symmetric form \eqref{s6aa},
which is convenient to work with.   We also give explicit
expressions for the coefficients $c_k$ of the quadratic
differential~\eqref{s6aa} as functions of the parameters of the
Rabi problem. In Section~3, we collect definitions and results
from the theory of quadratic differentials needed for the study of
the Stokes graphs of equation~(\ref{eq:1}).

Section~4 contains classification of possible critical graphs and
domain configurations of the quadratic differential (\ref{s6aa})
assuming that  its coefficients $c_k$ are real numbers not
necessarily related to the parameters $\Delta$, $E$, and $g$ of
the Rabi problem. To avoid having too many cases and figures in
just one paper, we also assume in what follows that the quadratic
differential $Q_0(z)\,dz^2$ defined by (\ref{s6aa}) has \emph{full
set of critical points}, which means that the zeros of the
numerator in (\ref{s6aa}) are distinct from the poles $\pm 1$. The
description of the Stokes graphs and domain configurations of the
so-called \emph{depressed quadratic differential of the Rabi
problem}, when cancellation of zeros and poles happens and
$Q_0(z)\,dz^2$ has at most simple pole at least one of the points
$z=\pm 1$, will be postponed for a sequel paper.

Our classification of Stokes graphs and domain configurations of
$Q_0(z)\,dz^2$ is based on the number of real zeros, on types of
domains present in the domain configuration of this quadratic
differential, and on the positions of zeros on boundaries of these
domains.

As it is expected, the classification includes many cases.
We want to mention here the analogy between classifications of
critical graphs of quadratic differentials with prescribed number
of critical points and classifications of real algebraic curves of
a given degree. As it is well known, the classification of cubic
curves, first suggested by I.~Newton in the seventeenth century
and completed later, contains 78 types of curves. A different,
more topological classification of cubics, was discussed in
\cite{Weinberg1988}. Interestingly enough, the modern approach to
the classification problem for real algebraic curves, initiated by
J.C. Langer and D.A. Singer in \cite{LangerSinger 2007} and then
used in \cite{SolyninSolynin2022}, reveals that shapes of these
curves can be identified as critical graphs of appropriate
quadratic differentials defined on compact Riemann surfaces.

 In Section~5, we describe the set of coefficients
 $(c_3,c_2,c_1,c_0)\in \mathbb{R}^4$ of the quadratic differential
 $Q_0(z)\,dz^2$, which correspond to the real values of the Rabi
 parameters $\Delta$, $E$, and $g^2$.
 In our last Section~\ref{sec:Asymptotic}, we study
the limit
 domain configurations, when the boson-fermion coupling $g$ tends to $\infty$, assuming that the parameters $\Delta$
and $E$ are certain functions of $g$. The latter study is
motivated by the isomonodromic problem associated to the Rabi
model~\cite{Cunha2016}, when one tries to relate the parameters of
the model providing the same monodromy data for the ODE. This
problem is closely related to the tau-function of the Painlev\'e V
equation and it could be helpful in the study of the quantized
spectrum of the Rabi model.

Appendix A contains notations consistently used throughout the
paper. Appendix B, that is the ``Zoo'' of Stokes graphs and domain
configurations, contains examples of possible Stokes
graphs and domain configurations of the  quadratic
differential  $Q_0(z)\,dz^2$ that is the symmetrized form of the quadratic differential (\ref{eq:2}). %
These Stokes graphs and domain configurations are described in
details in Section~\ref{sec:RealZeros}.

\section{Rabi model and associated quadratic differential}\label{sec:Rabi-model}


Before describing the Rabi model, we present general facts from
the theory of linear ODE needed for our study.
Consider the matrix linear differential equation
\begin{equation}\label{eq:GMODE}
    \frac{d\Psi(z)}{dz}=A(z)\Psi(z) 
\end{equation}
in a domain $\Omega \subset\overline{\mathbb{C}}$. Here %
$$ %
A(z)= \begin{pmatrix}a_{11}&a_{21}\\a_{12}&a_{22}\end{pmatrix}
\quad {\mbox{and}} \quad \Psi(z)=\begin{pmatrix}f_{11}&f_{21}\\f_{12}&f_{22}\end{pmatrix}%
$$ %
are $2\times 2$ matrices with coefficients, depending on $z\in
\Omega$. 
Each of the columns $f_1=\begin{pmatrix}f_{11}\\
f_{12}\end{pmatrix}$ and $f_2=\begin{pmatrix}f_{21}\\
f_{22}\end{pmatrix}$ of the unknown matrix valued function
$\Psi(z)$ is a solution to the equation
$$
\frac{d f_k(z)}{dz}=A(z)f_k(z),\quad k=1,2.
$$
If $A(z)$ is holomorphic in $\Omega$, then for every $z_0\in
\Omega$ there is a unique  fundamental solution $\Psi(z)$
to~\eqref{eq:GMODE} holomorphic at $z_0$, satisfying the
normalization condition $\Psi(z_0)=I$, where $I$ is the identity
matrix. Any other fundamental solution $\widetilde{\Psi}(z)$,
which is holomorphic at $z_0$, has the form $
\widetilde{\Psi}(z)=\Psi(z) C$, where $C$ is a constant matrix. If
$A(z)$ is meromorphic in $\Omega$, then the behavior of solutions
to (\ref{eq:GMODE}) near singular points of $A(z)$ is more
complicated; see, for instance,~\cite{Fokas2006}.

The functions $f_{11}$ and $f_{21}$ are solutions to the second
order ODE 
\begin{equation}\label{eq:2ODE}
    \frac{d^2 f}{d z^2}+p(z)\frac{df}{d z}+q(z)f(z)=0
\end{equation}
with
$$
p(z)=-{\rm Tr} A-\frac{1}{a_{12}}\frac{da_{12}(z)}{d z}=-{\rm Tr}
A-\frac{d}{dz}\log a_{12},
$$
and
$$
q(z)=\frac{a_{11}}{a_{12}}\frac{da_{12}(z)}{d
z}-\frac{da_{11}(z)}{dz}+\det A=a_{11}\frac{d}{dz}\log
a_{12}-\frac{da_{11}(z)}{dz}+\det A.
$$
Note that if $f_{11}(z)$ and $f_{21}(z)$ are linearly independent
solutions of~\eqref{eq:2ODE}, then
$$
\begin{pmatrix}
    f_{11}(z)&f_{21}(z)
    \\
    f'_{11}(z)&f'_{12}(z)
\end{pmatrix}
$$
is the fundamental solution to~\eqref{eq:GMODE}.

Furthermore, changing variable in~\eqref{eq:2ODE} via $f(z)=\psi(z)y(z)$ with %
\begin{equation} \label{s3} %
    \psi(z)={\rm exp}\left(-\frac{1}{2}\,\int_{z_0}^z
    p(\tau)\,d\tau\right),
\end{equation} %
we rewrite~\eqref{eq:2ODE} in the following equivalent form: %
\begin{equation} \label{s4} %
    y''(z)+Q(z)y(z)=0,
\end{equation} %
where %
\begin{equation} \label{s5} %
    Q(z)=q(z)-\frac{1}{4}p^2(z)-\frac{1}{2}p'(z).
\end{equation} %

\bigskip


Now we turn to the Rabi model that is a physical model describing
a simple harmonic oscillator, or two level quantum
system~\cite{Cunha2016, Zhong2013}. The Rabi model in the standard
Garnier form~\cite{Iwasaki1991} is governed 
by the
differential
equation (\ref{eq:GMODE}) with the matrix 
\begin{equation}\label{eq:3}
    A(z)=\frac{\sigma_3}{2}+\frac{A_0}{z}+\frac{A_t}{z-t},
\end{equation}
where $t=-4g^2$, $z\in\overline{\mathbb C}$,
$$
\sigma_3=
\begin{pmatrix}
    1&0
    \\
    0&-1
\end{pmatrix},
\quad A_0=\begin{pmatrix}
    E+g^2&-\Delta
    \\
    0&0
\end{pmatrix},
\quad A_t=\begin{pmatrix}
    0&0
    \\
    -\Delta&E+g^2
\end{pmatrix}.
$$
Here $\sigma_3$ is one of the Pauli matrices, $2\Delta$ is the
energy difference between the two fermion levels, $g$ is the
boson-fermion coupling, and $E$ is the eigenvalue of a Hamiltonian
defined by the physical problem. In a general setting of the Rabi
problem, the energy difference $\Delta$ is real but the
boson-fermion coupling  $g$ and the eigenvalue $E$ may be complex
numbers. As we already mentioned in the Introduction, below in
this paper we always assume that the parameters  $\Delta$ and $E$
are real numbers and that $g\not=0$ is either real or pure
imaginary nonzero number. In this case, the parameter
$t=-4g^2\not=0$ is real and therefore the function $A(z)$ defined
by (\ref{eq:3}) has a pole at the point $z=t$, $t\not=0$, of the
real axis.

With the matrix $A(z)$ defined in (\ref{eq:3}),
system~(\ref{eq:GMODE}) has two regular singular points
$z=0$, $z=t$, and one irregular singular point
at $z\in\infty$ of Poisson rank 1; see~\cite{Fokas2006}.

Reducing the matrix differential equation to the second order ODE,
as it was described earlier in this section, we obtain the
following equation:
\begin{equation}\label{s11}
    f''(z)+p(z)f'(z)+q(z)f(z)=0,
\end{equation}
where
$$
p(z,t)=\frac{1-\theta}{z}-\frac{\theta}{z-t},
$$
$$
q(z)=-\frac{1}{4}+\frac{1}{z}\Big(-\frac{1}{2}+\frac{\Delta^2}{t}-\frac{\theta^2}{t}-\frac{\theta}{2}\Big)+
\frac{1}{z-t}\Big(-\frac{\Delta^2}{t}+\frac{\theta^2}{t}+\frac{\theta}{2}\Big),
$$
and 
$$ %
\theta={\rm Tr}A_0={\rm Tr}A_t=E+g^2=E-\frac{t}{4}. %
$$ %

Changing variable in~\eqref{s11} via~\eqref{s3}, %
we obtain equation~\eqref{s4} %
with the function $Q(z)=Q(z,t)$, where %
\begin{equation*}\label{eq:Q}
    Q(z,t)=
    -\frac{1}{4}\frac{z^4+a_3z^3+a_2z^2+a_1z+a_0}{z^2(z-t)^2},
\end{equation*}
with the coefficients $a_k$, $k=0,1,2,3$, 
given by
\begin{eqnarray*}
   a_3&=&-2t+2, \quad \quad     a_2= t^2-t(2\theta+4)+4\Delta^2-1,
    \\
    a_1&=& t^2(2\theta+2)-t(4\Delta^2-2\theta-2), \quad \quad
    a_0= t^2(\theta^2-1).
\end{eqnarray*}

The goal now is to describe possible Stokes graphs of the
equation~\eqref{s4} or, equivalently, describe possible structures
of the critical trajectories of the quadratic differential
$Q(z,t)\,dz^2$. Changing variables in the quadratic differential
$Q(z,t)\,dz^2$  via the linear transformation $z\to
\frac{t}{2}(1-z)$ and then multiplying the resulting quadratic
differential by $4$, we obtain the following more symmetric form
of this quadratic differential, which is easier to work with:
\begin{equation} \label{s6aa} %
    Q_0(z)\,dz^2
    =-\frac{P_0(z)}{(z-1)^2(z+1)^2}\,dz^2=-\frac{z^4+c_3z^3+c_2z^2+c_1z+c_0}{(z-1)^2(z+1)^2}\,dz^2,
\end{equation} %
where the coefficients of the numerator
$P_0(z)=z^4+c_3z^3+c_2z^2+c_1z+c_0$ are given by
\begin{eqnarray}
  \label{eq2.9} &c_3&=g^{-2}, \quad
    c_2=\frac{1}{4g^4}(8Eg^2+4\Delta^2+4g^2-1),
    \\
 \label{eq2.10} &c_1&= -\frac{1}{2g^4}(4g^2+2E+1), %
\quad
    c_0=-\frac{1}{4g^4}(4\Delta^2-4E^2-4E+1). %
\end{eqnarray}
We note here that critical trajectories and domain configuration
of the resulting quadratic differential $Q_0(z)\,dz^2$ coincide
with those of $Q(z,t)\,dz^2$ up to scaling, reflection with
respect to the imaginary axis, and translation in the horizontal
direction. Thus, in what follows we work with the rescaled
quadratic differential $Q_0(z)\,dz^2$ given by
equation~(\ref{s6aa}) assuming that it has full set of critical
points, i.e. we assume that $P_0(\pm 1)\not=0$ and therefore in
the
cases under consideration $Q_0(z)\,dz^2$ has double poles at the points $\pm 1$. %



\section{Basics on quadratic differentials}\label{sec:StokesGraph} 


In this section, we first recall definitions and some basic facts
about quadratic differentials. Some notations, needed for future
use, will also be introduced. After that we discuss basic
characteristics of the quadratic differential (\ref{s6aa}),
assuming that all its coefficients $c_k$ are real. In particular,
few basic properties of the Stokes graph of this quadratic
differential and its domain configuration, which do not depend on
the positions of the zeros of $Q_0(z)\,dz^2$, will be mentioned.
The detailed description of  properties of the Stokes graphs and
domain configurations, which depend on the positions of these
zeros, will be given in the next section.

In this paper, we deal with quadratic differentials defined on the
Riemann sphere $\overline{\mathbb{C}}$. For more general theory of
quadratic differentials defined on Riemann surfaces the interested
reader may consult classical monographs by J.~Jenkins
\cite{Jenkins} and
 K.~Strebel~\cite{Strebel1984}.

A quadratic differential on a domain $D\subset
\overline{\mathbb{C}}$ is a differential form $Q(z)\,dz^2$ with
meromorphic $Q(z)$ and with the conformal transformation rule
\begin{equation} \label{2.1} %
Q_1(\zeta)\,d\zeta^2=Q(\varphi(z))\left(\varphi'(z)\right)^2\,dz^2,
\end{equation}
where $\zeta=\varphi(z)$ is a conformal map from $D$ onto a domain
$G$ in the extended plane of the parameter $\zeta$.

The zeros and poles of $Q(z)$ are critical points of $Q(z)\,dz^2$,
in particular, zeros and simple poles are finite critical points
and poles of order greater than $1$ are infinite critical points
of $Q(z)\, dz^2$.

A trajectory (respectively, orthogonal trajectory) of $Q(z)\,dz^2$
is a closed analytic Jordan curve or maximal open analytic arc
$\gamma\subset D$ such
that %
$$ 
Q(z)\,dz^2>0 \quad {\mbox{along $\gamma$}} \quad \quad
({\mbox{respectively}}, Q(z)\,dz^2<0 \quad {\mbox{along
$\gamma$}}).
$$ 
A trajectory $\gamma$ is called \emph{critical} if at least one of
its end points is a finite critical point of $Q(z)\,dz^2$.

In the theory of ODE's, critical trajectories of the quadratic
differential $Q(z)\,dz^2$ associated with  equation (\ref{s4}) are
known as the \emph{Stokes lines} of this equation. Accordingly, a
finite critical point of $Q(z)\,dz^2$, that is an end point of a
Stokes line, is called  \emph{turning point} of the equation
(\ref{s4}). We have to stress here, that terminology concerning
quadratic differentials used in this paper may differ from the one
used by some other authors. For instance, in the Fedoryuk's
monograph \cite{Fedor1993} the Stokes lines are defined as our
orthogonal critical trajectories. In this work, we will stick with
definitions used consistently in the classical publications of
J.~Jenkins~\cite{Jenkins,Jenkins2}, K.~Strebel~\cite{Strebel1984}, and others.

Let $\Phi_Q$, $\overline{\Phi}_Q$, and $G_Q=\partial
\overline{\Phi}_Q$ denote the union of points of critical
trajectories of a quadratic differential $Q(z)\,dz^2$ defined on
$\overline{\mathbb{C}}$, the closure of this union of points, and
the boundary of this closure, respectively. It is known, see
\cite[Theorem~3.5]{Jenkins}, that the set $G_Q$ is either empty or
it consists of a finite number of critical trajectories of
$Q(z)\,dz^2$. Furthermore, each of these critical trajectories in
each direction has an end point situated at some point in
$\overline{\mathbb{C}}$ and at least one of these end points is a
finite critical point of $Q(z)\,dz^2$. If $G_Q\not=\emptyset$, it
is called the \emph{critical graph} of $Q(z)\,dz^2$. Interestingly
enough, the study initiated in \cite{Solynin2009} and continued in
\cite {Solynin2020} shows that every weighted planar graph $G$
with simply connected and doubly connected faces can be realized
as the critical graph of a certain quadratic differential defined
on $\overline{\mathbb{C}}$.

In relation with ODE's, the critical graph $G_Q$ is also known as
the \emph{Stokes graph} of  equation~(\ref{s4}), see
\cite{Fedor1993}. As usual, the critical trajectories that are
components of $G_Q$ are edges of $G_Q$, their end points are
vertices of $G_Q$, and the connected components $\Omega_k$ of
$\overline{\mathbb{C}}\setminus G_Q$ are faces of $G_Q$.

The set $\{\Omega_k\}$ of all faces of $Q(z)\,dz^2$ is called the
domain configuration of $Q(z)\,dz^2$. The domains $\Omega_k$,
which are bounded by the Stokes lines, play an essential role in
the asymptotic analysis of ODE's.   According to the Basic
Structure Theorem of J.~Jenkins, \cite[Theorem~3.5]{Jenkins}, the
set $\{\Omega_k\}$ consists of a finite number of domains
$\Omega_k$, each of which belongs to one of the
following 5 types: %
\begin{enumerate} %
 \item[$\bullet$]  $\Omega_k$ is called a circle domain of $Q(z)\,dz^2$, if it is a simply connected domain
 bounded by a finite number of critical trajectories, which end points are finite critical points of
 $Q(z)\,dz^2$, and such that $\Omega_k$ contains exactly one critical point of $Q(z)\,dz^2$,
 called the \emph{center} of $\Omega_k$, which is a pole of order~$2$.

   \item[$\bullet$] $\Omega_k$ is called a ring domain of $Q(z)\,dz^2$, if it is a doubly connected
   domain, free of critical points and critical trajectories, such
   that
 each of two boundary components of $\Omega_k$ consists of a finite number of critical trajectories
with  end points in the set of finite critical points of
 $Q(z)\,dz^2$.

 \item[$\bullet$]  $\Omega_k$ is called an end domain of $Q(z)\,dz^2$, if it is a simply connected domain,
 free of critical points and critical trajectories,
which boundary consists of a finite number of critical
trajectories, such that two of them have a common end point at a
pole of order $\ge 2$, called the \emph{vertex} of $\Omega_k$,
while all other end points of these critical trajectories are
finite critical points of $Q(z)\,dz^2$.

  \item[$\bullet$]   $\Omega_k$ is called a strip domain of $Q(z)\,dz^2$, if it is a simply connected domain,
  free of critical points and critical trajectories,
which boundary, consisting of four or more critical trajectories,
has exactly two distinct boundary points, called \emph{vertices},
which belong to the set of infinite critical points of
$Q(z)\,dz^2$. The boundary arcs joining vertices of a strip domain
are called the \emph{sides} of this domain. It is also required
that each side consists of two or more critical trajectories.

\item[$\bullet$] $\Omega_k$ is called a density domain of
$Q(z)\,dz^2$ if every trajectory, which crosses $\Omega_k$, is
dense in $\Omega_k$.
\end{enumerate} %

\smallskip

Below in this paper, we use the following notations. By $(a,b)$
and $[a,b]$ we denote, respectively, open and closed intervals
having end points at $z=a$ and $z=b$. Notations $(-\infty,a)$,
$(b,\infty)$, etc, will be used to denote infinite intervals on
the real axis. If $\gamma$ is a rectifiable arc in a domain $D$,
where a quadratic differential $Q(z)\,dz^2$ is defined, then its
$Q$-length is defined as $|\gamma|_Q=\int_\gamma
|Q(z)|^{1/2}\,|dz|$. Also, if $a,b\in \overline{\mathbb{C}}$ are
not infinite critical points of $Q(z)\,dz^2$ and the open interval
$(a,b)$ does not contain
critical points of $Q(z)\,dz^2$, then we define $[a,b]_Q$ as follows: %
\begin{equation} \label{[a,b]-Integral}%
[a,b]_Q=\int_a^b\sqrt{Q(z)}\,dz,
\end{equation} %
where the integration is taken along the interval $[a,b]$. In what
follows, we mostly work with the real or imaginary parts of the
integrals defined as in (\ref{[a,b]-Integral}). In such cases, we
assume that the branch of the square root in
(\ref{[a,b]-Integral}) is chosen such that these real or imaginary
parts are non-negative.

An important property of quadratic differentials is that the
transformation rule (\ref{2.1}) respects trajectories, orthogonal
trajectories, and their $Q$-lengths, as well as it respects domain
configurations and critical points together with their
multiplicities and trajectory structure nearby.

\smallskip

In the classification of domain configurations of $Q_0(z)\,dz^2$, presented in Section~4, we routinely
use a few simple properties of critical graphs of quadratic
differentials, which, for convenience of the reader, are collected
in the following lemma.

\begin{lemma} \label{Lemma 1} %
Let $Q(z)\,dz^2$ be a quadratic differential on
$\overline{\mathbb{C}}$, which does not have density domains among
its faces. Then the following properties hold: %

\begin{enumerate} %
\item[1.] Let $D\subset \mathbb{C}$ be a simply connected domain,
which boundary consists of a finite number of critical
trajectories of $Q(z)\,dz^2$ and their end points. Let $P$ and $N$
denote the number of poles and zeros of $Q(z)\,dz^2$ on
$\overline{D}$, where poles and zeros in $D$ are counted with
their multiplicities and poles and zeros on $\partial D$,
considered as boundary critical points of $Q(z)\,dz^2$,
are counted with half of their multiplicities. Then %
$$ %
P-N=2.
$$ %
\item[2.] %
Let $\Gamma$ be a connected boundary component of a face $\Omega$.
If the connected component of $\overline{\mathbb{C}}\setminus
\Gamma$, containing $\Omega$, also contains at least one zero of
$Q(z)\,dz^2$, then $\Omega$ is a ring domain.

\item[3.] %
Let $\Gamma$ be a Jordan arc consisting of a finite number of
critical trajectories of $Q(z)\,dz^2$, which end points are
infinite critical points of $Q(z)\,dz^2$, and there is no other
infinite critical point on $\Gamma$. If $\Gamma$ is a proper
boundary arc on the boundary of a face $\Omega$ of the quadratic
differential $Q(z)\,dz^2$, then $\Omega$ is a strip domain.

\item[4.] %
If $Q(z)\,dz^2$ has $n\ge 1$ zeros, counting multiplicity, in a
Jordan domain $D\subset \mathbb{C}$ and does not have other
critical points on $\overline{D}$, then there are at least $n+2$
critical trajectories of $Q(z)\,dz^2$ crossing the boundary of
$D$.
\end{enumerate} %
\end{lemma} %

\noindent %
 \emph{Proof.} The formula in part 1 of this lemma is
just a simplest special case of the relation between zeros and
poles given in \cite[Lemma 3.2]{Jenkins}. Parts 2 and 3
immediately follow from the Basic Structure Theorem, see
\cite[Theorem 3.5]{Jenkins}, and from the definitions of end,
circle, ring, and strip domains given above.

Part 4 is a property known in the Graph theory without relation to
quadratic differentials. It can be proved as follows. Consider the
graph $G_Q(D)$ that is the restriction of the critical graph $G_Q$
onto the domain $D$. Since $Q(z)\,dz^2$ does not have poles in a
simply connected domain $D$, it follows from part 1 of Lemma~3.1
that $G_Q(D)$ does not have cycles.

It is enough to prove the property required in this part assuming
that the graph $G_Q(D)$ is connected. Indeed, if the result holds
for every connected component of $G_Q(D)$, then adding the numbers
of trajectories exiting $D$ for each connected subgraph of
$G_Q(D)$, we obtain the required result for $G_Q(D)$.  Thus, we
assume that $G_Q(D)$ is connected and without cycles;  in the
Graph theory such graphs are known as \emph{trees}. We may start
with the case when all zeros of $Q(z)\,dz^2$, which are in the
domain $D$, are simple. Then the degree of each vertex of $G_Q(D)$
is $3$. Thus, the sum of all these degrees is $3n$. If the
vertices $v_1$ and $v_2$ of $G_Q(D)$ are connected by an edge of
$G_Q(D)$, we merge these vertices along this edge to obtain a
graph $G^1$ with $n-1$ vertices, with total degree of vertices
equal $3n-2$, and with the same number of edges having end points
on $\partial D$ as for the graph $G_Q(D)$. This merging procedure
can be done $n-1$ times until we obtain a graph $G^{n-1}$ with a
single vertex of degree $n+2$ such that every edge of $G^{n-1}$
has its end point on $\partial D$. Since the number of the end
points on $\partial D$ for the graph $G^{n-1}$ is the same as for
the graph $G_Q(D)$, the required result is proved.
 \hfill  $\Box$



\bigskip


 Next, we present a list of basic properties of the quadratic differential $Q_0(z)\,dz^2$ defined in equation (\ref{s6aa}),
which always hold when all its coefficients $c_k$ are real.

\begin{enumerate} %
    \item[(1)] The quadratic differential $Q_0(z)\,dz^2$ has at most three
    distinct poles and therefore it follows from Jenkins's \emph{Three
        poles lemma} that the domain configuration of $Q_0(z)\,dz^2$ does not
    contain density domains~\cite{Jenkins}.

    \item[(2)] Since all the coefficients of the rational function
    $Q_0(z)$ are real, the complex zeros of $Q_0(z)\,dz^2$ are
    in conjugate pairs, the number of real zeros (counting multiplicity) is even, and the
    critical graph and domain configuration of $Q_0(z)\,dz^2$ are
    symmetric with respect to the real line.

    \item[(3)] Since $Q_0(z)\,dz^2=-(1+o(1))\,dz^2$ as $z\to \infty$, it
    follows that $Q_0(z)\,dz^2$ has a pole of order four at $z=\infty$
    with two critical directions defined by condition
    $-1\cdot\,dz^2>0$. Thus, $d_1=i$ and $d_2=-i$ are the critical
    directions of $Q_0(z)\,dz^2$ at $z=\infty$. Furthermore, the domain
    configuration of $Q_0(z)\,dz^2$ always includes exactly two end
    domains, the ``left domain'' $\Omega^l_e$ and the ``right domain''
    $\Omega^r_e$, such that $\Omega^l_e\supset (-\infty,-a)$ and
    $\Omega^r_e\supset (a,+\infty)$, for all $a>0$ big enough.
    This,
    together with the symmetry property, imply that, if $a>0$ is big
    enough, then  the intervals $(-\infty,-a)$ and $(a,+\infty)$ lie
    on orthogonal trajectories of $Q_0(z)\,dz^2$.

    \item[(4)] Let $e_k$, $k=1,2,3,4$, denote the zeros of the numerator $P_0(z)$ of the quadratic differential (\ref{s6aa}).
    In the case $e_k\not= \pm 1$, $k=1,2,3,4$, the quadratic differential
    $Q_0(z)\,dz^2$ has two second order poles and therefore it may have
    at most two circle domains centered at the poles $z=-1$ and
    $z=1$. If such circle domains exist, we denote them as $\Omega_c(-1)\ni -1$  and $\Omega_c(1)\ni 1$.
    Furthermore, $Q_0(z)\,dz^2$ may have at most one ring domain
    $\Omega_r$, which, if exists, must separate the poles $z=-1$
    and $z=1$ from the pole  $z=\infty$.

    \end{enumerate}

    \medskip

        For the long classification of possible domain configurations
        of the quadratic differential~(\ref{s6aa}) presented in the next section,
        it is convenient to introduce necessary notations and fix some terminology.

Everywhere below, $\gamma_{a,b}$ stands for a critical trajectory,
including its end points, which starts at $a$ and ends at $b$.
Thus, $\gamma_{b,a}$ is the same critical trajectory as
$\gamma_{a,b}$ but with opposite orientation. If a critical
trajectory $\gamma_{a,b}$ is symmetric with respect to the real
axis, then we add superscripts ``l'', ``c'', and ``r'', like
$\gamma_{a,b}^l$, $\gamma_{a,b}^c$, $\gamma_{a,b}^r$,  to indicate
when $\gamma_{a,b}$ crosses the real axis to the left of the pole
$z=-1$, in between the poles $z=-1$ and $z=1$, or to the right of
the pole $z=1$. In the case, when $a=b$, we use a shorter notation
$\gamma_a^l=\gamma_{a,a}^l$, etc, assuming counter clock-wise
orientation of $\gamma_a^l$, etc. An additional superscript
``$-$'', like $\gamma_a^{l-}$, etc, will be used to indicate that
this critical trajectory is oriented clock-wise in the case under
consideration.

By $\Gamma_e^l$, $\Gamma_e^r$, $\Gamma_c(-1)$ and $\Gamma_c(1)$,
we denote the boundaries of the domains $\Omega_e^l$,
$\Omega_e^r$, $\Omega_c(-1)$ and $\Omega_c(1)$ assuming positive
orientation of these boundaries, with respect to corresponding
domains. Also, $\Gamma_r^{(out)}$ and $\Gamma_r^{(inn)}$ will
denote the outer and inner boundary components of the ring domain
$\Omega_r$, where we assume that $\Gamma_r^{(out)}$ is oriented in
the positive direction with respect to $\Omega_r$ and
$\Gamma_r^{(inn)}$ is oriented in the negative direction with
respect to $\Omega_r$.

By $\Omega_s(a,b)$ we denote a strip domain with vertices $a$ and
$b$. This notation with $a=-i\infty$ and/or $b=i\infty$ means that
$\Omega_s(a,b)$ approaches its vertex $a=\infty$ along the
negative direction of the imaginary axis and it approaches its
vertex $b=\infty$ along the positive direction of the imaginary
axis. When $a=-i\infty$, and/or $b=i\infty$, by $\Gamma_s^l(a,b)$
and $\Gamma_s^r(a,b)$ we denote the left and right sides of
$\Omega_s(a,b)$, respectively. In the case when $a=b$, which can
happen when $a=\pm 1$, by $\Gamma_s^{(out)}(a,b)$ and
$\Gamma_s^{(inn)}(a,b)$ we denote the outer and inner sides of
$\Omega_s(a,b)$, respectively. Furthermore, $\Gamma_s^+(-1,1)$ and
$\Gamma_s^-(-1,1)$ will denote the sides of $\Omega_s(-1,1)$ lying
in the upper half-plane and in the lower half-plane, respectively.

\smallskip

      To characterize the behavior of $Q_0(z)\,dz^2$ near its poles $z=\pm 1$, we introduce the following notations.    Let
    \begin{equation*} \label{q1.01} %
        \alpha_{-1}=-\frac{1}{4}(1+c_3+c_2+c_1+c_0), \quad \quad
        \alpha_1=-\frac{1}{4}(1-c_3+c_2-c_1+c_0). %
    \end{equation*} %
    If $\alpha_k\not=0$, $k=-1,1$, then $\alpha_k$ is the leading coefficient of
    the Laurent expansion of $Q_0(z)$ at the pole $z=k$.  Therefore,  $Q_0(z)\,dz^2$ has second order pole at $z=k$
    if $\alpha_k\not=0$. Since $c_k$, $k=0,1,2,3$, are real, it follows that, if
    $\alpha_k\not=0$, then there is a trajectory or orthogonal trajectory
    of $Q_0(z)\,dz^2$ surrounding
    the point $z=k$, which $Q_0$-length 
    will be
    denoted by $\delta_k>0$; i.e. %
    \begin{equation} \label{q1.02} %
        \delta_k= \left| \int_{|z-k|=\epsilon}
        \sqrt{Q_0(z)}\,dz\right|=2\pi\,|\alpha_k|^{1/2}, \quad k=-1,1, %
    \end{equation} %
    where $\epsilon>0$ is small enough.

\smallskip

We already mentioned that all the domain configurations of the
quadratic differential $Q_0(z)\,dz^2$ with real coefficients are
symmetric with respect to the real axis. Also, in many cases, we
will have pairs of domain configurations, which are symmetric to
each other with respect to the imaginary axis. This happens, for
instance, when positions of zeros  in one configuration are
symmetric with respect to the imaginary axis to positions of zeros
in the other configuration. In cases like this, we will describe
with details one of these configurations and then mention that the
other one is the \emph{mirror configuration} of the configuration
described above.



        \section{Stokes graphs and domain configurations of $Q_0(z)\,dz^2$}\label{sec:RealZeros}


        In this section, we assume that the coefficients $c_k$, $k=0,1,2,3$,
        of the numerator $P_0(z)$ of the quadratic differential $Q_0(z)\,dz^2$
        are real and that $Q_0(z)\,dz^2$ has full set of critical points. Under these assumptions,
     $P_0(z)$ has four zeros (counting
        multiplicity), which are either real, not equal to $\pm 1$, or in conjugate pairs.
        Below, we describe how the Stokes graphs and domain configurations of $Q_0(z)\,dz^2$ depend on the
        positions of these zeros. 
        In particular,
        how they depend on the number of real zeros of $P_0(z)$.
        Thus, we will distinguish  between three main cases:
        case I, when there are no  real zeros, case II with two real zeros, and case III with four real zeros.
        In what follows, we describe in details possible domain
        configurations in the \emph{generic cases}; i.e. when the zeros $e_k$, $k=1,2,3,4$, of $P_0(z)$
        are distinct. 
       The remaining
         cases, which we call \emph{degenerate cases}, will be also mentioned but most details will
        be left to the interested reader.


\bigskip

        \textbf{I.} Suppose that there are no real zeros. Let
        $e_1=\alpha_1+i\beta_1$ and $e_2=\alpha_2+i\beta_2$, with
        $\beta_1>0$, and $\beta_2>0$, be zeros of $Q_0(z)\,dz^2$. Then their
        complex conjugates $e_3=\overline{e_1}$ and $e_4=\overline{e_2}$
        are also zeros of $Q_0(z)\,dz^2$.
        In this case, the intervals $(-\infty,-1)$,
        $(-1,1)$, and $(1,\infty)$ are orthogonal trajectories of
        $Q_0(z)\,dz^2$. This implies that $Q_0(z)\,dz^2$ has two circle
        domains $\Omega_c(-1)$ and $\Omega_c(1)$ with respective boundaries $\Gamma_c(-1)$ and $\Gamma_c(1)$,
        each of which contains at least one of the pairs $\{e_1,e_3\}$ and
        $\{e_2,e_4\}$ of the zeros of $Q_0(z)\,dz^2$. Changing numeration, if necessary, we may assume
        that $e_1,e_3\in \Gamma_c(-1)$.

        Depending on the positions of $e_1$ and $e_2$ on the boundaries, we have the
        following subcases.

\medskip

            1. Suppose that $e_1\in \Gamma_c(-1)$ and $e_2\not\in
            \Gamma_c(-1)\cup \Gamma_c(1)$. Then we also must have $e_1\in \Gamma_c(1)$.
            This happens if and only if the following
            inequalities hold:
            \begin{equation} \label{eqI(1)1}
                \lim_{\varepsilon\to +0}\left(\im [-1+\varepsilon,e_1]_{Q_0} -\im
                [-1+\varepsilon,e_2]_{Q_0}\right)<0,
            \end{equation} 
            \begin{equation} \label{eqI(1)2}
                \lim_{\varepsilon\to +0}\left(\im [1+\varepsilon,e_1]_{Q_0} -\im
                [1+\varepsilon,e_2]_{Q_0}\right)<0.
            \end{equation} 
            Roughly speaking, inequalities~\eqref{eqI(1)1} and~\eqref{eqI(1)2}  mean that  the zeros $e_1$ and $e_2$ are not connected by critical trajectories
            having end points at the finite critical points of $Q_0(z)\,dz^2$ and that the zero $e_1$ is closer,
            in terms of the $Q_0$-metric, to the poles $\pm 1$
            than the zero $e_2$.

            Since $e_2\not\in  \Gamma_c(-1)\cup
            \Gamma_c(1)$, it follows that each of the boundaries
            $\Gamma_c(-1)$ and $\Gamma_c(1)$ consists of two
            critical trajectories joining $e_1$ and $e_3$.
            Precisely,
            $\Gamma_c(-1)=\gamma_{e_1,e_3}^l\cup\gamma_{e_3,e_1}^c$,
            where $\gamma_{e_1,e_3}^l$ intersects the interval
            $(-\infty,-1)$ and $\gamma_{e_3,e_1}^c$ intersects the
            interval $(-1,1)$ and $\Gamma_c(1)=\gamma_{e_1,e_3}^c\cup\gamma_{e_3,e_1}^r$,
            where $\gamma_{e_1,e_3}^c$ coincides up to orientation with $\gamma_{e_3,e_1}^c$, and $\gamma_{e_3,e_1}^r$ intersects the
            interval $(1,\infty)$.

            It follows also that the set $\Gamma_r^{(inn)}=\gamma_{e_1,e_3}^l\cup
            \gamma_{e_3,e_1}^r$ is a closed Jordan curve that is a boundary component of one of
            the faces of $Q_0(z)\,dz^2$. By part 2 of Lemma~3.1, in the case under consideration, this face must
            be a ring domain $\Omega_r$. Furthermore, the second
            boundary component $\Gamma_r^{(out)}$ of the ring domain $\Omega_r$
            must consist of two critical trajectories joining
            $e_2$ and $e_4$. Thus, $\Gamma_r^{(out)}=\gamma_{e_2,e_4}^l\cup \gamma_{e_4,e_2}^r$,  where $\gamma_{e_2,e_4}^l$ intersects the interval
            $(-\infty,-1)$ and $\gamma_{e_4,e_2}^r$ intersects the
            interval $(1,\infty)$.

            The two remaining critical
            trajectories $\gamma_{e_2,i\infty}\subset \mathbb{H}_+$  and
            $\gamma_{-i\infty,e_4}\subset \mathbb{H}_-$ are arcs on the boundaries $\Gamma_e^l=\gamma_{-i\infty,e_4}\cup \gamma_{e_4,e_2}^l\cup
            \gamma_{e_2,i\infty}$ and  $\Gamma_e^r=\gamma_{i\infty,e_2}\cup
            \gamma_{e_2,e_4}^r\cup \gamma_{e_4,-i\infty}$ of the end domains $\Omega^l_e$ and $\Omega^r_e$, respectively.
            Figure~\ref{fig:I-1} Case I-1 gives an example of the domain
            configuration.

            \medskip

            2. Suppose that $e_1\in \Gamma_c(-1)$ but  $e_1\not\in \Gamma_c(1)$ and $e_2\in
            \Gamma_c(1)$ but  $e_2\not\in \Gamma_c(-1)$. This position of zeros happens if and only if
            the limit in inequality~\eqref{eqI(1)1} is negative
            as in the case 1,
            but the limit in inequality~\eqref{eqI(1)2} is positive.
 The latter conditions imply that  the zeros $e_1$ and $e_2$ are not connected by critical trajectories
            having end points at the finite critical points of $Q_0(z)\,dz^2$ and that $e_1$ is, in terms of the $Q_0$-metric,
            closer to the pole $-1$ than $e_2$, and $e_2$ is closer to the pole $1$ than~$e_1$.

            Same argument as in part I-1 above, implies that
            there are critical trajectories $\gamma_{e_1,e_3}^c$, crossing the interval $(-1,1)$ at $x_1$, and
            $\gamma_{e_2,e_4}^c$, crossing the interval $(x_1,1)$ such that
            $\Gamma_c(-1)=\gamma_{e_1,e_3}^l\cup \gamma_{e_3,e_1}^c$ is the boundary of the circle domain $\Omega_c(-1)$ and
            $\Gamma_c(1)=\gamma_{e_2,e_4}^c\cup \gamma_{e_4,e_2}^r$ is the boundary of
            $\Omega_c(1)$.

            Since $e_1\in \Gamma_c(-1)\setminus \Gamma_c(1)$ and $e_2\in
            \Gamma_c(1)\setminus \Gamma_c(-1)$, it follows that there is a certain ``gap'' between circle domains $\Omega_c(-1)$ and
            $\Omega_c(1)$, which must be a strip domain
            $\Omega_s(-i\infty,i\infty)$ with the vertices at $-i\infty$ and $i\infty$ and with sides
           $\Gamma_s^l=\gamma_{-i\infty,e_3}\cup
            \gamma_{e_3,e_1}^c\cup \gamma_{e_1,i\infty}$ and  $\Gamma_s^r=\gamma_{-i\infty,e_4}\cup
            \gamma_{e_4,e_2}^c\cup \gamma_{e_2,i\infty}$.

            The boundaries of the end domains $\Omega^l_e$ and $\Omega^r_e$ are
            $\Gamma_e^l=\gamma_{-i\infty,e_3}\cup \gamma_{e_3,e_1}^l\cup
            \gamma_{e_1,i\infty}$ and
            $\Gamma_e^r=\gamma_{i\infty,e_2}\cup
            \gamma_{e_2,e_4}^r\cup \gamma_{e_4,-i\infty}$, respectively.
            Figure~\ref{fig:I-1}  Case I-2  gives an example of the domain
            configuration.

            \medskip

       3.  Suppose now that the boundary of one of the domains
       $\Omega_c(-1)$ and $\Omega_c(1)$ contains both zeros $e_1$
       and $e_2$, but the boundary of the other domain contains
       only one of these zeros. Assume, without loss of
       generality,
       that  $e_1,e_2\in \Gamma_c(-1)$ and $e_2\in
            \Gamma_c(1)$ but  $e_1\not\in \Gamma_c(1)$. This implies that
            $e_1\not=e_2$ and that $e_1$ and $e_2$ are connected by the critical
            trajectory $\gamma_{e_1,e_2}\subset \mathbb{H}_+$ of $Q_0(z)\,dz^2$.  By symmetry,
            $e_3$ and $e_4$ are connected by the critical
            trajectory $\gamma_{e_3,e_4}\subset \mathbb{H}_-$.
            Furthermore, it implies that $\Omega_c(-1)$ has a common
            boundary arc with each of the domains $\Omega_c(1)$,
            $\Omega_e^l$, and $\Omega_e^r$. The latter implies, in
            turn, that $\Omega_c(-1)$, $\Omega_c(1)$,
            $\Omega_e^l$, and $\Omega_e^r$ are the only domains in
            the domain configuration of $Q_0(z)\,dz^2$. This
            configuration occurs if and only if the limit in each of inequalities~\eqref{eqI(1)1} and~\eqref{eqI(1)2} is zero and,
            additionally, the following inequality holds:
           \begin{equation}\label{I.3a} %
                \re\, [b,e_1]_{Q_0}+\re\,
                [a,e_1]_{Q_0}>
                \re\, [b,e_2]_{Q_0}+\re\,
                [a,e_2]_{Q_0},
            \end{equation} %
where the points $-1<a<1$ and $b>1$ are chosen so that the points
$e_1,e_2,a$ and the points $e_1,e_2,b$ do not lie on a straight
line.

              In this
            case, the boundary of the circle domain $\Omega_c(1)$
            is 
            $\Gamma_c(1)=\gamma_{e_2,e_4}^c\cup\gamma_{e_4,e_2}^r$,
            and the boundary of $\Omega_c(-1)$ is
            $\Gamma_c(-1)=\gamma_{e_1,e_3}^l\cup \gamma_{e_3,e_4}\cup
            \gamma_{e_4,e_2}^c\cup\gamma_{e_2,e_1}$.

            The boundaries of the end domains $\Omega_e^l$ and $\Omega_e^l$ are
            $\Gamma_e^l=\gamma_{-i\infty,e_3}\cup \gamma_{e_3,e_1}^l\cup
            \gamma_{e_1,i\infty}$
            and
            $\Gamma_e^r=\gamma_{i\infty,e_1}\cup
            \gamma_{e_1,e_2} \cup \gamma_{e_2,e_4}^r\cup \gamma_{e_4,e_3}\cup
            \gamma_{e_3,-i\infty}$, respectively. See Figure~\ref{fig:I-3} Case I-3.

            In the case when $e_1,e_2\in \Gamma_c(1)$ and $e_2\in
            \Gamma_c(-1)$ but  $e_1\not\in \Gamma_c(-1)$,  the domain configuration
            is the mirror configuration 
            to the configuration described above as it is shown in Figure~\ref{fig:I-3} Case I-3-m.
\smallskip

  The remaining case, when the boundary of each of the  domains
       $\Omega_c(-1)$ and $\Omega_c(1)$ contains both zeros $e_1$
       and $e_2$, is our first degenerate case. Indeed, the argument used earlier in part I-3 shows
       that, if $e_1\not= e_2$, then there is a critical trajectory
       $\gamma_{e_1,e_2}$, which belongs to the boundaries of
       both $\Omega_c(-1)$ and $\Omega_c(1)$. The latter easily
       leads to a contradiction. Hence, this case happens if and
       only if $e_1=e_2$. Thus, $Q_0(z)\,dz^2$ has only two zeros $e_{1,2}$ and $e_{3,4}$ of
       order two each.  The Stokes graph and domain configuration
       for this degenerate case are shown in Figure~\ref{fig:I-3-deg}.


        \bigskip

        \textbf{II.} Suppose that there are two real zeros $e_1$ and $e_2$ and two complex conjugate zeros
        $e_3=\alpha_3+i\beta_3$  with
        $\beta_3>0$ and $e_4=\overline{e_3}$. Below, we classify possible
        domain configurations, first,  depending on positions of the real zeros with respect to the poles $z=-1$ and $z=1$,
        and then by
        positions of zeros on the boundaries of domains present in the domain configurations of $Q_0(z)\,dz^2$.

         In the generic cases discussed in this part and illustrated in Figures 6--24 in Appenix~B,
        we assume that all these zeros are distinct and that $e_k\not= \pm 1$,
        $k=1,2$.

        \medskip

        1. Let $e_1<e_2<-1$. In this case, the interval
            $(e_1,e_2)$ is  a trajectory  of $Q_0(z)\,dz^2$
            and the intervals $(-\infty,e_1)$, $(e_2,-1)$, $(-1,1)$, and
            $(1,\infty)$ are orthogonal trajectories of $Q_0(z)\,dz^2$. This implies that the quadratic differential $Q_0(z)\,dz^2$ has two circle domains
            $\Omega_c(-1)$, $\Omega_c(1)$, two end domains $\Omega_e^l$, $\Omega_e^r$, and, possibly, some other domains.

            One more conclusion, which easily follows from the assumption $e_1<e_2<-1$
            is that none of the points of the interval $(e_1,e_2]$  belongs to the boundary of
            $\Omega_e^l$.
Indeed, if $x_0\in \Gamma_e^l$, $e_1<x_0\le e_2$,  then there is a
Jordan arc $\gamma\subset \Omega_e^l$ symmetric with respect to
the real axis, which has both its end points at $x_0$ and crosses
the real axis at some point $x_1<e_1$. Thus, the zero $e_1$ is the
only critical point of $Q_0(z)\,dz^2$ that is inside the domain
bounded by $\gamma$. Now, the required result follows from part 4
of Lemma~3.1 or, alternatively, it can be proved as follows. Since
the critical trajectories, different from the interval
$(e_1,e_2)$, each of which  has one end point at $e_1$, must have
a second end point at some critical point of $Q_0(z)\,dz^2$, these
trajectories must intersect the curve $\gamma$. Since $\gamma$
lies in the end domain $\Omega_e^l$, the latter is not possible.
 Similar argument shows that none of the points of the interval $[e_1,e_2)$ belongs to the boundary of $\Omega_c(-1)$.

 Thus, each of the boundaries $\Gamma_e^l$ and $\Gamma_c(-1)$ may
 contain $1$, $2$, or $3$ zeros of $Q_0(z)\,dz^2$. Accordingly, we have the following
            subcases.

        \smallskip

(a)     Suppose that $e_1$ is the only zero on the boundary of
$\Omega_e^l$ and $e_2$ is the only zero on the boundary of
$\Omega_c(-1)$. In this case, the boundary of the end domain
$\Omega^l_e$ is
$\Gamma^l_e=\gamma_{-i\infty,e_1}\cup\gamma_{e_1,i\infty}$ and the
boundary of the circle domain $\Omega_c(-1)$ is
$\Gamma_c(-1)=\gamma_{e_2}^c$, where the critical trajectory
$\gamma_{e_2}^c$ has both its end points at $e_2$ and intersects
the interval $(-1,1)$ at some point $x_1$. The latter also implies
that $e_3,e_4\in \Gamma_c(1)$ but $e_1\not\in \Gamma_c(1)$. Thus,
in terms of the $Q_0$-metric, the zero $e_2$ is closer to the pole
$-1$ than $e_3$ and the zero $e_3$ is closer to the pole $1$ than
$e_2$. This means that the assumptions in this part hold  if and
only if the following inequalities hold:
\begin{equation} \label{eqII-1(a)-1} 
                \lim_{\varepsilon\to +0}\left(\im [-1-\varepsilon,e_2]_{Q_0}
                -\im [-1-\varepsilon,e_3]_{Q_0}\right)<0,
     \end{equation} 
  \begin{equation} \label{eqII-1(a)-2} 
                \lim_{\varepsilon\to +0}\left(\im [1+i\varepsilon,e_2]_{Q_0}
                -\im [1+i\varepsilon,e_3]_{Q_0}\right)>0,
     \end{equation} 

Since $e_3,e_4$ are the only zeros on $\Gamma_c(1)$ it follows
that $\Gamma_c(1)=\gamma_{e_3,e_4}^c\cup \gamma_{e_4,e_3}^r$,
where $\gamma_{e_3,e_4}^c$ intersects $(-1,1)$ at some point
$x_2$, $x_1<x_2<1$. The remaining critical trajectories are
$\gamma_{e_3,i\infty}$ and $\gamma_{e_4,-i\infty}$. Under these
circumstances, there is one more face of the Stokes graph of
$Q_0(z)\,dz^2$ that is a strip domain $\Omega_s(-i\infty,i\infty)$
with sides $\Gamma_s^l(-i\infty,i\infty)=\gamma_{-i\infty,e_1}\cup
[e_1,e_2]\cup \gamma_{e_2}^c\cup[e_2,e_1]\cup\gamma_{e_1,i\infty}$
and $\Gamma_s^r(-i\infty,i\infty)=\gamma_{-i\infty,e_4}\cup
\gamma_{e_4,e_3}^c\cup\gamma_{e_3,i\infty}$. Figure~\ref{fig:II-1-a} Case II-1-a
gives an
                example of the domain configuration.

The mirror configuration, shown in Figure~\ref{fig:II-1-a} Case
II-1-a-m, occurs when $1<e_2<e_1$, $e_1$ is the only zero on
$\Gamma_e^r$, and $e_2$ is the only zero on $\Gamma_c(1)$. This
case happens if and only if
\begin{equation} \label{eqII-1(a)-1-m} 
                \lim_{\varepsilon\to +0}\left(\im [1+\varepsilon,e_2]_{Q_0}
                -\im [1+\varepsilon,e_3]_{Q_0}\right)<0,
     \end{equation} 
  \begin{equation} \label{eqII-1(a)-2-m} 
                \lim_{\varepsilon\to +0}\left(\im [-1+i\varepsilon,e_2]_{Q_0}
                -\im [-1+i\varepsilon,e_3]_{Q_0}\right)>0.
     \end{equation} %

      \smallskip


                (b) Suppose that $e_1$ is the only zero on $\Gamma_e^l$ and that $e_3,e_4\in \Gamma_c(-1)$ but
                $e_1,e_2\not\in \Gamma_c(-1)$. This assumption
                holds if and only if the limits in inequalities~\eqref{eqII-1(a)-1} and~\eqref{eqII-1(a)-2} are positive.
                It is immediate from the latter assumption that $\Gamma_e^l=\gamma_{-i\infty,e_1}\cup \gamma_{e_1,i\infty}$,
                $\Gamma_c(-1)=\gamma_{e_3,e_4}^l\cup \gamma_{e_4,e_3}^c$
                and $\Gamma_c(1)=\gamma_{e_3,e_4}^c\cup
                \gamma_{e_4,e_3}^r$.

                Furthermore, the set $\Gamma_r^{(inn)}=\gamma_{e_3,e_4}^l\cup
                \gamma_{e_4,e_3}^r$ is a boundary component of a
                face of the Stokes graph of $Q_0(z)\,dz^2$, which
                in this case must be a ring domain $\Omega_r$ of
                $Q_0(z)\,dz^2$, see Lemma~3.1. Under these circumstances, the
                only possibility for the outer boundary component
                of $\Omega_r$ is that $\Gamma_r^{(out)}=\gamma_{e_2}^r$, where $\gamma_{e_2}^r$
                has both end points at $e_2$ and crosses the interval $(1,\infty)$.

                The boundary of the end domain $\Omega_e^r$ is
                $\Gamma_e^r=\gamma_{i\infty,e_1}\cup [e_1,e_2]\cup
                \gamma_{e_2}^r\cup[e_2,e_1]\cup \gamma_{e_1,-i\infty}$.
                Figure~\ref{fig:II-1-b} Case II-1-b gives an example of the domain
                configuration.

                The mirror configuration shown in
Figure~\ref{fig:II-1-b} Case II-1-b-m, occurs when $1<e_2<e_1$,
$e_1$ is the only zero on $\Gamma_e^r$,  $e_3,e_4\in \Gamma_c(1)$
but
                $e_1,e_2\not\in \Gamma_c(1)$. This case happens if
and only if the limits in inequalities~\eqref{eqII-1(a)-1-m} and~\eqref{eqII-1(a)-2-m} are positive.

                 \smallskip

               (c) Let $e_1$ be the only zero on $\Gamma_e^l$ and $e_2,e_3,e_4\in \Gamma_c(-1)$.
               As we have mentioned earlier, $e_1\not \in \Gamma_c(-1)$.
                This happens if and only if the limits in inequalities~\eqref{eqII-1(a)-1} and~\eqref{eqII-1(a)-2} are
                zero. Since $e_2,e_3,e_4\in \Gamma_c(-1)$, it
                follows that there are critical trajectories
                $\gamma_{e_2,e_3}\subset \mathbb{H}_+$ and $\gamma_{e_2,e_4}\subset \mathbb{H}_-$,
                which belong to the boundary of $\Omega_c(-1)$.
                Therefore in this case,
                $\Gamma_c(-1)=\gamma_{e_2,e_4}\cup
                \gamma_{e_4,e_3}^c\cup \gamma_{e_3,e_2}$ and
                $\Gamma_c(1)=\gamma_{e_3,e_4}^c\cup\gamma_{e_4,e_3}^r$.
                Under these circumstances, the only possibilities
                for the boundaries of the end domains are the
                following:
                $\Gamma_e^l=\gamma_{-i\infty,e_1}\cup\gamma_{e_1,i\infty}$
                and $\Gamma_e^r=\gamma_{i\infty,e_1}\cup[e_1,e_2]\cup\gamma_{e_2e_3}\cup
                \gamma_{e_3,e_4}^r\cup \gamma_{e_4,e_2}\cup[e_2,e_1]\cup
                \gamma_{e_1,-i\infty}$. This also implies that $\Omega_c(-1)$, $\Omega_c(1)$, $\Omega_e^l$, and
$\Omega_e^r$ are the only domains in the domain configuration of
$Q_0(z)\,dz^2$. The case is illustrated in Figure~\ref{fig:II-1-c} Case II-1-c.

The mirror configuration,  shown in Figure~\ref{fig:II-1-c} Case
II-1-c-m, occurs when $1<e_2<e_1$, $e_1$ is the only zero on
$\Gamma_e^r$,  $e_2,e_3,e_4\in \Gamma_c(1)$.
 This case happens if
and only if the limits in inequalities~\eqref{eqII-1(a)-1-m} and~
\eqref{eqII-1(a)-2-m} are zero.

\smallskip


                (d) Suppose that $\Gamma_e^l$ contains two zeros, which in this case are $e_3$ and
                $e_4$. In terms of the $Q_0$-metric, the latter means that the zeros $e_1$ and
                $e_2$ are closer to the poles $z=\pm 1$ than the
                zeros $e_3$ and $e_4$.
                This assumption holds if and only if the limits in
inequalities~\eqref{eqII-1(a)-1} and~\eqref{eqII-1(a)-2} are negative.
In this case, there are critical trajectories
$\gamma_{e_3,e_4}^l$, crossing the interval
                $(-\infty,e_1)$, and $\gamma_{e_3,e_4}^r$, crossing the interval
                $(1,\infty)$ at some point $x_2$, and therefore we also have that
                $e_3,e_4\in \Gamma_e^r$, but $e_1,e_2\not\in
                \Gamma_e^r$. The boundaries of the end domains $\Omega_e^l$ and $\Omega_e^r$ are
                $\Gamma_e^l=\gamma_{-i\infty,e_4}\cup \gamma_{e_4,e_3}^l\cup\gamma_{e_3,i\infty}$
                and
                $\Gamma_e^r=\gamma_{i\infty,e_3}\cup \gamma_{e_3,e_4}^r\cup
                \gamma_{e_4,-i\infty}$.
It follows also that the set
$\Gamma_r^{(out)}=\gamma_{e_3,e_4}^l\cup\gamma_{e_4,e_3}^r$ is a
boundary component of one of
            the faces of $Q_0(z)\,dz^2$. Since the interior of $\Gamma_r^{(out)}$
            contains more than one critical points, by Lemma~3.1, this face must be a ring domain  $\Omega_r$.
            The latter implies that  there is a critical trajectory $\gamma_{e_1}^r$ having both end points at $e_1$, which crosses the
                interval $(1,\infty)$ at some point $x_1$, $1<x_1<x_2$. Then,
                $\Gamma_r^{(inn)}=\gamma_{e_1}^r$.

                Under these conditions, we must have one more
                critical trajectory $\gamma_{e_2}^c$, which has
                both its end points at $e_2$ and crosses the
                interval $(-1,1)$. Therefore, the boundary of the
                circle domain $\Omega_c(-1)$ is
                $\Gamma_c(-1)=\gamma_{e_2}^c$ and the boundary of the
                circle domain $\Omega_c(1)$ is
                $\Gamma_c(1)=\gamma_{e_1}^r\cup [e_1,e_2]\cup \gamma_{e_2}^c\cup[e_2,e_1]$.
                Figure~\ref{fig:II-1-d} Case II-1-d gives an example of the domain
                configuration described above.

The mirror configuration, shown in Figure~\ref{fig:II-1-d} Case
II-1-d-m, occurs when $1<e_2<e_1$, and when $e_3$ and $e_4$ are
the only zero on $\Gamma_e^r$.
 This case happens if
and only if the limits in~\eqref{eqII-1(a)-1-m} and
~\eqref{eqII-1(a)-2-m} are negative.

                \smallskip

            (e) Suppose now that $e_1,e_3,e_4\in \Gamma_e^l$.
            This case happens if and only if
the limits in~\eqref{eqII-1(a)-1} and~\eqref{eqII-1(a)-2} are
zero. Then the boundary of the end domain $\Omega_e^l$ must
contain critical trajectories $\gamma_{e_1,e_3}$ and
$\gamma_{e_1,e_4}$ and therefore
$\Gamma_e^l=\gamma_{-i\infty,e_4}\cup \gamma_{e_4,e_1}\cap
\gamma_{e_1,e_3}\cup\gamma_{e_3,i\infty}$. The latter implies, in
turn, that
$\Gamma_e^r=\gamma_{i\infty,e_3}\cup\gamma_{e_3,e_4}^r\cup
\gamma_{e_4,-i\infty}$. Under these circumstances the remaining
possibility is that there is a critical trajectory
$\gamma_{e_2}^c$, which has both its end points at $e_2$ and
crosses the interval $(-1,1)$. The latter implies that
$\Gamma_c(-1)=\gamma_{e_2}^c$,
$\Gamma_c(1)=\gamma_{e_1,e_4}\cup\gamma_{e_4,e_3}^r\cup\gamma_{e_3,e_1}\cup
[e_1,e_2]\cup \gamma_{e_1}^c\cup[e_2,e_1]$ and that
$\Omega_c(-1)$, $\Omega_c(1)$, $\Omega_e^l$, and $\Omega_e^r$ are
the only domains in the domain configuration of $Q_0(z)\,dz^2$ in
the case under consideration.
Figure~\ref{fig:II-1-e} Case II-1-e illustrates the domain
                configuration.

The mirror configuration, shown in Figure~\ref{fig:II-1-e} Case
II-1-e-m, occurs when $1<e_2<e_1$, and when
$e_1,e_3,e_4\in\Gamma_e^r$.
 This case happens if
and only if the limits in~\eqref{eqII-1(a)-1-m} and~\eqref{eqII-1(a)-2-m} are zero.

\medskip

2. Let $-1<e_1<e_2<1$. Then, the interval
            $(e_1,e_2)$ is  a trajectory  of $Q_0(z)\,dz^2$
            and the intervals $(-\infty,-1)$, $(-1,e_1)$, $(e_2,1)$, and
            $(1,\infty)$ are orthogonal trajectories of $Q_0(z)\,dz^2$.
            As in the case~1 above, this implies that the quadratic differential $Q_0(z)\,dz^2$ has two circle domains
            $\Omega_c(-1)$, $\Omega_c(1)$, two end domains $\Omega_e^l$, $\Omega_e^r$, and, possibly, some other domains.
            Also, it is not difficult to see that either $[e_1,e_2]\subset \Gamma_e^l$ or
            $[e_1,e_2]\cap \Gamma_e^l=\emptyset$. Similarly, we have that either $[e_1,e_2]\subset \Gamma_e^r$ or
            $[e_1,e_2]\cap \Gamma_e^r=\emptyset$.

 The latter implies that each of the boundaries $\Gamma_e^l$ and $\Gamma_e^r$ may
 contain $2$ or $4$ zeros of $Q_0(z)\,dz^2$. Thus, we have the following
            subcases.

           \smallskip

           (a) Suppose that $e_3,e_4$ are the only zeros on each
           of the boundaries $\Gamma_e^l$ and $\Gamma_e^r$.     
                This means that the zeros $e_1$ and $e_2$ are closer, in terms of the $Q_0$-metric, to the poles $z=\pm 1$ than the zeros $e_3$ and
                $e_4$. The latter happens if and only the following
                inequalities hold:
               \begin{equation} \label{eqII-2(a)-1} %
                               \lim_{\varepsilon\to +0}\left(\im [-1+\varepsilon,e_1]_{Q_0}
                -\im [-1+\varepsilon,e_3]_{Q_0}\right)<0
                \end{equation} 
                \begin{equation} \label{eqII-2(a)-2} 
                \lim_{\varepsilon\to +0}\left(\im [1-\varepsilon,e_2]_{Q_0}
                -\im [1-\varepsilon,e_3]_{Q_0}\right)<0.
                \end{equation} 
 Since $e_3,e_4$ are the only zeros on $\Gamma_e^l$ and on $\Gamma_e^r$, it follows that there are critical
 trajectories $\gamma_{e_3,e_4}^l$ that crosses the interval
                $(-\infty,-1)$ at some point $x_1$ and
                $\gamma_{e_3,e_4}^r$ crossing the interval
                $(1,\infty)$ an some point $x_2$. The boundaries of the end domains $\Omega_e^l$ and $\Omega_e^r$  are
                $\Gamma_e^l=\gamma_{-i\infty,e_4}\cup \gamma_{e_4,e_3}^l\cup\gamma_{e_3,i\infty}$
                and
                $\Gamma_e^r=\gamma_{i\infty,e_3}\cup \gamma_{e_3,e_4}^r\cup
                \gamma_{e_4,-i\infty}$.

                The set $\Gamma_r^{(out)}=\gamma_{e_3,e_4}^l\cup
                \gamma_{e_4,e_3}^r$ is a Jordan curve that is an
                outer
                boundary component of a face of the Stokes graph of
                $Q_0(z)\,dz^2$, which in this case must be a ring
                domain $\Omega_r$ by Lemma~3.1. Under these circumstances, there are critical trajectories $\gamma_{e_1}^l$ that crosses the
                interval $(x_1,-1)$  and $\gamma_{e_2}^r$ crossing
                the interval $(1,x_2)$. Then the
                inner boundary component of $\Omega_r$ is
                $\Gamma^{(inn)}=\gamma_{e_1}^l\cup [e_1,e_2]\cup\gamma_{e_2}^r\cup[e_2,e_1]$.
  The boundaries of the circle domains $\Omega_c(-1)$ and $\Omega_c(1)$ are $\Gamma_c(-1)=\gamma_{e_1}^l$ and
                $\Gamma_c(1)=\gamma_{e_2}^r$.
                Figure~\ref{fig:II-2-a} Case II-2-a gives an example of the domain
                configuration.

\smallskip

(b) Suppose that $e_3,e_4$ are the only zeros on the boundary $\Gamma_e^l$ and $e_1,e_2$ are the only zeros on the boundary $\Gamma_e^r$.
  The latter happens if and only the limit in~\eqref{eqII-2(a)-1} is positive and the limit in~\eqref{eqII-2(a)-2}
  is negative. The boundary of the
                end domain $\Omega_e^l$ is 
                $\Gamma_e^l=\gamma_{-i\infty,e_4}\cup\gamma_{e_4,e_3}^l\cup
                \gamma_{e_3,i\infty}$. Under these assumptions, the domain configuration must contain
                a strip domain $\Omega_s(-i\infty,i\infty)$ with the vertices at $\pm i\infty$,
  which separates the end domains $\Omega_e^l$ and $\Omega_e^r$.   It has sides
                $\Gamma_s^l(-i \infty,i\infty)=\gamma_{-i\infty,e_4}\cup\gamma_{e_4,e_3}^c\cup\gamma_{e_3,i\infty}$
                and
                $\Gamma_s^r(-i\infty,i\infty)=\gamma_{-i\infty,e_1}\cup\gamma_{e_1,i\infty}$.

                The circle domains have the boundaries
                $\Gamma_c(-1)=\gamma_{e_3,e_4}^l\cup \gamma_{e_4,e_3}^c$,
                $\Gamma_c(1)=\gamma_{e_2}^r$.

                In this case, the boundary of the
                end domain $\Omega_e^r$ is $\Gamma_e^r=\gamma_{i\infty,e_1}\cup [e_1,e_2]\cup
                \gamma_{e_2}^r\cup[e_2,e_1]\cup \gamma_{e_1,-i\infty}$.
                Figure~\ref{fig:II-2-b} Case II-2-b gives an example of such a domain
                configuration.



The mirror configuration, shown in Figure~\ref{fig:II-2-b} Case
II-2-b-m, occurs when $-1<e_2<e_1<1$, and when  $e_3,e_4$ are the
only zeros on the boundary $\Gamma_e^r$ and $e_1,e_2$ are the only
zeros on the boundary $\Gamma_e^l$. This case happens if and only
if the limit in~\eqref{eqII-2(a)-1} is negative and the limit
in~\eqref{eqII-2(a)-2} is positive.


\smallskip

(c) Suppose that $e_3,e_4$ are the only zeros on $\Gamma_e^l$ and
all zeros $e_1,e_2,e_3,e_4$ belong to the boundary $\Gamma_e^r$.
This happens  if and only if the limit in~\eqref{eqII-2(a)-1} is zero (then the limit in~\eqref{eqII-2(a)-2}
is also zero) and
$\re [e_1,e_3]_{Q_0}<\re [e_2,e_3]_{Q_0}$. %
 The boundary of the end domain $\Omega_e^l$ is the same as in
the cases (a) and (b) above,
$\Gamma_e^l=\gamma_{-i\infty,e_4}\cup\gamma_{e_4,e_3}^l\cup
                \gamma_{e_3,i\infty}$. Since $\Gamma_e^r$ contains
                all zeros, it follows that there are critical
                trajectories $\gamma_{e_1,e_3}\subset \mathbb{H}_+$ and
                $\gamma_{e_1,e_4}\subset \mathbb{H}_-$. In this case,
                $\Gamma_e^r=\gamma_{i\infty,e_3}\cup
                \gamma_{e_3,e_1}\cup [e_1,e_2]\cup
                \gamma_{e_2}^r\cup[e_2,e_1]\cup\gamma_{e_1,e_4}\cup
                \gamma_{e_4,-i\infty}$. The boundary of the circle
                domain $\Omega_c(-1)$ is
                $\Gamma_c(-1)=\gamma_{e_3,e_4}^l\cup
                \gamma_{e_4,e_1}\cup \gamma_{e_1,e_3}$ and  the boundary of the circle
                domain $\Omega_c(1)$ is
                $\Gamma_c(1)=\gamma_{e_2}^r$.  Figure~\ref{fig:II-2-c} Case II-2-c shows the Stokes graph and
                domain configuration.



The mirror configuration, shown in Figure~\ref{fig:II-2-c} Case
II-2-c-m, occurs when $-1<e_2<e_1<1$, and when  $e_3,e_4$ are the
only zeros on the boundary $\Gamma_e^r$ and all zeros
$e_1,e_2,e_3,e_4$ belong to the boundary $\Gamma_e^l$. This case
happens if and only if the limits in~\eqref{eqII-2(a)-1}
and~\ref{eqII-2(a)-2} are zero and
$\re [e_1,e_3]_{Q_0}<\re [e_2,e_3]_{Q_0}$. %

\medskip

3. Let  $e_1<-1< e_2<1$. The intervals
            $(-\infty,e_1)$, $(e_2,1)$ and $(1,\infty)$ are orthogonal
            trajectories of $Q_0(z)\,dz^2$ and the intervals $(e_1,-1)$,
            $(-1,e_2)$ are trajectories of $Q_0(z)\,dz^2$. Therefore, there is
            only one circle domain $\Omega_c(1)$ and there is at least one
            strip domain having one or both its vertices at $z=-1$.
             Notice that the intervals $[e_1,-1]$ and $[-1,e_2]$ can not lie on the boundary of the end domain $\Omega_e^r$.
            The latter implies that $e_3,e_4$ are the only zeros of $Q_0(z)\,dz^2$, which belong to $\Gamma_e^r$.
            Therefore, $\Gamma_e^r=\gamma_{i\infty,e_3}\cup \gamma_{e_3,e_4}^r\cup \gamma_{e_4,-i\infty}$ in all subcases considered below.
Also, as in the previous cases, it is not difficult to see that
            $e_2$ can not belong to the boundary of the end domain
            $\Omega_e^l$.
             Thus, $\Gamma_e^l$ may contain $1$, $2$,
            or $3$ zeros of $Q_0(z)\,dz^2$. Accordingly, we have
            the following subcases.

\smallskip

(a) Let $e_1$ be the only zero on $\Gamma_e^l$. Then
$\Gamma^l_e=\gamma_{-i\infty,e_1}\cup \gamma_{e_1,i\infty}$. This
together with part 3 of Lemma~3.1 imply that there are strip
domains $\Omega_s(-1,i\infty)$ and $\Omega_s(-i\infty,-1)$,
symmetric to each other with respect to the real axis, which left
sides are $\Gamma_s^l(-1,i\infty)=[-1,e_1]\cup
\gamma_{e_1,i\infty}$ and
$\Gamma_s^l(-i\infty,-1)=\gamma_{-i\infty,e_1}\cup [e_1,-1]$,
respectively. The right side of each of the strip domains
$\Omega_s(-1,i\infty)$ and $\Omega_s(-1,-i\infty)$ must contain at
least one zero of $Q_0(z)\,dz^2$. Thus, we have three
possibilities.

\smallskip

($\alpha$) Let $e_3$ be the only zero on $\Gamma_s^r(-1,i\infty)$.
Then, by symmetry, $e_4$ is the only zero on
$\Gamma_s^r(-i\infty,-1)$. Under the assumptions of this case, the
latter happens if and only if the following holds: 
\begin{equation}\label{eqII3-a-alpha} %
2\im [e_1,e_3]_{Q_0}+ 2\im [e_2,e_3]_{Q_0} = \delta_{-1},
\end{equation} %
where $\delta_{-1}$ is defined in~\eqref{q1.02}.  Under these
conditions, there are critical trajectories
$\gamma_{-1,e_3}\subset \mathbb{H}_+$, which joins the pole $-1$
and $e_3$, and  $\gamma_{-1,e_4}\subset \mathbb{H}_-$, which joins
$-1$ and $e_4$. Therefore,
$\Gamma_s^r(-1,i\infty)=\gamma_{-1,e_3}\cup \gamma_{e_3,i\infty}$
and $\Gamma_s^r(-i\infty,-1)=
\gamma_{-i\infty,e_4}\cup\gamma_{e_4,-1}$.

Under these circumstances, the remaining critical trajectory is
$\gamma_{e_2}^r$ and the boundary of the circle domain
$\Omega_c(1)$ is $\Gamma_c(1)=\gamma_{e_2}^r$.

It follows from our discussion above that there is one more face
of the Stokes graph of $Q_0(z)\,dz^2$, which in this case must be
a strip domain $\Omega_s(-1-1)$ with both its vertices at $-1$
having $\Gamma_s^{(out)}(-1,-1)=\gamma_{-1,e_4}\cup
\gamma_{e_4,e_3}^r\cup \gamma_{e_3,-1}$ as its outer side and
$\Gamma_s^{(inn)}(-1,-1)=[-1,e_2]\cup\gamma_{e_2}^r\cup[e_2,-1]$
as its inner side. Figure~\ref{fig:II-3-a-alpha} Case II-3-a-$\alpha$  shows an
example of the Stokes graph and domain configuration.

The mirror configuration, shown in Figure~\ref{fig:II-3-a-alpha}
Case II-3-a-$\alpha$-m, occurs when $-1<e_2<1<e_1$,
 $e_1$ is the only zero on $\Gamma_e^r$, and $e_3$ is the only zero on $\Gamma_s^l(1,i\infty)$.
 This
case happens if and only if the following holds: 
\begin{equation}
2\im [e_1,e_3]_{Q_0}+ 2\im [e_2,e_3]_{Q_0} = \delta_{1}.
\end{equation} %

\smallskip %

($\beta$) Let $e_2$ be the only zero on each of the sides
$\Gamma_s^r(-1,i\infty)$ and $\Gamma_s^r(-i\infty,-1)$. This case
happens if and only if  %
\begin{equation}\label{eqII3-a-beta} %
2\im[e_1,e_3]_{Q_0}= 2\im [e_2,e_3]_{Q_0}+ \delta_{-1}.
\end{equation} %
Under these assumptions, there are critical trajectories
$\gamma_{e_2,i\infty}$ and $\gamma_{e_2,-i\infty}$ and therefore
$\Gamma_s^r(-1,i\infty)=[-1,e_2]\cup \gamma_{e_2,i\infty}$ and
$\Gamma_s^r(-i\infty,e_2)=\gamma_{-i\infty,e_2}\cup [e_2,-1]$.

Hence, there is one more face of the Stokes
graph of $Q_0(z)\,dz^2$, which is a strip domain
$\Omega_s(-i\infty,i\infty)$ with its vertices at $i\infty$ and
$-i\infty$. The sides of this strip domain are
$\Gamma_s^l(-i\infty,i\infty)=\gamma_{-i\infty,e_2}\cup
\gamma_{e_2,i\infty}$ and
$\Gamma_s^r(-i\infty,i\infty)=\gamma_{-i\infty,e_4}\cup\gamma_{e_4,e_3}^c\cup
\gamma_{e_3,i\infty}$. In this case, the boundary of the circle
domain $\Omega_c(1)$ is $\Gamma_c(1)=\gamma_{e_3,e_4}^c\cup
\gamma_{e_4,e_3}^r$, see Figure~\ref{fig:II-3-a-beta} Case II-3-a-$\beta$.

The mirror configuration, shown in Figure~\ref{fig:II-3-a-beta}
Case II-3-a-$\beta$-m, occurs when $-1<e_2<1<e_1$,
 $e_1$ is the only zero on $\Gamma_e^r$, and
$e_2$ is the only zero on each of the sides
$\Gamma_s^l(1,i\infty)$ and $\Gamma_s^l(-i\infty,1)$. This case
happens if and only if  %
\begin{equation}
2\im[e_1,e_3]_{Q_0}= 2\im [e_2,e_3]_{Q_0}+ \delta_{1}.
\end{equation} %

\smallskip %

($\gamma$) Suppose now that $e_2,e_3\in \Gamma_s^r(-1,i\infty)$
and $e_2,e_4\in\Gamma_s^r(-1,-i\infty)$. This case
happens if and only if  %
$\im[e_1,e_3]_{Q_0}=\delta_{-1}$.

Since zeros $e_2$ and $e_3$ belong to the same side of the strip
domain $\Omega_s(-1,i\infty)$, we have that
$\Gamma_s^r(-1,i\infty)=[-1,e_2]\cup
\gamma_{e_2,e_3}\cup\gamma_{e_3,i\infty}$. By symmetry,
$\Gamma_s^r(-i\infty,-1)=\gamma_{-i\infty,e_4}\cup
\gamma_{e_4,e_2}\cup[e_2,-1]$. The latter implies, in turn, that
the boundary of the circle domain $\Omega_c(1)$ is
$\Gamma_c(1)=\gamma_{e_2,e_4}\cup \gamma_{e_4,e_3}^r\cup
\gamma_{e_3,e_2}$. The Stokes graph and the domain configuration
are shown in Figure~\ref{fig:II-3-a-gamma} Case II-3-a-$\gamma$.

The mirror configuration, shown in Figure~\ref{fig:II-3-a-gamma}
Case II-3-a-$\gamma$-m, occurs when $-1<e_2<1<e_1$,
 $e_1$ is the only zero on $\Gamma_e^r$, and
$e_2,e_3\in \Gamma_s^l(1,i\infty)$. This case
happens if and only if  $\im[e_1,e_3]_{Q_0}=\delta_{1}$.%

\smallskip

(b) Suppose that $\Gamma_e^l$ contains two zeros, which are
$e_3,e_4$. This case
happens if and only if  %
$$ 
2\im[e_2,e_3]_{Q_0}= 2\im [e_1,e_3]_{Q_0}+ \delta_{-1}.
$$ 
Then $\Gamma_e^l=\gamma_{-i\infty,e_4}\cup\gamma_{e_4,e_3}^l\cup
\gamma_{e_3,i\infty}$. This also implies that $e_3,e_4\in
\Gamma_e^r$ and, therefore,
$\Gamma_e^r=\gamma_{i\infty,e_3}\cup\gamma_{e_3,e_4}^r\cup
\gamma_{e_4,-i\infty}$, where $\gamma_{e_3,e_4}^r$ intersects
$(1,\infty)$ at some point $x_3$.

Then, the set $\Gamma_r^{(out)}=\gamma_{e_3,e_4}^r \cup
\gamma_{e_4,e_3}^l$ is an outer boundary component of a face of
the Stokes graph of $Q_0(z)\,dz^2$, which must be a ring domain
$\Omega_r$, see Lemma~3.1. The inner boundary component of this
ring domain contains just one zero $e_1$ and therefore
$\Gamma_r^{(inn)}=\gamma_{e_1}^r$, where $\gamma_{e_1}^r$
intersects the real axis at some point $x_2$, $1<x_2<x_3$. Since
$\Omega_r$ separates $e_1,e_2$ from $e_3,e_4$, this case happens
if and only if $2\im[e_2,e_3]_{Q_0}>\delta_{-1}$.

Under these circumstances, the only possibility for the boundary
of the circle domain $\Omega_c(1)$ is that
$\Gamma_c(1)=\gamma_{e_2}^r$, where $\gamma_{e_2}^r$ intersects
the real axis at some point $x_1$, $1<x_1<x_2$.

The latter implies that  the remaining face of the Stokes graph is
the strip domain $\Omega_s(-1,-1)$ with both vertices at $-1$,
which outer and inner sides are
$\Gamma_s^{(out)}(-1,-1)=[-1,e_1]\cup \gamma_{e_1}^r\cup[e_1,-1]$
and $\Gamma_s^{(inn)}(-1,-1)=[-1,e_2]\cup
\gamma_{c_2}^r\cup[e_2,-1]$, see Figure~\ref{fig:II-3-b} Case II-3-b.

The mirror configuration, shown in Figure~\ref{fig:II-3-b} Case
II-3-b-m, occurs when $-1<e_2<1<e_1$ and when $e_3$ and $e_4$ are
the only zeros on $\Gamma_e^r$.  This case
happens if and only if  %
$$ 
2\im[e_2,e_3]_{Q_0}= 2\im [e_1,e_3]_{Q_0}+ \delta_{1}.
$$ 

\smallskip

(c) Suppose now that $\Gamma_e^l$ contains zeros $e_1,e_3,e_4$.
This case
happens if and only if  %
$ 
2\im[e_2,e_3]_{Q_0}=  \delta_{-1}.
$ 
This implies that
$\Gamma_e^l=\gamma_{-i\infty,e_4}\cup\gamma_{e_4,e_1}\cup\gamma_{e_1,e_3}\cup
\gamma_{e_3,i\infty}$. This also implies that $e_3,e_4\in
\Gamma_e^r$ and, therefore,
$\Gamma_e^r=\gamma_{i\infty,e_3}\cup\gamma_{e_3,e_4}^r\cup
\gamma_{e_4,-i\infty}$, where $\gamma_{e_3,e_4}^r$ intersects
$(1,\infty)$ at some point $x_2$.

In this case, the set $[-1,e_1]\cup \gamma_{e_1,e_4}\cup
\gamma_{e_4,e_3}^r\cup\gamma_{e_3,e_1}\cup[e_1,-1]$ is connected
and therefore, by Lemma~3.1, it is an outer side
$\Gamma_s^{(out)}(-1,-1)$ of the strip domain $\Omega_s(-1,-1)$.
The inner side of this strip domain is
$\Gamma_s^{(inn)}(-1,-1)=[-1,e_2]\cup \gamma_{e_2}^r\cup[e_2,-1]$,
where $\gamma_{e_2}^r$ intersects the real axis at some point
$x_1$, $1<x_1<x_2$.

Under the made assumptions,  the remaining possibility for the
boundary of the circle domain $\Omega_c(1)$ is that
$\Gamma_c(1)=\gamma_{e_2}^r$, where $\gamma_{e_2}^r$. The Stokes
graph has four faces only, as it is shown in Figure~\ref{fig:II-3-c} Case II-3-c.

The mirror configuration, shown in Figure~\ref{fig:II-3-c} Case
II-3-c-m, occurs when $-1<e_2<1<e_1$ and
 $\Gamma_e^r$ contains zeros $e_1,e_3,e_4$. This case
happens if and only if  $2\im[e_2,e_3]_{Q_0}=\delta_{1}$.%

\medskip

4. Consider the case when  $e_1<-1$ and $e_2>1$. In this case the
intervals
            $(-\infty,e_1)$ and $(e_2,\infty)$ are orthogonal
            trajectories of $Q_0(z)\,dz^2$ and the intervals $(e_1,-1)$,
            $(-1,1)$ and $(1,e_2)$ are trajectories of $Q_0(z)\,dz^2$. Therefore, there
            are no circle domains in this case and, in all possible subcases, there is a strip domain $\Omega_s(-1,1)$
            with vertices at the poles $-1$ and $1$, which is symmetric with respect to the real axis.

As in the previous cases, each of the boundaries $\Gamma_e^l$ and
$\Gamma_e^r$ of the end domains may contain $1$, $2$, or $3$ zeros
and $e_2\not\in \Gamma_e^l$, $e_1\not\in \Gamma_e^r$. Thus, we
consider the following subcases.

\smallskip

(a) Let $e_1$ be the only zero on $\Gamma_e^l$. Then
$\Gamma_e^l=\gamma_{-i\infty,e_1}\cup \gamma_{e_1,i\infty}$. This
combined with Lemma~3.1  imply that there are strip domains
$\Omega_s(-1,i\infty)$ and $\Omega_s(-i\infty,-1)$, symmetric to
each other with respect to the real axis, which left and right
sides are $\Gamma_s^l(-1,i\infty)=[-1,e_1]\cup
\gamma_{e_1,i\infty}$ and
$\Gamma_s^r(-i\infty,-1)=\gamma_{-i\infty,e_1}\cup [e_1,-1]$,
respectively.

In its turn, the boundary of $\Omega_e^r$ also may have $1$, $2$,
or $3$ zeros. Thus, we have three possibilities.

\smallskip

($\alpha$) Let $e_2$ be the only zero on $\Gamma_e^r$.  This
happens if and only if $2\im[e_1,e_3]_{Q_0}<\delta_{-1}$ and
$2\im[e_2,e_3]_{Q_0}<\delta_{1}$. Then,
$\Gamma_e^r=\gamma_{i\infty,e_2}\cup \gamma_{e_2,-i\infty}$. This
also implies that there are strip domains $\Omega_s(1,i\infty)$
and $\Omega_s(-i\infty,1)$, symmetric to each other with respect
to the real axis, which right sides are
$\Gamma_s^r(1,i\infty)=[1,e_2]\cup \gamma_{e_2,i\infty}$ and
$\Gamma_s^r(-i\infty,1)=\gamma_{-i\infty,e_2}\cup [e_2,1]$,
respectively. The right sides of each of the strip domains
$\Omega_s(-1,i\infty)$ and $\Omega_s(-i\infty,-1)$ and the left
sides of each of the strip domains $\Omega_s(1,i\infty)$ and
$\Omega_s(-i\infty,1)$ must contain at least one zero of
$Q_0(z)\,dz^2$. This implies that
$\Gamma_s^r(-1,i\infty)=\gamma_{-1,e_3}\cup \gamma_{e_3,i\infty}$,
$\Gamma_s^r(-i\infty,-1)=\gamma_{-i\infty,e_4}\cup
\gamma_{e_4,-1}$, $\Gamma_s^l(1,i\infty)=\gamma_{1,e_3}\cup
\gamma_{e_3,i\infty}$, and
$\Gamma_s^l(-i\infty,1)=\gamma_{-i\infty,e_4}\cup \gamma_{e_4,1}$.
Under these circumstances, the sides of the strip domain
$\Omega_s(-1,1)$ are $\Gamma_s^+(-1,1)=\gamma_{-1,e_3}\cup
\gamma_{e_3,1}$ and $\Gamma_s^-(-1,1)=\gamma_{-1,e_4}\cup
\gamma_{e_4,1}$. The corresponding domain configuration is shown
in Figure~\ref{fig:II-4-a-alpha} Case II-4-a-$\alpha$.

\smallskip

 ($\beta$) Suppose that $\Gamma_e^r$ contains two zeros,
which in this case are $e_3$ and $e_4$. This case happens if and
only if $2\im[e_1,e_3]_{Q_0}+2\im [e_2,e_3]+\delta_1=\delta_{-1}$.
 Then, $\Gamma_e^r=\gamma_{i\infty,e_3}\cup \gamma_{e_3,e_4}^r\cup
\gamma_{e_4,-i\infty}$. This also implies that the right sides of
 the strip domains $\Omega_s(-1,i\infty)$ and
$\Omega_s(-i\infty,-1)$ are
$\Gamma_s^r(-1,i\infty)=\gamma_{-1,e_3}\cup \gamma_{e_3,i\infty}$
and $\Gamma_s^r(-i\infty,-1)=\gamma_{-i\infty,e_4}\cup
\gamma_{e_4,-1}$, respectively. Furthermore, the assumptions imply
that $e_2$ is the only zero on the boundary of the strip domain
$\Omega_s(-1,1)$. Therefore, there are critical trajectories
$\gamma_{-1,e_2}^+\subset \mathbb{H}_+$ and
$\gamma_{-1,e_2}^-\subset \mathbb{H}_-$ and the sides of
$\Omega_s(-1,1)$ are $\Gamma_s^+(-1,1)=\gamma_{-1,e_2}^+\cup
[e_2,1]$ and $\Gamma_s^-(-1,1)=\gamma_{-1,e_2}^-\cup [e_2,1]$. The
remaining face of $Q_0(z)\,dz^2$ in this case is the strip domain
$\Omega_s(-1,-1)$, symmetric with respect to the real axis, with
both its vertices at $-1$ and sides
$\Gamma_s^{(out)}(-1,-1)=\gamma_{-1,e_4}\cup
\gamma_{e_4,e_3}^r\cup \gamma_{e_3,-1}$ and
$\Gamma_s^{(inn)}(-1,-1)=\gamma_{-1,e_2}^-\cup \gamma_{e_2,-1}^+$.
Figure~\ref{fig:II-4-a-beta} Case II-4-a-$\beta$ gives an example of a domain
configuration.

The mirror configuration, shown in Figure~\ref{fig:II-4-a-beta}
Case II-4-a-$\beta$-m, occurs when $e_1>1$, $e_2<-1$ and when
$e_3$, $e_4$ are the only zeros of  $\Gamma_e^l$.  This case
happens if and only if $2\im[e_1,e_3]_{Q_0}+2\im
[e_2,e_3]+\delta_{-1}=\delta_1$.

\smallskip

 ($\gamma$) Suppose now that $e_2,e_3,e_4\in \Gamma_e^r$. This happens if and
only if $2\im[e_1,e_3]_{Q_0}+\delta_1=\delta_{-1}$. Then,
$\Gamma_e^r=\gamma_{i\infty,e_3}\cup \gamma_{e_3,e_2}\cup
\gamma_{e_2,e_4}\cup \gamma_{e_4,-i\infty}$. This also implies
that the right sides of
 the strip domains $\Omega_s(-1,i\infty)$ and
$\Omega_s(-i\infty,-1)$ are
$\Gamma_s^r(-1,i\infty)=\gamma_{-1,e_3}\cup \gamma_{e_3,i\infty}$
and $\Gamma_s^r(-i\infty,-1)=\gamma_{-i\infty,e_4}\cup
\gamma_{e_4,-1}$, respectively.

Furthermore, our assumptions in this case imply that $e_2$, $e_3$,
and $e_4$ belong to the boundary of the strip domain
$\Omega_s(-1,1)$. Therefore, $\Gamma_s^+(-1,1)=\gamma_{-1,e_3}\cup
\gamma_{e_3,e_2}\cup [e_2,1]$ and
$\Gamma_s^-(-1,1)=\gamma_{-1,e_4}\cup \gamma_{e_4,e_2}\cup
[e_2,1]$. There are no other domains in the domain configuration
of $Q_0(z)\,dz^2$, see Figure~\ref{fig:II-4-a-gamma} Case II-4-a-$\gamma$.

The mirror configuration, shown in Figure~\ref{fig:II-4-a-gamma}
Case II-4-a-$\gamma$-m, occurs when $e_1>1$, $e_2<-1$ and when
$e_2,e_3,e_4\in \Gamma_e^l$. This case happens if and only if
$2\im[e_1,e_3]_{Q_0}+\delta_{-1}=\delta_1$.

\smallskip

(b) Suppose that each of the boundaries $\Gamma_e^l$ and
$\Gamma_e^r$ contains two zeros, which are $e_3$ and $e_4$ for
each of these boundaries.  Then there are  critical trajectories
$\gamma_{e_3,e_4}^l$, which crosses the real axis at some point
$x_1<e_1$ and $\gamma_{e_3,e_4}^r$, which crosses the real axis at
some point $x_2>e_2$. This immediately implies that
$\Gamma_e^l=\gamma_{-i\infty,e_4}\cup \gamma_{e_4,e_3}^l\cup
\gamma_{e_3,i\infty}$ and $\Gamma_e^r=\gamma_{i\infty,e_3}\cup
\gamma_{e_3,e_4}^r\cup \gamma_{e_4,-i\infty}$. %
In this case, the closed Jordan curve
$\Gamma_r^{(out)}=\gamma_{e_3,e_4}^l\cup \gamma_{e_4,e_3}^r$ is an
outer boundary component of a face of the quadratic differential
$Q_0(z)\,dz^2$, which must be a ring domain $\Omega_r$. The inner
boundary component $\Gamma_r^{(inn)}$ of $\Omega_r$ may contain
one or both of the zeros $e_1$, $e_2$. Thus, we have the following
subcases.

\smallskip %

($\alpha$) Let $e_1\in \Gamma_r^{(inn)}$, but $e_2\not\in
\Gamma_r^{(inn)}$. This case happens if and only if $\im
[e_1,e_3]_{Q_0}<\im[e_2,e_3]_{Q_0}$ and $\delta_{-1}>\delta_{1}$.
Then, $\Gamma_r^{(inn)}=\gamma_{e_1}^r$, where $\gamma_{e_1}^r$ is
the critical trajectory with both end points at $e_1$, which
crosses the real axis at some point $x_3$, $e_2<x_3<x_2$. Under
the assumptions, there are critical trajectories
$\gamma_{e_2,-1}^+\subset \mathbb{H}_+$ and
$\gamma_{e_2,-1}^-\subset \mathbb{H}_-$ having their end points at
the points $-1$ and $e_2$. This implies that
$\Gamma_s^+(-1,1)=\cup \gamma_{-1,e_2}^+\cup [e_2,1]$ and
$\Gamma_s^-(-1,1)=\gamma_{-1,e_2}^-\cup [e_1,1]$.

Under these circumstances, there is one more face of the Stokes
graph of $Q_0(z)\,dz^2$, which is a strip domain $\Omega_s(-1,-1)$
symmetric with respect to the real axis having both its vertices
at the pole $-1$. In this case,
$\Gamma_s^{(out)}(-1,-1)=[-1,e_1]\cup \gamma_{e_1}^r\cup [e_1,-1]$
and $\Gamma_s^{(inn)}(-1,-1)=\gamma_{-1,e_2}^+\cup
\gamma_{e_2,-1}^-$. Figure~\ref{fig:II-4-b-alpha} Case II-4-b-$\alpha$ gives an
example of the Stokes graph and domain configuration.

The mirror configuration, shown in Figure~\ref{fig:II-4-b-alpha} Case II-4-b-$\alpha$-m, occurs when $e_1>1$, $e_2<-1$, when $e_3,e_4$ are the
only zeros on  $\Gamma_e^l$ and on $\Gamma_e^r$, and when  $e_1\in
\Gamma_r^{(inn)}$ but $e_2\not\in \Gamma_r^{(inn)}$.  This case
happens if and only if $\im [e_1,e_3]_{Q_0}<\im[e_2,e_3]_{Q_0}$
and $\delta_{-1}<\delta_{1}$.

\smallskip



 ($\beta$) The points $e_1,e_2$ belong to $\Gamma_r^{(inn)}$ if and only if $\im
[e_1,e_3]_{Q_0}=\im[e_2,e_3]_{Q_0}$ and $\delta_{-1}=\delta_{1}$.
In this case, there are critical
 trajectories $\gamma_{e_1,e_2}^+\subset \mathbb{H}_+$ and $\gamma_{e_1,e_2}^-\subset
 \mathbb{H}_-$ having their end points at $e_1$ and $e_2$ and, therefore,  $\Gamma_r^{(inn)}=\gamma_{e_1,e_2}^-\cup
 \gamma_{e_2,e_1}^+$. This also implies that $\Gamma_s^+(-1,1)=[-1,e_1]\cup \gamma_{e_1,e_2}^+\cup [e_2,1]$ and
$\Gamma_s^-(-1,1)=[-1,e_1]\cup \gamma_{e_1,e_2}^-\cup [e_2,1]$,
see Figure~\ref{fig:II-4-b-beta} Case II-4-b-$\beta$.

\smallskip






(c) Suppose now that $e_1,e_3,e_4\in\Gamma_e^l$.  Then
$\Gamma_e^l=\gamma_{-i\infty,e_4}\cup \gamma_{e_4,e_1}\cup
\gamma_{e_1,e_3}\cup \gamma_{e_3,i\infty}$. In its turn, the
boundary of $\Omega_e^r$  may have $1$, $2$, or $3$ zeros and
$e_1\not\in \Gamma_s^r$. In the case, when $e_2$ is the only zero
on $\Gamma_e^r$, the domain configuration is the mirror
configuration mentioned in part~II-4-a-$\gamma$ above. The
remaining subcases are the following.

\smallskip


\smallskip

($\alpha$) Suppose that $e_3$ and $e_4$ are the only zeros on
$\Gamma_e^r$. This case happens if and only if
$\im[e_1,e_3]_{Q_0}=0$ and $\delta_{-1}>\delta_{1}$. In this case
there are critical trajectories $\gamma_{e_1,e_3}\subset
\mathbb{H}_+$, $\gamma_{e_1,e_4}\subset \mathbb{H}_-$,
$\gamma_{-1,e_2}^+\subset \mathbb{H}_+$ and
$\gamma_{-1,e_2}^-\subset \mathbb{H}_-$. This implies that
$\Gamma_e^r=\gamma_{i\infty,e_3}\cup \gamma_{e_3,e_4}^r\cup
\gamma_{e_4,-i\infty}$ and that
$\Gamma_s^+(-1,1)=\gamma_{-1,e_2}^+\cup [e_2,1]$ and
$\Gamma_s^-(-1,1)= \gamma_{-1,e_2}^-\cup [e_2,1]$. The remaining
face of $Q_0(z)\,dz^2$ in this case is the strip domain
$\Omega_s(1,1)$ symmetric with respect to the real axis with both
its vertices at $1$. Therefore,
$\Gamma_s^{(out)}(1,1)=[-1,e_1]\cup \gamma_{e_1,e_4}\cup
\gamma_{e_4,e_3}^r\cup \gamma_{e_3,e_1}\cup [e_1,-1]$ and
$\Gamma_s^{(inn)}(1,1)= \gamma_{-1,e_2}^-\cup \gamma_{e_2,-1}^+$.
Figure~\ref{fig:II-4-c-alpha} Case II-4-c-$\alpha$ gives an example of a domain
configuration.

The mirror configuration for this case, shown in
Figure~\ref{fig:II-4-c-alpha} Case II-4-c-$\alpha$-m, occurs when
$e_1>1$, $e_2<-1$, when $e_1,e_3,e_4\in \Gamma_e^r$ and  when
$e_3$ and $e_4$ are the only zeros on $\Gamma_e^l$. This case
happens if and only if $\im[e_1,e_3]_{Q_0}=0$ and
$\delta_{-1}<\delta_{1}$.

\smallskip

($\beta$) Suppose that $\Gamma_e^r$ contains three zeros, which,
in this case,  are $e_2$, $e_3$ and $e_4$. This case is
self-mirrored, it  happens if and only if $\im[e_1,e_3]_{Q_0}=\im
[e_2,e_3]_{Q_0}=0$. In this case,
$\Gamma_e^r=\gamma_{i\infty,e_3}\cup \gamma_{e_3,e_2}\cup
\gamma_{e_2,e_4}\cup \gamma_{e_4,-i\infty}$. The latter implies
that $\Omega_e^l$, $\Omega_e^r$ and $\Omega_s(-1,1)$ are the only
domains in the domain configuration of $Q_0(z)\,dz^2$ and that the
sides of $\Omega_s(-1,1)$ are $\Gamma_s^+(-1,1)=[-1,e_1]\cup
\gamma_{e_1,e_3}\cup \gamma_{e_3,e_2}\cup[e_2,1]$ and
$\Gamma_s^-(-1,1)=[-1,e_1]\cup \gamma_{e_1,e_4}\cup
\gamma_{e_4,e_2}\cup[e_2,1]$, see Figure~\ref{fig:II-4-b-beta}
Case II-4-c-$\beta$.


        \bigskip

        {\textbf{ III.}}         Suppose that all zeros of $P_0(z)$ are
        real. In the generic cases, illustrated in Figures 25--45 in Appenix~B,
        we assume that all these zeros are distinct and that $e_k\not= \pm 1$,
        $k=1,2,3,4$. Possible degenerate cases appeared from this part
        are shown in Figures~46--50 in Appendex~B.
        Depending on the number of zeroes on each of the intervals
        $(-\infty,-1)$, $(-1,1)$ and $(1,\infty)$, we consider the
        following subcases.

        \medskip

        1. Suppose that $e_1<e_2<e_3<e_4<-1$.  In
            this case, the intervals $(-\infty,e_1)$, $(e_2,e_3)$, $(e_4,-1)$, $(-1,1)$, and $(1,\infty)$ are
            orthogonal trajectories of $Q_0(z)\,dz^2$ and the intervals $(e_1,e_2)$ and $(e_3,e_4)$ are  trajectories of $Q_0(z)\,dz^2$.
This implies that there are two circle domains $\Omega_c(-1)$ and
$\Omega_c(1)$. Furthermore,  the topological argument based on
Lemma~3.1 shows that $e_1$ is the only zero on the boundary of
$\Omega_e^l$ and $e_4$ is the only zero on the boundary of
$\Omega_c(-1)$. Therefore, $\Gamma_e^l=\gamma_{-i\infty,e_1}\cup
\gamma_{e_1,i\infty}$ and $\Gamma_c(-1)=\gamma_{e_4}^c$.
            Under the assumptions of this case, there are closed critical trajectories $\gamma_{e_3}^r$
            intersecting $(1,\infty)$ at $x_1$ and $\gamma_{e_2}^r$ intersecting
            $(x_1,\infty)$. This implies that the circle domain $\Omega_c(1)$
            has the boundary $\Gamma_c(1)=\gamma_{e_3}^r\cup[e_3,e_4]\cup\gamma_{e_4}^{r-}\cup[e_4,e_3]$ and the end domain $\Omega_e^r$ has the boundary
  $\Gamma_e^r=\gamma_{-i\infty,e_1}\cup[e_1,e_2]\cup\gamma_{e_2}^r\cup[e_2,e_1]\cup\gamma_{e_1,i\infty}$.
  Under these circumstances, there is one more face of $Q_0(z)\,dz^2$, which is the ring domain $\Omega_r$
  with the boundary components $\Gamma_r^{(out)}=\gamma_{e_2}^r$ and $\Gamma_r^{(inn)}=\gamma_{e_3}^r$.

             The Stokes graph and domain configuration are shown in Figure~\ref{fig:III-1} Case III-1.

             The mirror configuration, shown in Figure~\ref{fig:III-1} Case III-1-m, occurs if $1<e_4<e_3<e_2<e_1$.









        \medskip


        \medskip

        2. Suppose that $-1<e_1<e_2<e_3<e_4<1$. In this case, the intervals
        $(-\infty,-1)$, $(-1,e_1)$, $(e_2,e_3)$, $(e_4,-1)$, and $(1,\infty)$ are
            orthogonal trajectories of $Q_0(z)\,dz^2$ and the intervals $(e_1,e_2)$ and $(e_3,e_4)$ are  trajectories of $Q_0(z)\,dz^2$.
As before, this implies that there are two circle domains
$\Omega_c(-1)$ and $\Omega_c(1)$. Furthermore, using Lemma~3.1 we
conclude that
 $e_1$ is the only zero on the boundary of $\Omega_c(-1)$
and $e_4$ is the only zero on the boundary of $\Omega_c(1)$.
Therefore, $\Gamma_c(-1)=\gamma_{e_1}^l$ and
$\Gamma_c(1)=\gamma_{e_4}^r$.
            Under the assumptions of this case, there are critical trajectories
            $\gamma_{e_2,i\infty}$, $\gamma_{e_2,-i\infty}$,
            $\gamma_{e_3,i\infty}$, and $\gamma_{e_3,-i\infty}$.
            This implies that in this case there is a strip domain
            $\Omega_s(-i\infty,i\infty)$ with sides
            $\Gamma_s^l(-i\infty,i\infty)=\gamma_{-i\infty,e_2}\cup
            \gamma_{e_2,i\infty}$ and $\Gamma_s^r(-i\infty,i\infty)=\gamma_{-i\infty,e_3}\cup
            \gamma_{e_3,i\infty}$.

             The domain configuration in this case is shown in Figure~\ref{fig:III-2} Case III-2.

             \medskip

3. Suppose that $e_1<e_2<e_3<-1<e_4<1$. In this case, the
intervals
        $(-\infty,e_1)$, $(e_2,e_3)$, $(e_4,1)$, and $(1,\infty)$ are
            orthogonal trajectories of $Q_0(z)\,dz^2$ and the intervals $(e_1,e_2)$, $(e_3,-1)$,
            and $(-1,e_4)$ are  trajectories of $Q_0(z)\,dz^2$.
Thus, there is only one circle domain $\Omega_c(1)$ in this case.

A topological argument similar to the one used in the proof of
Lemma~3.1, which is based on the information obtained from the
Basic Structure Theorem \cite[Theorem 3.5]{Jenkins}, shows that,
in the case under consideration, there are three critical
trajectories $\gamma_{e_k}^r$, $k=2,3,4$, such that
$\gamma_{e_k}^r$ crosses the interval $(1,\infty)$ at the point
$x_k$, $1<x_4<x_3<x_2$. Moreover, $\gamma_{e_3}^r$ is contained in
the Jordan domain bounded by $\gamma_{e_2}^r$, and
$\gamma_{e_4}^r$ is contained in a Jordan domain bounded by
$\gamma_{e_3}^r$. This implies that there exist critical
trajectories $\gamma_{e_1,i\infty}$ and $\gamma_{e_1,-i\infty}$.
Now, when the Stokes graph of $Q_0(z)\,dz^2$ is identified, one
can easily see that the domain configuration of $Q_0(z)\,dz^2$
consists of end domains $\Omega_e^l$ and $\Omega_e^r$, circle
domain $\Omega_c(1)$, ring domain $\Omega_r$, and strip domain
$\Omega_s(-1,-1)$. The corresponding boundaries, boundary
components, and sides are the following:
$\Gamma_e^l=\gamma_{-i\infty,e_1}\cup \gamma_{e_1,i\infty}$,
$\Gamma_e^r=\gamma_{i\infty,e_1}\cup [e_1,e_2]\cup
\gamma_{e_2}^r\cup [e_2,e_1]\cup
\gamma_{e_1,-i\infty}$, %
$\Gamma_c(1)=\gamma_{e_4}^r$, $\Gamma_r^{(out)}=\gamma_{e_2}^r$,
$\Gamma_r^{(inn)}=\gamma_{e_3}^r$, $\Gamma_s^+(-1,1)=[-1,e_3]\cup
\gamma_{e_3}^r\cup [e_3,-1]$,
$\Gamma_s^-(-1,1)=[-1,e_4]\gamma_{e_4}^r\cup [e_4,-1]$.

The Stokes graph and domain configuration are shown in
Figure~\ref{fig:III-3} Case III-3.

The mirror configuration, shown in Figure~\ref{fig:III-3} Case
III-3-m, occurs when $-1<e_4<1<e_3<e_2<e_1$.

4. Suppose that $e_1<e_2<e_3<-1<1<e_4$. In this case, the
intervals
        $(-\infty,e_1)$, $(e_2,e_3)$, and $(e_4,\infty)$ are
            orthogonal trajectories of $Q_0(z)\,dz^2$ and the intervals $(e_1,e_2)$, $(e_3,-1)$, $(-1,1)$,
             and $(1,e_4)$ are  trajectories of $Q_0(z)\,dz^2$.
Thus, there are no circle domains in this case and there is a
strip domain $\Omega_s(-1,1)$, which is symmetric with respect to
the real axis.

Using the topological argument based on the Basic Structure
Theorem \cite[Theorem 3.5]{Jenkins}, we conclude that there are
critical trajectories $\gamma_{e_1,i\infty}\subset
\overline{\mathbb{H}}_+$ and $\gamma_{e_1,-i\infty}\subset
\overline{\mathbb{H}}_-$. The latter implies that
$\Gamma_e^l=\gamma_{-i\infty,e_1}\cup \gamma_{e_1,i\infty}$.

Since the strip domain $\Omega_s(-1,1)$ is symmetric with respect
to the real axis, its boundary $\partial \Omega_s(-1,1)$ must
contain at least one of the zeros $e_3$, $e_4$. Thus, we consider
the following subcases.

\smallskip

(a) Let $e_3$ be the only zero on $\partial \Omega_s(-1,1)$. This
case happens if and only if $\delta_{-1}<\delta_1$. This
inequality implies that there are critical trajectories
$\gamma_{e_3,1}^+\subset \overline{\mathbb{H}}_+$ and
$\gamma_{e_3,1}^-\subset \overline{\mathbb{H}}_-$. Therefore, the
upper and lower sides of the strip domain $\Omega_s(-1,1)$ are,
respectively, $\Gamma_s^+(-1,1)=[-1,e_3]\cup \gamma_{e_3,1}^+$ and
$\Gamma_s^-(-1,1)=[-1,e_3]\cup \gamma_{e_3,1}^-$.

As concerns critical trajectories, different from the interval
$(e_1,e_2)$, which  have at least one end point at $e_2$, there
are three possibilities.

\smallskip

($\alpha$) There are critical trajectories
$\gamma_{e_2,1}^+\subset \overline{\mathbb{H}}_+$ and
$\gamma_{e_2,1}^-\subset \overline{\mathbb{H}}_-$. This subcase
happens if and only if $\delta_{-1}+2[e_2,e_3]_{Q_0}<\delta_1$.
Under these assumptions, there is a strip domain $\Omega_s(1,1)$,
the inner and the outer sides of which are
$\Gamma_s^{(inn)}=\gamma_{1,e_3}^-\cup \gamma_{e_3,1}^+$ and
$\Gamma_s^{(out)}=\gamma_{1,e_2}^-\cup \gamma_{e_2,1}^+$. Under
these circumstances, the set
$\Gamma_s^l(1,i\infty)=\gamma_{1,e_2}^+\cup [e_2,e_1]\cup
\gamma_{e_1,i\infty}$ is a boundary arc of one of the faces of the
Stokes graph of $Q_0(z)\,dz^2$, which in this case must be a strip
domain $\Omega_s(1,i\infty)$. Thus, $\Gamma_s^l(1,i\infty)$ is the
left side of $\Omega_s(1,i\infty)$. Since the right side of
$\Omega_s(1,i\infty)$ must contain at least one zero of
$Q_0(z)\,dz^2$, the only possibility is that
$\Gamma_s^r(1,i\infty)=[1,e_4]\cup \gamma_{e_4,i\infty}$, where
$\gamma_{e_4,i\infty} \subset \overline{\mathbb{H}}_+$ is the
critical trajectory of $Q_0(z)\,dz^2$ joining $e_4$ and $\infty$.
Since the trajectory structure of $Q_0(z)\,dz^2$ is symmetric with
respect to the real axis, it follows that there is a strip domain
$\Omega_s(-i\infty,1)$ with left and right sides
$\Gamma_s^l(-i\infty,1)=\gamma_{-i\infty,e_1}\cup [e_1,e_2]\cup
\gamma_{e_2,1}^-$ and
$\Gamma_s^r(-i\infty,1)=\gamma_{-i\infty,e_4}\cup [e_4,1]$,
 respectively. The latter also implies that the boundary of the end
domain $\Omega_e^r$ is
$\Gamma_e^r=\gamma_{i\infty,e_4}\cup\gamma_{e_4,-i\infty}$. There
are no other domains in the domain configuration of $Q_0(z)\,dz^2$
in this case.
The Stokes graph and domain configuration are shown in
Figure~\ref{fig:III-4-a-alpha} Case III-4-a-$\alpha$.

The mirror configuration, shown in Figure~\ref{fig:III-4-a-alpha}
Case III-4-a-$\alpha$-m, occurs when $e_4<-1<1<e_3<e_2<e_1$ and it
occurs if and only if $\delta_{1}+2[e_2,e_3]_{Q_0}<\delta_{-1}$.

\smallskip

($\beta$) There is a critical trajectory $\gamma_{e_2}^r$, which
intersects the interval $(e_4,\infty)$.  This subcase happens if
and only if $\delta_{-1}<\delta_1<\delta_{-1}+2[e_2,e_3]_{Q_0}$.
In this case the boundary of the end domain $\Omega_e^r$ is
$\Gamma_e^r=\gamma_{i\infty,e_1}\cup [e_1,e_2] \cup
\gamma_{e_2}^{r-}\cup [e_2,e_1]\cup \gamma_{e_1,-i\infty}$.

Furthermore, the curve $\gamma_{e_2}^r$ is an outer boundary
component of one of the faces of the Stokes graph of
$Q_0(z)\,dz^2$, which  must be a ring domain $\Omega_r$.
Therefore, $\Gamma_r^{(out)}=\gamma_{e_2}^r$. Under these
circumstances there is a critical trajectory $\gamma_{e_4}^l$
intersecting the interval $(e_2,e_3)$, which is the inner boundary
component of $\Omega_r$, i.e. $\Gamma_r^{(inn)}=\gamma_{e_4}^l$.
The remaining face of the Stokes graph in this case is a strip
domain $\Omega_s(1,1)$ symmetric with respect to the real axis,
the inner and outer sides of which are
$\Gamma_s^{(inn)}(1,1)=\gamma_{1,e_3}^+\cup \gamma_{e_3,1}^-$ and
$\Gamma_s^{(out)}(1,1)=[1,e_4]\cup \gamma_{e_4}^l \cup [e_4,1]$.
The Stokes graph and domain configuration are shown in
Figure~\ref{fig:III-4-a-beta} Case III-4-a-$\beta$.

The mirror configuration for this, shown in
Figure~\ref{fig:III-4-a-beta} Case III-4-a-$\beta$-m, occurs when
$e_4<-1<1<e_3<e_2<e_1$ and it occurs if and only if
$\delta_{1}<\delta_{-1}<\delta_{1}+2[e_2,e_3]_{Q_0}$.

\smallskip

($\gamma$) There are critical trajectories $\gamma_{e_2,e_4}^+$
and $\gamma_{e_2,e_4}^-$. This subcase happens if and only if
$\delta_1=\delta_{-1}+2[e_2,e_3]_{Q_0}$. The boundary of the end
domain $\Omega_e^r$ is $\Gamma_e^r=\gamma_{i\infty,e_1}\cup
[e_1,e_2] \cup \gamma_{e_2,e_4}^+\cup \gamma_{e_4,e_2}^-\cup
[e_2,e_1]\cup\gamma_{e_1,-i\infty}$. The remaining face of the
Stokes graph in this case is the strip domain  $\Omega_s(1,1)$
symmetric with respect to the real axis, the inner and outer sides
of which are $\Gamma_s^{(inn)}(1,1)=\gamma_{1,e_3}^+\cup
\gamma_{e_3,1}^-$ and $\Gamma_s^{(out)}(1,1)=[1,e_4]\cup
\gamma_{e_4,e_2}^+ \cup \gamma_{e_2,e_4}^-\cup [e_4,1]$.
The Stokes graph and domain configuration are shown in
Figure~\ref{fig:III-4-a-gamma} Case III-4-a-$\gamma$.

The mirror configuration, shown in Figure~\ref{fig:III-4-a-gamma}
Case III-4-a-$\gamma$-m, occurs when $e_4<-1<1<e_3<e_2<e_1$ and it
occurs if and only if $\delta_{-1}=\delta_{1}+2[e_2,e_3]_{Q_0}$.

\smallskip

(b) Let $e_4$ be the only zero on $\partial \Omega_s(-1,1)$. This
case happens if and only if $\delta_{1}<\delta_{-1}$. This
inequality implies that there are critical trajectories
$\gamma_{-1,e_4}^+\subset \overline{\mathbb{H}}_+$ and
$\gamma_{-1,e_4}^-\subset \overline{\mathbb{H}}_-$. Therefore, the
upper and lower sides of the strip domain $\Omega_s(-1,1)$ are
 $\Gamma_s^+(-1,1)=\gamma_{-1,e_4}^+\cup [e_4,1]$ and
$\Gamma_s^-(-1,1)=\gamma_{-1,e_4}^-\cup [e_4,1]$, respectively. In
this case, the set $\Gamma_s^{(inn)}(-1,-1)=\gamma_{-1,e_4}^-\cup
\gamma_{e_4,-1}^+$ must be an inner side of the strip domain
$\Omega_s(-1,-1)$. Under these circumstances, there is a critical
trajectory $\gamma_{e_3}^r$ intersecting the interval
$(e_4,\infty)$ at some point $x_1$. The outer side of
$\Omega_s(-1,-1)$ in this case is
$\Gamma_s^{(out)}(-1,-1)=[-1,e_3]\cup \gamma_{e_3}^r\cup
[e_3,-1]$.

Furthermore, the curve $\Gamma_r^{(inn)}=\gamma_{e_3}^r$ is an
inner component of some face of the quadratic differential
$Q_0(z)\,dz^2$, which must be a ring domain $\Omega_r$. The latter
implies that there is a critical trajectory $\gamma_{e_2}^r$
intersecting the interval $(x_1,\infty)$. In this case,
$\Gamma_r^{(out)}=\gamma_{e_2}^r$. Now, when all critical
trajectories of $Q_0(z)\,dz^2$ are identified, the boundary of the
end domain $\Omega_e^r$ is $\Gamma_e^r=\gamma_{i\infty,e_1}\cup
[e_1,e_2]\cup \gamma_{e_2}^{r-}\cup [e_2,e_1]\cup
\gamma_{e_1,-i\infty}$.
The Stokes graph and domain configuration are shown in
Figure~\ref{fig:III-4-b} Case III-4-b.

The mirror configuration for this case, shown in
Figure~\ref{fig:III-4-b} Case III-4-b-m, occurs when
$e_4<-1<1<e_3<e_2<e_1$ and it occurs if and only if
$\delta_{-1}<\delta_{1}$.

\smallskip

(c) Let $e_3,e_4\in \partial \Omega_s(-1,1)$. This case happens if
and only if $\delta_{-1}=\delta_{1}$. This equality implies that
there are critical trajectories $\gamma_{e_3,e_4}^+\subset
\overline{\mathbb{H}}_+$ and $\gamma_{e_3,e_4}^-\subset
\overline{\mathbb{H}}_-$. Therefore, the upper and lower sides of
the strip domain $\Omega_s(-1,1)$ are
 $\Gamma_s^+(-1,1)=[-1,e_3]\cup \gamma_{e_3,e_4}^+\cup [e_4,1]$ and
$\Gamma_s^-(-1,1)=[-1,e_3]\cup \gamma_{e_3,e_4}^-\cup [e_4,1]$,
respectively. In this case, the set
$\Gamma_r^{(inn)}=\gamma_{e_3,e_4}^-\cup \gamma_{e_4,e_3}^+$ must
be an inner boundary component of the ring domain $\Omega_r$.
Under these circumstances, there is a critical trajectory
$\gamma_{e_2}^r$ intersection the interval $(e_4,\infty)$. The
outer side of $\Omega_r$ in this case is
$\Gamma_r^{(out)}=\gamma_{e_2}^r$.

Now, when all critical trajectories of $Q_0(z)\,dz^2$ are
identified, the boundary of the end domain $\Omega_e^r$ is
$\Gamma_e^r=\gamma_{i\infty,e_1}\cup [e_1,e_2]\cup
\gamma_{e_2}^{r-}\cup [e_2,e_1]\cup \gamma_{e_1,-i\infty}$.
The Stokes graph and domain configuration are shown in
Figure~\ref{fig:III-4-c} Case III-4-c.

The mirror configuration, shown in Figure~\ref{fig:III-4-c} Case
III-4-c-m, occurs when $e_4<-1<1<e_3<e_2<e_1$ and it occurs if and
only if $\delta_{-1}=\delta_{1}$.

\medskip

5. Suppose that $e_1<-1<e_2<e_3<e_4<1$. In this case, the
intervals
        $(-\infty,e_1)$, $(e_2,e_3)$, $(e_4,1)$, and $(1,\infty)$ are
            orthogonal trajectories of $Q_0(z)\,dz^2$ and the intervals
$(e_1,-1)$, $(-1,e_2)$ and  $(e_3,e_4)$ are  trajectories of
$Q_0(z)\,dz^2$. Thus, in this case there is only one circle domain
$\Omega_c(1)$.

The topological restrictions from the Basic Structure Theorem
\cite[Theorem~3.5]{Jenkins} imply that  there exist critical
trajectories $\gamma_{e_1,i\infty}$, $\gamma_{e_1,-i\infty}$,
$\gamma_{e_2,i\infty}$, $\gamma_{e_2,-i\infty}$,
$\gamma_{e_3,i\infty}$, and $\gamma_{e_3,-i\infty}$. Moreover, the
latter implies that there is a critical trajectory
$\gamma_{e_4}^r$ that is the boundary of the circle domain
$\Omega_c(1)$; i.e. $\Gamma_c(1)=\gamma_{e_4}^r$. In this case,
the boundaries of the end domains $\Omega_e^l$ and $\Omega_e^r$
are $\Gamma_e^l=\gamma_{-i\infty,e_1}\cup \gamma_{e_1,i\infty}$
and $\Gamma_e^r=\gamma_{i\infty,e_3}\cup [e_3,e_4]\cup
\gamma_{e_4}^r\cup [e_4,e_3]\cup \gamma_{e_3,-i\infty}$.

Furthermore, there exist three strip domains
$\Omega_s(-1,i\infty)$, $\Omega_s(-1,-i\infty)$ and
$\Omega_s(-i\infty,i\infty)$. The corresponding sides of these
strip domains are the following:
$\Gamma_s^l(-1,i\infty)=[-1,e_1]\cup \gamma_{e_1,i\infty}$ and
$\Gamma_s^r(-1,i\infty)=[-1,e_2]\cup \gamma_{e_2,i\infty}$,
$\Gamma_s^l(-i\infty,-1)=\gamma_{-i\infty,e_1}\cup [e_1,-1]$ and
$\Gamma_s^r(-i\infty,-1)=\gamma_{-i\infty,e_2}\cup [e_2,-1]$,
$\Gamma_s^l(-i\infty,i\infty)=\gamma_{-i\infty,e_2}\cup
\gamma_{e_2,i\infty}$ and
$\Gamma_s^r(-i\infty,i\infty)=\gamma_{-i\infty,e_3}\cup
\gamma_{e_3,i\infty}$.
The Stokes graph and domain configuration are shown in
Figure~\ref{fig:III-5} Case III-5.

The mirror configuration, shown in Figure~\ref{fig:III-5} Case
III-5-m, occurs when $-1<e_4<e_3<e_2<1<e_1$.

             \medskip

6. Suppose that $e_1<e_2<-1<e_3<e_4<1$. In this case, the
intervals
        $(-\infty,e_1)$, $(e_2,-1)$, $(-1,e_3)$, $(e_4,1)$, and $(1,\infty)$ are
            orthogonal trajectories of $Q_0(z)\,dz^2$ and the intervals $(e_1,e_2)$ and $(e_3,e_4)$ are  trajectories of $Q_0(z)\,dz^2$.
This implies that there are two circle domains $\Omega_c(-1)$ and
$\Omega_c(1)$. As in the previous cases, the topological argument
based on  the Basic Structure Theorem \cite[Theorem~3.5]{Jenkins}
and Lemma~3.1 implies that
 $e_1$ is the only zero on the boundary of $\Omega_e^l$
and $e_4$ is the only zero on the boundary of $\Omega_c(1)$.
Therefore, $\Gamma_e^l=\gamma_{-i\infty,e_1}^l\cup
\gamma_{e_1,i\infty}$ and $\Gamma_c(1)=\gamma_{e_4}^r$, where
$\gamma_{e_4}^r$ crosses $(1,\infty)$ at some point $x_1$.

            Under the assumptions of this case, the boundary of $\Omega_c(-1)$ may contain 1 or 2 zeros.
            Thus, to identify the remaining domains, we consider three subcases.

\smallskip

(a) Let $e_2$ be the only zero on $\Gamma_c(-1)$. Under the
assumptions of Case~III-6,  the latter happens if and only if
            \begin{equation} \label{III-4-a} %
            \lim_{\varepsilon\to
            +0}\left([e_2,-1-\varepsilon]_{Q_0}
            -[-1+\varepsilon, e_3]_{Q_0}\right)<0.
            \end{equation} %
            In this case, $\Gamma_c(-1)=\gamma_{e_2}^c$, where $\gamma_{e_2}^c$ crosses the interval $(-1,e_3)$.
Thus, there is one more face of
$Q_0(z)\,dz^2$ that is the  strip
            domain $\Omega_s(-i\infty,i\infty)$ with the sides
            $\Gamma_s^l(-i\infty,i\infty)=\gamma_{-i\infty,e_1}\cup[e_1,e_2]\cup\gamma_{e_2}^c\cup[e_2,e_1]\cup\gamma_{e_1,i\infty}$
            and
            $\Gamma_s^r=\gamma_{-i\infty,e_3}\cup\gamma_{e_3,i\infty}$.
            Finally the boundary of the end domain $\Omega_e^r$ is
$\Gamma_e^r=\gamma_{-i\infty,e_3}\cup[e_3,e_4]\cup\gamma_{e_4}^r\cup[e_4,e_3]\cup\gamma_{e_3,i\infty}$.
The Stokes graph and domain configuration  are shown in
Figure~\ref{fig:III-6-a} Case III-6-a.

 The mirror configuration, shown in Figure~\ref{fig:III-6-a} Case III-6-a-m, occurs when
 $-1<e_4<e_3<1<e_2<e_1$ and with these assumptions it happens if and only if
  \begin{equation} \label{III-4-a-m} %
            \lim_{\varepsilon\to
            +0}\left([e_2,1+\varepsilon]_{Q_0}
            -[1-\varepsilon, e_3]_{Q_0}\right)<0.
            \end{equation} %

\smallskip

(b) Let $e_3$ be the only zero on $\Gamma_c(-1)$. Under the
assumptions of the case~III-6, the latter
happens if and only if the limit in~\eqref{III-4-a} is positive. %
            In this case, $\Gamma_c(-1)=\gamma_{e_3}^l$, where $\gamma_{e_3}^l$ crosses the interval $(e_2,-1)$.
Under these circumstances, the set
$\Gamma_r^{(inn)}=\gamma_{e_3}^l\cup[e_3,e_4]\cup
\gamma_{e_4}^r\cup [e_4,e_3]$ is a boundary component of a face of
the Stokes graph of $Q_0(z)\,dz^2$, which  must be a ring domain
$\Omega_r$ by Lemma~3.1.  This implies that there is a critical
trajectory $\gamma_{e_2}^r$, which crosses the interval
$(x_4,\infty)$, where $x_4$ is defined earlier in the case 3, and
therefore $\Gamma_r^{(out)}=\gamma_{e_2}^r$.

            Finally the boundary of the end domain $\Omega_e^r$ is
$\Gamma_e^r=\gamma_{i\infty,e_1}\cup[e_1,e_2]\cup\gamma_{e_2}^r\cup[e_2,e_1]\cup\gamma_{e_1,-i\infty}$.
The Stokes graph and domain configuration  are shown in
Figure~\ref{fig:III-6-b} Case III-6-b.

The mirror configuration, shown in Figure~\ref{fig:III-6-b} Case
III-6-b-m, occurs when
 $1<e_4<e_3<1<e_2<e_1$ and if and only if the limit in~\eqref{III-4-a-m} is positive.

\smallskip

(c) Suppose now that $e_2,e_3\in \Gamma_c(-1)$.  Under the
assumptions of Case III-6, the latter
happens if and only if the limit in~\eqref{III-4-a} is zero. %
            In this case, $\Gamma_c(-1)=\gamma_{e_2,e_3}^+\cup \gamma_{e_3,e_2}^-$ and
            $\Gamma_e^r=\gamma_{i\infty,e_1}\cup[e_1,e_2]\cup\gamma_{e_2,e_3}^+\cup[e_3,e_4]
            \cup\gamma_{e_4}^r\cup [e_4,e_3]\cup \gamma_{e_3,e_2}^-\cup [e_2,e_1]\cup \gamma_{e_1,-i\infty}$.
The Stokes graph and domain configuration  are shown in
Figure~\ref{fig:III-6-c} Case III-6-c.

The mirror configuration, shown in Figure~\ref{fig:III-6-c} Case
III-6-c-m, occurs when $-1<e_4<e_3<1<e_2<e_1$ and  if and only if
the limit in~\eqref{III-4-a-m} is zero.

\medskip

7. Suppose that $e_1<e_2<-1<1<e_3<e_4$. In this case, the
intervals
        $(-\infty,e_1)$, $(e_2,-1)$, $(-1,1)$, $(1,e_3)$, and $(e_4,\infty)$ are
            orthogonal trajectories of $Q_0(z)\,dz^2$ and the intervals $(e_1,e_2)$ and $(e_3,e_4)$ are  trajectories of $Q_0(z)\,dz^2$.
Thus, there are two circle domains $\Omega_c(-1)$ and
$\Omega_c(1)$.

First, we mention few topological obstructions for the critical
trajectories starting at zeros $e_2$ and $e_3$. There is no
trajectories with one end point at one of these zeros and second
end point at $\infty$. Indeed, if such a trajectory, say
$\gamma_{e_2,i\infty}$, exists (then $\gamma_{e_2,-i\infty}$
exists as well),  then the zero $e_1$ is the only critical point
of $Q_0(z)\,dz^2$ in the simply connected domain $D\ni e_1$
bounded by the curve $\gamma_{-i\infty,e_2}\cup
\gamma_{e_2,i\infty}$, which is impossible, see part 1 of
Lemma~3.1. Similar argument shows that in the case under
consideration there are no trajectories joining the zeros $e_2$
and $e_3$, and there are no trajectories with both end points at
$e_2$ or at $e_3$, which cross the interval $(1,e_3)$ or the
interval $(e_2,-1)$, respectively. Thus, we are left with the
following subcases.

\smallskip

(a) Suppose that there are critical trajectories $\gamma_{e_2}^c$
and $\gamma_{e_3}^c$ crossing $(-1,1)$ at some points $x_1$ and
$x_2$, respectively, such that $x_1<x_2$. This case happens if and
only if the following inequalities hold:
\begin{equation} \label{III-4-a-1}
           \lim_{\varepsilon\to +0}\left( \im [e_2,-1+i\varepsilon]_{Q_0}-\im
           [e_3,-1+i\varepsilon]_{Q_0}\right)<0,
\end{equation} %
\begin{equation} \label{III-4-a-2}
           \lim_{\varepsilon\to +0}\left( \im [e_3,1+i\epsilon]_{Q_0}-\im
           [e_2,1+i\epsilon]_{Q_0}\right)<0.
\end{equation} %

Under these circumstances, $\Gamma_c(-1)=\gamma_{e_2}^c$,
$\Gamma_c(1)=\gamma_{e_3}^c$ and there are critical trajectories
$\gamma_{e_1,i\infty}$, $\gamma_{e_1,-i\infty}$,
$\gamma_{e_4,i\infty}$, and $\gamma_{e_4,-i\infty}$. This implies
that $\Gamma_e^l=\gamma_{-i\infty,e_1}\cup \gamma_{e_1,i\infty}$
and $\Gamma_e^r=\gamma_{i\infty,e_4}\cup \gamma_{e_4,-i\infty}$.
Furthermore, this implies that there is one more face of the
Stokes graph of $Q_0(z)\,dz^2$, which is the strip domain
$\Omega_s(-i\infty,i\infty)$ with the sides
$\Gamma_s^l(-i\infty,i\infty)=\gamma_{-i\infty,e_1}\cup
[e_1,e_2]\cup \gamma_{e_2}^c \cup [e_2,e_1] \cup
\gamma_{e_1,i\infty}$ and
$\Gamma_s^r(-i\infty,i\infty)=\gamma_{-i\infty,e_4}\cup
[e_4,e_3]\cup \gamma_{e_3}^c \cup [e_3,e_4] \cup
\gamma_{e_4,i\infty}$, see Figure~\ref{fig:III-7-a} Case III-7-a.

\smallskip

(b) Suppose that there are critical trajectories $\gamma_{e_2}^r$
crossing $(e_4,\infty)$ and $\gamma_{e_3}^c$ crossing $(-1,1)$.
This case happens if and only if the limit in~\eqref{III-4-a-1} is
positive and the limit in~\eqref{III-4-a-2} is negative. These
conditions imply that there is a critical trajectory
$\gamma_{e_4}^l$, which crosses the interval $(e_2,-1)$. The
remaining two critical trajectories in this case are
$\gamma_{e_1,i\infty}$ and $\gamma_{e_1,-i\infty}$. Now, when all
critical trajectories are identified, the domain configuration of
$Q_0(z)\,dz^2$ consists of end domains $\Omega_e^l$, $\Omega_e^r$,
circle domains $\Omega_c(-1)$, $\Omega_c(1)$, and a ring domain
$\Omega_r$. The corresponding boundaries are the following:
$\Gamma_e^l=\gamma_{-i\infty,e_1}\cup \gamma_{e_1,i\infty}$,
$\Gamma_e^r=\gamma_{i\infty,e_1}\cup [e_1,e_2]\cup \gamma_{e_2}^r
\cup [e_2,e_1]\cup\gamma_{e_1,-i\infty}$,
$\Gamma_c(-1)=\gamma_{e_4}^l\cup [e_4,e_3]\cup \gamma_{e_3}^c\cup
[e_3,e_4]$, $\Gamma_c(1)=\gamma_{e_3}^c$,
$\Gamma_r^{(out)}=\gamma_{e_2}^r$, and
$\Gamma_r^{(inn)}=\gamma_{e_4}^l$.
The Stokes graph and domain configuration  are shown in
Figure~\ref{fig:III-7-b} Case III-7-b.

The mirror configuration, shown in Figure~\ref{fig:III-7-b} Case
III-7-b-m, occurs when $e_4<e_3<-1<1<e_2<e_1$ and  if and only if
the following
inequalities hold: %
\begin{equation} \label{III-4-b-1}
           \lim_{\varepsilon\to +0}\left( \im [e_2,1+i\varepsilon]_{Q_0}-\im
           [e_3,1+i\varepsilon]_{Q_0}\right)>0,
\end{equation} %
\begin{equation} \label{III-4-b-2}
           \lim_{\varepsilon\to +0}\left( \im [e_3,-1+i\epsilon]_{Q_0}-\im
           [e_2,-1+i\epsilon]_{Q_0}\right)<0.
\end{equation} %

\smallskip

(c) Suppose that there are critical trajectories
$\gamma_{e_2,e_4}^+$ and  $\gamma_{e_2,e_4}^-$. This case happens
if and only if the limits in~\eqref{III-4-a-1} and~\eqref{III-4-a-2} are zero. The remaining critical trajectories in
this case are $\gamma_{e_1,i\infty}$, $\gamma_{e_1,-i\infty}$, and
$\gamma_{e_3}^c$. Under these conditions, the domain configuration
of $Q_0(z)\,dz^2$ consists of end domains $\Omega_e^l$,
$\Omega_e^r$ and circle domains $\Omega_c(-1)$, $\Omega_c(1)$. The
corresponding boundaries are the following:
$\Gamma_e^l=\gamma_{-i\infty,e_1}\cup \gamma_{e_1,i\infty}$,
$\Gamma_e^r=\gamma_{i\infty,e_1}\cup [e_1,e_2]\cup
\gamma_{e_2,e_4}^+ \cup \gamma_{e_4,e_2}^- \cup
[e_2,e_1]\cup\gamma_{e_1,-i\infty}$,
$\Gamma_c(-1)=\gamma_{e_2,e_4}^+\cup [e_4,e_3]\cup
\gamma_{e_3}^c\cup [e_3,e_4]\cup \gamma_{e_4,e_2}^-$, and
$\Gamma_c(1)=\gamma_{e_3}^c$, see Figure~\ref{fig:III-7-c} Case III-7-c.

The mirror configuration, shown in Figure~\ref{fig:III-7-c} Case
III-7-c-m, occurs when $e_4<e_3<-1<1<e_2<e_1$ and if and only if
the limits in~\eqref{III-4-b-1} and~\eqref{III-4-b-2} are zero.

\medskip

8. Suppose that $e_1<e_2<-1<e_3<1<e_4$. In this case, the
intervals
        $(-\infty,e_1)$, $(e_2,-1)$, $(-1,e_3)$, $(e_4,\infty)$ are
            orthogonal trajectories of $Q_0(z)\,dz^2$ and the intervals $(e_1,e_2)$, $(e_3,1)$ and  $(1,e_4)$
            are  trajectories of $Q_0(z)\,dz^2$.
Thus,  there is only one circle domain $\Omega_c(-1)$.

Applying the topological argument based on the  Basic Structure
Theorem \cite[Theorem~3.5]{Jenkins} and Lemma~3.1 once more, we
conclude that
 $e_1$ is the only zero on the boundary of $\Omega_e^l$
and therefore, $\Gamma_e^l=\gamma_{-i\infty,e_1}^l\cup
\gamma_{e_1,i\infty}$. Furthermore, similar topological argument
implies that $e_1,e_4\not\in \Gamma_c(-1)$. Therefore,
$\Gamma_c(-1)$ may contain one of the zeros $e_2$, $e_3$ or both
these zeros. Thus, we have the following subcases.

\smallskip

(a) Suppose that $e_2\in \Gamma_c(-1)$ but $e_3\not\in
\Gamma_c(-1)$. This happens if and only if the limit in~\eqref{III-4-a-1} is negative. Then
$\Gamma_c(-1)=\gamma_{e_2}^c$, where $\gamma_{e_2}^c$ intersects
the interval $(-1,e_3)$. In this case,
$\Gamma_s^l(-i\infty,i\infty)=\gamma_{-i\infty,e_1}\cup[e_1,e_2]\cup
\gamma_{e_2}^c\cup [e_2,e_1]\cup \gamma_{e_1,i\infty}$ is a
boundary arc of one of the faces of the Stokes graph of
$Q_0(z)\,dz^2$, which in this case must be a strip domain
$\Omega_s(-i\infty,i\infty)$ having $\Gamma_s^l(-i\infty,i\infty)$
as its left side. The right side of $\Omega_s(-i\infty,i\infty)$
is
$\Gamma_s^r(-i\infty,i\infty)=\gamma_{-i\infty,e_3}\cup\gamma_{e_3,i\infty}$.
Under these circumstances, there are critical trajectories
$\gamma_{e_4,i\infty}\subset \mathbb{H}_+$ and
$\gamma_{e_4,-i\infty}\subset \mathbb{H}_-$ and, hence, the
boundary of the end domain $\Omega_e^r$ is
$\Gamma_e^r=\gamma_{i\infty,e_4}\cup \gamma_{e_4,-i\infty}$. This
also implies that there are strip domains
$\Omega_s(1,i\infty)\subset \mathbb{H}_+$ and
$\Omega_s(-i\infty,1)\subset \mathbb{H}_-$, which sides are
$\Gamma_s^l(1,i\infty)=[1,e_3]\cup \gamma_{e_3,i\infty}$,
$\Gamma_s^r(1,i\infty)=[1,e_4]\cup \gamma_{e_4,i\infty}$ and
$\Gamma_s^l(-i\infty,1)=\gamma_{-i\infty,e_3}\cup [e_3,1]$,
$\Gamma_s^r(-i\infty,1)=\gamma_{-i\infty,e_4}\cup [e_4,1]$,
 respectively.
The Stokes graph and domain configuration are shown in
Figure~\ref{fig:III-8-a} Case III-8-a.

The mirror configuration for this case, shown in
Figure~\ref{fig:III-8-a} Case III-8-a-m, occurs if and only if
$e_4<-1<e_3<1<e_2<e_1$ and the limit in~\eqref{III-4-a-1} is
negative.

\smallskip

(b) Suppose that $e_3\in \Gamma_c(-1)$ but $e_2\not\in
\Gamma_c(-1)$. This happens if and only if the limit in~\eqref{III-4-a-1} is positive. Then $\Gamma_c(-1)=\gamma_{e_3}^l$,
where $\gamma_{e_3}^l$ intersects the interval $(e_2,-1)$ at some
point $x_1$. In this case,
$\Gamma_s^{(inn)}(1,1)=[1,e_3]\cup\gamma_{e_3}^l\cup [e_1,1]$ is a
boundary arc of one of the faces of the Stokes graph of
$Q_0(z)\,dz^2$, which in this case must be a strip domain
$\Omega_s(1,1)$ having $\Gamma_s^{(inn)}(1,1)$ as its inner side.
The outer side $\Gamma_s^{(out)}(1,1)$ of $\Omega_s(1,1)$ must
contain at least one of the zeros $e_2$, $e_4$. Therefore, we have
the following three subcases.

\smallskip

($\alpha$) Let $e_2\in \Gamma_s^{(out)}(1,1)$ but $e_4\not\in
\Gamma_s^{(out)}(1,1)$. This subcase happens if and only if
$2\im\,\left([e_2,i]_{Q_0}+[i,e_3]_{Q_0}\right)<\delta_1$. In this
subcase, there are critical trajectories $\gamma_{e_2,1}^+\subset
\overline{\mathbb{H}}_+$ and $\gamma_{e_2,1}^-\subset
\overline{\mathbb{H}}_-$ such that
$\Gamma_s^{(out)}(1,1)=\gamma_{1,e_2}^+\cup \gamma_{e_2,1}^-$.

Under these circumstances, there are critical trajectories
$\gamma_{e_4,i\infty}\subset \mathbb{H}_+$ and
$\gamma_{e_4,-i\infty}\subset \mathbb{H}_-$ and, hence, the
boundary of the end domain $\Omega_e^r$ is
$\Gamma_e^r=\gamma_{i\infty,e_4}\cup \gamma_{e_4,-i\infty}$. This
also implies that there are strip domains
$\Omega_s(1,i\infty)\subset \mathbb{H}_+$ and
$\Omega_s(-i\infty,1)\subset \mathbb{H}_-$, which sides are
$\Gamma_s^l(1,i\infty)=\gamma_{1,e_2}^+\cup[e_2,e_1]\cup
\gamma_{e_1,i\infty}$, $\Gamma_s^r(1,i\infty)=[1,e_4]\cup
\gamma_{e_4,i\infty}$ and
$\Gamma_s^l(-i\infty,1)=\gamma_{-i\infty,e_1}\cup [e_1,e_2]\cup
\gamma_{e_2,1}^-$,
$\Gamma_s^r(-i\infty,1)=\gamma_{-i\infty,e_4}\cup [e_4,1]$,
respectively.
The Stokes graph and domain configuration are shown in
Figure~\ref{fig:III-8-b-alpha} Case III-8-b-$\alpha$.

The mirror configuration, shown in Figure~\ref{fig:III-8-b-alpha}
Case III-8-b-$\alpha$-m, occurs if and only if
$e_4<-1<e_3<1<e_2<e_1$ and
$2\im\,\left([e_2,i]_{Q_0}+[i,e_3]_{Q_0}\right)<\delta_{-1}$.

\smallskip

($\beta$) Let $e_4\in \Gamma_s^{(out)}(1,1)$ but $e_2\not\in
\Gamma_s^{(out)}(1,1)$. This subcase happens if and only if
$2\im\,\left([e_2,i]_{Q_0}+[i,e_3]_{Q_0}\right)>\delta_1$. These
assumptions imply that $\Gamma_s^{(out)}(1,1)=[1,e_4]\cup
\gamma_{e_4}^l\cup [e_4,1]$, where $\gamma_{e_4}^l$ is  a critical
trajectory of $Q_0(z)\,dz^2$, which intersect the interval
$(e_2,-1)$ at some point $x_2<x_1$.

Under conditions of this subcase, the critical trajectory
$\gamma_{e_4}^l$ is an inner boundary component of a face of the
Stokes graph of $Q_0(z)\,dz^2$, which  must be a ring domain
$\Omega_r$. Hence, $\Gamma_r^{(inn)}=\gamma_{e_4}^l$. Under these
circumstances, the outer boundary component $\Gamma_r^{(out)}$ of
$\Omega_r$ must contain the zero $e_4$. Therefore, in this case
there is a critical trajectory $\gamma_{e_2}^r$ intersecting the
interval $(e_4,\infty)$ and $\Gamma_r^{(out)}=\gamma_{e_2}^r$.
There are no other critical trajectories in this case, which
implies that $\Gamma_e^r=\gamma_{-i\infty,e_1}\cup [e_1,e_2]\cup
\gamma_{e_2}^r\cup [e_2,e_1]\cup \gamma_{e_1,i\infty}$.
The Stokes graph and domain configuration  are shown in
Figure~\ref{fig:III-8-b-beta} Case III-8-b-$\beta$.

The mirror configuration, shown in Figure~\ref{fig:III-8-b-beta}
Case III-8-b-$\beta$-m, occurs if and only if
$e_4<-1<e_3<1<e_2<e_1$ and
$2\im\,\left([e_2,i]_{Q_0}+[i,e_3]_{Q_0}\right)>\delta_{-1}$.

\smallskip

($\gamma$) Let $e_2,e_4\in \Gamma_s^{(out)}(1,1)$. This subcase
happens if and only if the following equality holds:
$2\im\,\left([e_2,i]_{Q_0}+[i,e_3]_{Q_0}\right)=\delta_1$. Since
the zeros $e_2$ and $e_4$ both belong to $\Gamma_s^{(out)}(1,1)$
it follows that there are critical trajectories
$\gamma_{e_2,e_4}^+\subset \overline{\mathbb{H}}_+$ and
$\gamma_{e_2,e_4}^-\subset \overline{\mathbb{H}}_-$. In this case,
$\Gamma_s^{(out)}(1,1)=[1,e_4]\cup \gamma_{e_4,e_2}^+\cup
\gamma_{e_2,e_4}^-\cup [e_4,1]$. Now, when all critical
trajectories are identified, the boundary of the end domain
$\Omega_e^r$ is $\Gamma_e^r=\gamma_{-i\infty,e_1}\cup
[e_1,e_2]\cup \gamma_{e_2,e_4}^-\cup \gamma_{e_4,e_2}^+\cup
[e_2,e_1]\cup \gamma_{e_1,i\infty}$. There are no other domains in
the domain configuration of $Q_0(z)\,dz^2$.
The Stokes graph and domain configuration are shown in
Figure~\ref{fig:III-8-b-gamma} Case III-8-b-$\gamma$.

The mirror configuration, shown in Figure~\ref{fig:III-8-b-gamma}
Case III-8-b-$\gamma$-m, occurs if and only if
$e_4<-1<e_3<1<e_2<e_1$ and
$2\im\,\left([e_2,i]_{Q_0}+[i,e_3]_{Q_0}\right)=\delta_{-1}$.

\smallskip

(c) Suppose that $e_2,e_3\in \Gamma_c(-1)$. This happens if and
only if the limit in~\eqref{III-4-a-1} is zero. Since the
zeros $e_2$ and $e_3$ both belong to $\Gamma_c(-1)$ it follows
that there are critical trajectories $\gamma_{e_2,e_3}^+\subset
\overline{\mathbb{H}}_+$ and $\gamma_{e_2,e_3}^-\subset
\overline{\mathbb{H}}_-$. In this case, $\Gamma_c(-1)=
\gamma_{e_2,e_3}^-\cup \gamma_{e_3,e_2}^+$.

Furthermore,
$\Gamma_s^l(1,i\infty)=[1,e_3]\cup\gamma_{e_3,e_2}^+\cup
[e_2,e_1]\cup \gamma_{e_1,i\infty}$ is a boundary arc of one of
the faces of the Stokes graph of $Q_0(z)\,dz^2$, which  must be a
strip domain $\Omega_s(1,i\infty)$ having $\Gamma_s^l(1,i\infty)$
as its left side. The only possibility for the right side of
$\Omega_s(1,i\infty)$ is that $\Gamma_s^r(1,i\infty)=[1,e_4]\cup
\gamma_{e_4,i\infty}$. Similarly, we conclude that there is a
strip domain $\Omega_s(-i\infty,1)$ with sides
$\Gamma_s^l(-i\infty,1)=\gamma_{-i\infty,e_1}\cup[e_1,e_2]\cup\gamma_{e_2,e_3}^-\cup
[e_3,1]$ and $\Gamma_s^r(-i\infty,1)=\gamma_{-i\infty,e_4}\cup
[e_4,1]$.

The latter also implies that the boundary of the end domain
$\Omega_e^r$ is $\Gamma_e^r=\gamma_{-i\infty,e_4}\cup
\gamma_{e_4,i\infty}$. There are no other domains in the domain
configuration of $Q_0(z)\,dz^2$.
The Stokes graph and domain configuration are shown in
Figure~\ref{fig:III-8-c} Case III-8-c.

The mirror configuration, shown in Figure~\ref{fig:III-8-c} Case
III-8-c-m, occurs if and only if $e_4<-1<e_3<1<e_2<e_1$ and the
limit in~\eqref{III-4-a-2} is zero.

\medskip

9. Suppose that $e_1<-1<e_2<e_3<1<e_4$. In this case, the
intervals
        $(-\infty,e_1)$, $(e_2,e_3)$, $(e_4,\infty)$ are
            orthogonal trajectories of $Q_0(z)\,dz^2$ and the intervals $(e_1,-1)$, $(-1,e_2)$ and  $(e_3,1)$,
            $(1,e_4)$ are  trajectories of $Q_0(z)\,dz^2$.
There are no circle domains.

As in the previous cases, the topological constrains related to
the  Basic Structure Theorem \cite[Theorem~3.5]{Jenkins} and
Lemma~3.1 imply that there exist  eight critical trajectories,
each having one of its end points at $i\infty$ or $-i\infty$.
These critical trajectories are: $\gamma_{e_k,i\infty}$,
$k=1,2,3,4$ and $\gamma_{e_k,-i\infty}$, $k=1,2,3,4$. Now, when
the Stokes graph of $Q_0(z)\,dz^2$ is identified, one can easily
see that the domain configuration of $Q_0(z)\,dz^2$ consists of
end domains $\Omega_e^l$, $\Omega_e^r$ and five strip domains,
which are $\Omega_s(-1,i\infty)$, $\Omega_s(-1,-i\infty)$,
$\Omega_s(1,i\infty)$, $\Omega_s(1,-i\infty)$, and
$\Omega_s(-i\infty,i\infty)$. The boundaries of the end domains
$\Omega_e^l$ and $\Omega_e^r$ are
$\Gamma_e^l=\gamma_{-i\infty,e_1}\cup \gamma_{e_1,i\infty}$ and
$\Gamma_e^r=\gamma_{i\infty,e_4}\cup \gamma_{e_4,-i\infty}$,
respectively.  The sides of the strip domains
$\Omega_s(-1,i\infty)$, $\Omega_s(-1,-i\infty)$,
$\Omega_s(1,i\infty)$, $\Omega_s(1,-i\infty)$,
$\Omega_s(-i\infty,i\infty)$ are
$\Gamma_s^l(-1,i\infty)=[-1,e_1]\cup \gamma_{e_1,i\infty}$ and
$\Gamma_s^r(-1,i\infty)=[-1,e_2]\cup \gamma_{e_2,i\infty}$,
$\Gamma_s^l(-i\infty,-1)=\gamma_{-i\infty,e_1}\cup [e_1,-1]$ and
$\Gamma_s^r(-i\infty,-1)=\gamma_{-i\infty,e_2}\cup [e_2,-1]$,
$\Gamma_s^l(1,i\infty)=[1,e_3]\cup \gamma_{e_3,i\infty}$ and
$\Gamma_s^r(1,i\infty)=[1,e_4]\cup \gamma_{e_4,i\infty}$,
$\Gamma_s^l(-i\infty,1)=\gamma_{-i\infty,e_3}\cup [e_3,1]$ and
$\Gamma_s^r(-i\infty,1)=\gamma_{-i\infty,e_4}\cup [e_4,1]$,
$\Gamma_s^l(-i\infty,i\infty)=\gamma_{-i\infty,e_2}\cup
\gamma_{e_2,i\infty}$ and
$\Gamma_s^r(-i\infty,i\infty)=\gamma_{-i\infty,e_3}\cup
\gamma_{e_3,i\infty}$, respectively.
The Stokes graph and domain configuration in this case are shown
in Figure~\ref{fig:III-9} Case III-9.

 \bigskip

        {\textbf{ IV.}} In this part, we discuss possible Stokes graphs and domain configurations for the degenerate cases.
        Precisely, we describe changes, which occur in
        the Stokes graphs and domain configurations when two or
        more zeros merge. In the case without real zeros, there is only one degenerate configuration
        described in Case I-3 and illustrated in Figure~\ref{fig:I-3-deg}.
        In the cases with two or four real zeros, we have the following
        possibilities.

        \smallskip

        1. In the case with two real zeros $e_1$ and $e_2$,  these
        zeros can merge if and only if both belong to one of the
        intervals $(-\infty,-1)$, $(-1,1)$, or $(1,\infty)$. In
        all these cases an interval $(e_1,e_2)$ is one of the critical trajectories of $Q_0(z)\,dz^2$
         and therefore $\gamma_{e_1,e_2}=[e_1,e_2]$.
        Thus, when $e_1$ and $e_2$ merge, the arc $\gamma_{e_1,e_2}$ shrinks to a point $e_{1,2}$,
        while the structure of domain configuration remains the same as in the corresponding generic
        case. Therefore, the Stokes graphs and domain
        configurations in the degenerate cases mentioned above are the same as in the generic cases II-1 and
        II-2 shown in Figures~\ref{fig:II-1-a}-\ref{fig:II-2-c} except that the interval
        $[e_1,e_2]$ shrinks to a single point.

        2. Similar situation occurs in the case with four real zeros when the intervals
        $(e_1,e_2)$ and/or  $(e_3,e_4)$ are critical trajectories of
        $Q_0(z)\,dz^3$ as it is illustrated in Figures~\ref{fig:III-1}-\ref{fig:III-8-c}. In
        these cases, if $e_1$ merges with $e_2$ and/or $e_3$
        merges with $e_4$, then the domain configuration remains the same as in the generic case shown in the corresponding figure
        except that the interval $[e_1,e_2]$ and/or the interval $[e_3,e_4]$ shrinks to a single point.

         3. In case III with four real zeros there are situations
         when a ring domain and/or strip domain collapses if two
         or more zeros merge. Precisely, in cases III-1, III-3, III-4-b, III-4-c and in their  mirror cases
         the ring domain $\Omega_r$ collapses, when the zeros
         $e_2$ and $e_3$ merge forming a double zero $e_{2,3}$. The corresponding degenerate
         domain configurations for these cases are shown in
         Figures~\ref{fig:III-1-deg},~\ref{fig:III-3-deg},~\ref{fig:III-4-b-deg}, and~\ref{fig:III-4-c-deg}.

         Furthermore, in cases III-2, III-5, III-9 and in their mirror cases
         the strip domain $\Omega_s(-i\infty,i\infty)$ collapses
         when $e_2$ and $e_3$ merge forming a double zero $e_{2,3}$
         as it is shown in Figures~\ref{fig:III-2-deg},~\ref{fig:III-5-deg},
         and~\ref{fig:III-9-deg}. Similarly, in cases III-4-a-$\alpha$ and
         III-4-a-$\gamma$ the strip domain $\Omega_s(1,1)$
         collapses and in the corresponding mirror cases
         III-4-a-$\alpha$-m and III-4-a-$\gamma$-m the strip
         domain $\Omega_s(-1,-1)$ collapses when the zeros $e_2$
         and $e_3$ merge to a double zero $e_{2,3}$; see Figures~\ref{fig:III-4-b-deg}, and~\ref{fig:III-4-c-deg}.

         In the remaining case, that is III-4-a-$\beta$, the ring
         domain $\Omega_r$ and the strip domain $\Omega_s(1,1)$
         both collapse when the zeros $e_2$ and $e_3$ merge to a double zero $e_{2,3}$.
         Similarly, in the mirror case III-4-a-$\beta$-m the
         domains $\Omega_r$ and $\Omega_s(-1,-1)$ both collapse
         when $e_2$ and $e_3$ merge forming a double zero $e_{2,3}$. The resulting degenerate
         domain configurations in these cases are those shown in
         Figures~\ref{fig:III-4-b-deg}, and~\ref{fig:III-4-c-deg}.

         Further merging of zeros, when a double zero merges
         with one or two single zeros, does not change the domain
         configurations. Possible cases are the following. In each
         of the cases III-1-deg, III-1-deg-m and III-2-deg shown in
         Figures~\ref{fig:III-1-deg} and~\ref{fig:III-2-deg}, the double zero $e_{2,3}$ can merge
         with $e_1$, then the edge
         $[e_1,e_{2,3}]$ shrinks to a point forming a triple zero,
         or $e_{2,3}$ can merge
         with $e_4$, then the edge
         $[e_4,e_{2,3}]$ shrinks to a point again forming a triple
         zero, or $e_{2,3}$ can merge
         with with both $e_1$ and $e_4$ forming a zero of order four, then
         both edges $[e_1,e_{2,3}]$ and $[e_{2,3}],e_4]$ shrink to this zero of order four.

         In all cases shown in Figures~\ref{fig:III-3-deg}-\ref{fig:III-5-deg}, the double zero
         $e_{2,3}$ can merge with $e_1$ forming a triple zero. In
         all these cases with triple zero, the domain
         configurations contain the same domains as shown in
         Figures~\ref{fig:III-3-deg}-\ref{fig:III-5-deg} and the Stokes graphs consists of the same
         edges as in these figures, except that the edge
         $[e_1,e_{2,3}]$ shrinks to a point forming a triple zero
         of the corresponding quadratic differential.


\section{Domain configurations for the Rabi model}\label{sec:DomConfRabi}

The existence and properties of solutions to Rabi problem depend
on the values of physical parameters $\Delta$, $E$, and $g$, while
our classification of the Stokes graphs for this problem is given
in terms of the number of real zeros and some other
characteristics of associated quadratic differential
$Q_0(z)\,dz^2$. Thus, to apply the results presented in Section~4
to Rabi problem, we have to identify which of the types I, II, or
III of Stokes graphs and domain configurations of $Q_0(z)\,dz^2$
correspond to  a particular choice of the Rabi parameters
$\Delta$, $E$, and $g$.

For given $\Delta$, $E$, and $g$, the coefficients $c_k$ of the
quartic polynomial $P_0(z)$, that is the numerator of $Q_0(z)$ in
formula~\eqref{s6aa}, are expressed explicitly by formulas~\eqref{eq2.9} and~\eqref{eq2.10}  as functions of  $\Delta$, $E$,
and $g$. Thus, to use our classification of Stokes graphs to study
the Rabi problem, we have to determine if the polynomial $P_0(z)$
with these coefficients  has no real zeros, has two real zeros, or
it has four real zeros.

The theory of quartic equations, which origin goes back to the
work of Lodovico Ferrari in the $16^{th}$ century, is well known
and contains all information on distribution of zeros of such
equations, which we need for our study. Precisely, in
Propositions~5.1 and 5.2 below, we present classical results of J.
L. Lagrange \cite[Chapitre V, Article III, Section 39, p.
67]{Lagrange} (see, also, Chapter IV, Section 7 in \cite{Dickson})
interpreted in terms of the parameters $\Delta$, $E$, and $g$ of
the Rabi problem. As it was shown by J. L. Lagrange, the number of
real roots of $P_0(z)$ depends on the signs of the discriminant
$\mathcal{D}_0$ and two additional characteristics,
$\mathcal{P}_0$ and $\mathcal{Q}_0$, of the polynomial $P_0$,
which
are defined as follows: %
\begin{eqnarray*} %
\mathcal{D}_0&=&
-27c_3^4c_0^2+18c_3^3c_2c_1c_0-4c_3^3c_1^3-4c_3^2c_2^3c_0+c_3^2c_2^2c_1^2+
144c_3^2c_2c_0^2
\\ %
&-&6c_3^2c_1^2c_0-80c_3c_2^2c_1c_0+18c_3c_2c_1^3+16c_2^4c_0-4c_2^3c_1^2
-192c_3c_1c_0^2 \\ %
&-&128c_2^2c_0^2+144c_2c_1^2c_0-27c_1^4+256c_0^3, %
\end{eqnarray*} %
$$ 
\mathcal{P}_0=8c_2-3c_3^2, \quad %
\mathcal{Q}_0=64c_0-16c_2^2+16c_3^2c_2-16c_3c_1-3c_3^4.%
$$ 

Using equations~\eqref{eq2.9} and~\eqref{eq2.10}, we express
 $\mathcal{D}_0$, $\mathcal{P}_0$, and  $\mathcal{Q}_0$, as the
following functions of the physical parameters $\Delta$, $E$, and
$g$: %
\begin{eqnarray*} %
\mathcal{D}_0&=& -\frac{1}{4g^{20}}\left(1024 \Delta^6 g^8-1024
\Delta^4 E^2 g^8+2048 \Delta^6 E g^6-2048 \Delta^4 E^3 g^6 \right.\\ %
&-& 1024 \Delta^4 E g^8 + 4608 \Delta^2 E g^{10}-4096 E^3
g^{10}+512 \Delta^8 g^4 +512 \Delta^6 E^2 g^4\\ %
&+&1024 \Delta^6 g^6 - 1024 \Delta^4 E^4 g^4-3072 \Delta^4 E^2
g^6+2048 \Delta^4 g^8+7680 \Delta^2 E^2 g^8 \\ %
&+& 2304 \Delta^2 g^{10} - 8192 E^4 g^8 - 6144 E^2 g^{10}+1728
g^{12}+512 \Delta^8 E g^2
\\%
&-& 512 \Delta^6 E^3 g^2+ 512 \Delta^6 E g^4- 2048 \Delta^4 E^3
g^4+3328 \Delta^4 E g^6  %
+ 1536 \Delta^2 E^3 g^6 \\%
&+& 7680 \Delta^2 E g^8-4096 E^5 g^6 - 16384 E^3 g^8+3840 E g^{10}
+64 \Delta^{10}-64 \Delta^8 E^2 \\ %
&+& 256 \Delta^8 g^2-768 \Delta^6 E^2 g^2+512 \Delta^6 g^4 +1024
\Delta^4 E^2 g^4+2176 \Delta^4 g^6 \\%
&-& 1536 \Delta^2 E^4 g^4+2304 \Delta^2 E^2 g^6 +2208 \Delta^2
g^8-10240 E^4 g^6-2688 E^2 g^8 \\%
&+& 2944 g^{10}-64 \Delta^8 E-224 \Delta^6 E g^2 +256 \Delta^4 E^3
g^2+2048 \Delta^4 E g^4 \\%
&-& 3072 \Delta^2 E^3 g^4+672 \Delta^2 E g^6 -3840 E^3 g^6+5504 E
g^8-64 \Delta^8+96 \Delta^6 E^2\\%
&+& 16 \Delta^6 g^2+384 \Delta^4 E^2 g^2 -52 \Delta^4 g^4-1440
\Delta^2 E^2 g^4-48 \Delta^2 g^6+960 E^4 g^4\\%
&+& 4480 E^2 g^6+1024 g^8 +96 \Delta^6 E-40 \Delta^4 E g^2+96
\Delta^2 E^3 g^2+96 \Delta^2 E g^4\\%
&+& 1920 E^3 g^4+1744 E g^6 +20 \Delta^6-52 \Delta^4 E^2-84
\Delta^4 g^2+144 \Delta^2 E^2 g^2\\%
&+& 60 \Delta^2 g^4+672 E^2 g^4 -216 g^6-52 \Delta^4 E+72 \Delta^2
E g^2-48 E^3 g^2-288 E g^4\\%
&-& 2 \Delta^4+12 \Delta^2 E^2 +12 \Delta^2 g^2-72 E^2 g^2-25
g^4+12 \Delta^2 E-26 E g^2-E^2\\%
&-&g^2-E),
\end{eqnarray*} %

$$ %
 \mathcal{P}_0=g^{-4}(16 E g^2+8 \Delta^2+8 g^2-5), %
 $$ %
 and
 $$ %
 \mathcal{Q}_0
 =-\frac{8}{g^8}(8 \Delta^2 g^4+8 \Delta^2 E g^2+2 \Delta^4+4 \Delta^2 g^2-8 E g^2-3 \Delta^2-4
 g^2+1).
$$ %

Now, having the functions
$\mathcal{D}_0=\mathcal{D}_0(\Delta,E,g)$,
$\mathcal{P}_0=\mathcal{P}_0(\Delta,E,g)$, and
$\mathcal{Q}_0=\mathcal{Q}_0(\Delta,E,g)$ depending on the
physical parameters of Rabi problem in hand, we can use J. L.
Lagrange theorem as it was stated in \cite[Chapter IV, Section
7]{Dickson}, to identify which of the types \textbf{I},
\textbf{II}, or \textbf{III} of Stokes graphs and domain
configurations of $Q_0(z)\,dz^2$ described in Section~4
corresponds to a given choice of the parameters $\Delta$, $E$, and
$g$.

\begin{prop}[Generic cases] %
Suppose that the parameters $\Delta$ and $E$ of the Rabi problem
are real and that $g^2\not=0$ is also real. Then the following
holds:
\begin{enumerate} %
\item[\textbf{I.}] Quadratic differential $Q_0(z)\,dz^2$ has four
distinct complex zeros, which are in conjugate pairs, if and only
if the discriminant  $\mathcal{D}_0(\Delta,E,g)$ is positive  and
at least one of the functions $\mathcal{P}_0(\Delta,E,g)$ and
$\mathcal{Q}_0(\Delta,E,g)$ is also positive. In this case,
possible Stokes graphs and domain configurations of $Q_0(z)\,dz^2$
are described in cases 1, 2, and 3 in part \textbf{I} of Section~4. %

\item[\textbf{II.}] Quadratic differential $Q_0(z)\,dz^2$ has a
pair of complex conjugate zeros and two distinct real zeros not
equal $\pm 1$ if and only if the discriminant
$\mathcal{D}_0(\Delta,E,g)$ is negative and $E\not=-g^2$,
$E\not=-g^2\pm 1$, $E\not=-g^2+2$.
In this case, possible Stokes graphs and domain
configurations of $Q_0(z)\,dz^2$ are described in part \textbf{II}
of Section~4.

\item[\textbf{III.}] Quadratic differential $Q_0(z)\,dz^2$ has
four distinct real zeros not equal $\pm 1$ if and only if the
discriminant $\mathcal{D}_0(\Delta,E,g)$ is positive, both
functions $\mathcal{P}_0(\Delta,E,g)$ and
$\mathcal{Q}_0(\Delta,E,g)$ are negative, and  $E\not=-g^2$,
$E\not=-g^2\pm 1$, $E\not=-g^2+2$.
In this case, possible Stokes
graphs and domain configurations of $Q_0(z)\,dz^2$
are described in part \textbf{III} of Section~4. %
\end{enumerate} %
\end{prop} %

\medskip


To specify possible positions of multiple zeros, we follow J. L.
Lagrange work cited above. For this, we need two more
characteristics, $\mathcal{R}_0$ and $\mathcal{S}_0$,  of the
polynomial $P_0(z)$, which can be expressed in terms of the
coefficients $c_k$ of $P_0(z)$
and in terms of the parameters of the Rabi problem as follows:%
$$
\mathcal{R}_0=c_3^3+8c_1-4c_3c_2=g^{-6}(1-16g^4-16Eg^2-4\Delta^2-7g^2)
$$ %
and %
\begin{eqnarray*} 
\mathcal{S}_0&=&c_2^2-3c_3c_1+12c_0
=\frac{1}{16g^8}(1-192\Delta^2g^4+256E^2
g^4+64\Delta^2Eg^2 \\ %
&+&256Eg^4+16\Delta^4+32\Delta^2g^2  %
+64g^4+32Eg^2-8\Delta^2+16g^2+1).
\end{eqnarray*} 

\begin{prop}[Degenerate cases with full set of critical points and multiple zeros of $P_0(z)$] %
Suppose that the parameters $\Delta$, $E$, and $g^2\not=0$ of the
Rabi problem are real such that $E\not=-g^2$, $E\not=-g^2\pm 1$,
$E\not=-g^2+2$. Then $P_0(z)$ has a multiple zero $\not= \pm 1$
if and only if $\mathcal{D}_0(\Delta,E,g)=0$. %

Furthermore, if $\mathcal{D}_0(\Delta,E,g)=0$, then the following
subcases refining positions of zeros of $P_0(z)$ happen:
\begin{enumerate} %
\item[1.] If $\mathcal{Q}_0(\Delta,E,g)=0$,
$\mathcal{P}_0(\Delta,E,g)>0$, and $\mathcal{R}_0(\Delta,E,g)=0$,
then there are two double complex conjugate zeros.  In this case,
possible Stokes graphs and domain configurations of $Q_0(z)\,dz^2$
are described in the case 4 in part~\textbf{I} of Section~4. %

\item[2.]  If $\mathcal{Q}_0(\Delta,E,g)=0$ and
$\mathcal{P}_0(\Delta,E,g)<0$, then there are two double real
zeros $\not= \pm 1$.

\item[3.] If $\mathcal{Q}_0(\Delta,E,g)=0$ and
$\mathcal{S}_0(\Delta,E,g)=0$, then $P_0(z)$ has a real zero of
order four at $z=-\frac{1}{4g^2}$.

\item[4.] If $\mathcal{P}_0(\Delta,E,g)<0$,
$\mathcal{Q}_0(\Delta,E,g)<0$, and
$\mathcal{S}_0(\Delta,E,g)\not=0$, then there is a double real
zero $\not= \pm 1$ and two simple real zeros  $\not= \pm 1$.

\item[5.] If $\mathcal{Q}_0(\Delta,E,g)>0$ or if
$\mathcal{P}_0(\Delta,E,g)>0$ and at least one of the quantities
$\mathcal{Q}_0(\Delta,E,g)$ and $\mathcal{R}_0(\Delta,E,g)$ is not
zero, then there are a double real zero  $\not= \pm 1$ and two
complex conjugate zeros.

\item[6.] If $\mathcal{S}_0(\Delta,E,g)=0$ and
$\mathcal{Q}_0(\Delta,E,g)\not=0$, then there are a triple real
zero  $\not= \pm 1$ and a simple real zero  $\not= \pm 1$.

\end{enumerate} %
\end{prop} %

\medskip

We stress here that Proposition~5.2 describes all possible cases
when $P_0(z)$ has multiple zeros and the quadratic differential
$Q_0(z)\,dz^2$ has double poles at the points $\pm 1$.

Equations~\eqref{eq2.9} and~\eqref{eq2.10} provide a parametric
description of the set of all quadruples $(c_3,c_2,c_1,c_0)\in
\mathbb{R}^4$, which coordinates are the coefficients of the
numerator $P_0(z)$ of the quadratic differential $Q_0(z)\,dz^2$
that appears in the framework of the
Rabi problem. Next, we prove two results, 
which provide more explicit description of this set.


 \begin{theorem} \label{Theorem 5.1} %
Let 
$
Q_0(z)\,dz^2=-\frac{z^4+c_3z^3+c_2z^2+c_1z+c_0}{(z-1)^2(z+1)^2}\,dz^2
$
be  a quadratic differential  associated with the Rabi problem for
some choice of the parameters $\Delta$, $E$, and $g$, such that
$\Delta,E,g^2\in
\mathbb{R}$, $g\not=0$. 
Then %
\begin{equation} \label{5.1} %
(c_3,c_2,c_1,c_0)\in \{a\}\times S(a), \quad {\mbox{where
$a=g^{-2}$,}} %
 \end{equation}%
and $S(a)$ denotes a parabolic cylinder in $\mathbb{R}^3$ defined by %
\begin{equation} \label{5.2} 
S(a)=\{(X,Y,Z)\in \mathbb{R}^3:\,
(Y+a)^2-a^2X-a^2Z-(1/4)a^2(4+3a^2)=0\}.
\end{equation}

Moreover, there is a constant $c\ge 0$ such that the coordinates
$X$, $Y$, $Z$ in~\eqref{5.2}, representing the coefficients $c_2$,
$c_1$, $c_0$, are related via the following
equations: %
\begin{equation} \label{5.4} 
X=-\frac{1}{4a}(8Y+a^3+16a)+c, \quad  \quad
Z=\frac{1}{2a^2}(2Y^2+8aY-a^4 + 8a^2)-c.
\end{equation} 
  \end{theorem}  %

\noindent %
 \emph{Proof.} To prove the first part of this theorem, we put $a=g^{-2}$ and suppose that the quadratic differential $Q_0(z)\,dz^2$
 is associated with the Rabi problem having parameters $\Delta$, $E$, and $g$, such that $\Delta,E,g^2\in
\mathbb{R}$, $g\not=0$. Since $c_3=a=g^{-2}$ by the first equation
in~\eqref{eq2.9}, we have to show that $(c_2,c_1,c_0)\in S(a)$.
 Substituting expressions for $c_2$, $c_1$, and $c_0$ given by
 equations~\eqref{eq2.9} and~\eqref{eq2.10} for the variables
 $X$, $Y$, and $Z$ in ({5.2}), we obtain %
 \begin{eqnarray*} %
(c_1+a)^2-a^2c_2-a^2c_0-\frac{a^2}{4}(4+3a^2)&=&\left(-\frac{1}{2g^4}(4g^2+2E+1)+g^{-2}
\right)^2\\ %
&-&\frac{1}{4g^8}(8Eg^2+4\Delta^2+4g^2-1)\\
&+&\frac{1}{4g^8}(4\Delta^2-4E^2-4E+1) \\ %
&-&\frac{1}{4g^8}(4g^4+3).
 \end{eqnarray*} %
 Simplifying the latter equation, we find that the right-hand side of this equality equals zero and therefore the
 point  $(c_2,c_1,c_0)$ lies on the surface of the parabolic
 cylinder~\eqref{5.2}.

To prove relations~\eqref{5.4}, we substitute $Y$ for $c_1$ in the
first equation in~\eqref{eq2.9} and then solve it for $E$ to get %
\begin{equation} \label{5.5}%
E=-\frac{1}{2a^2}(2Y+a^2+4a).
\end{equation}  %
Substituting this expression for $E$, $X$ for $c_2$, and $Z$ for
$c_0$ in  equations~\eqref{eq2.9} and~\eqref{eq2.10}, we obtain
relations~\eqref{5.4} with
$c=a^2\Delta^2\ge 0$.  \hfill $\Box$ %

\smallskip

Theorem~\ref{Theorem 5.1} shows that equality $a=g^{-2}$ and
relations~\eqref{5.2} and~\eqref{5.4} are necessary for the point
$(X,Y,Z)\in \mathbb{R}^3$ to represent coefficients $c_2$, $c_1$,
and $c_0$ of the quadratic differential $Q_0(z)\,dz^2$ associated
with the Rabi problem. Our next result shows that these conditions
are also sufficient.

 \begin{theorem} \label{Theorem 5.2} %
If $X$, $Y$, and $Z$ satisfy equations~\eqref{5.2} and~\eqref{5.4}
with some $a\not=0$ and $c\ge 0$, then there are parameters
$\Delta$, $E$, and $g$ of the Rabi problem such that
$g^{-2}=a$ and %
\begin{equation} \label{5.6} %
 X=c_2(\Delta,E,g) \quad Y=c_1(E,g), \quad Z=c_0(\Delta,E,g)
\end{equation} %
with $c_2(\Delta,E,g)$, $c_1(E,g)$, and $c_0(\Delta,E,g)$ defined
by~\eqref{eq2.9} and~\eqref{eq2.10}. %
  \end{theorem}  %

\noindent %
 \emph{Proof.} We can choose $g$ so that $a=g^{-2}$. Then, for a
 given $Y$, we choose $E$ as in~\eqref{5.5}. Solving~\eqref{5.5}
 for $Y$, we obtain %
 \begin{equation} \label{5.6.1}
Y=-(a/2)(2aE+a+4).
 \end{equation} %

 Substituting this expression for $Y$ and $g^{-2}$ for $a$ in
 equations~\eqref{5.4}, we find %
 \begin{equation} \label{5.6.2} %
X=\frac{1}{4g^4}(8g^2E+4g^2-1)+c, \quad
Z=\frac{1}{4g^4}(4E^2+4E-1)-c.
 \end{equation} %
 Choosing $\Delta$ so that $c=\Delta^2/g^4>0$, substituting this
in the latter equations, and taking into account equations~\eqref{eq2.9},~\eqref{eq2.10}, we obtain the desired
equations~\eqref{5.6}.~\hfill~$\Box$

 \smallskip

 \begin{figure} \label{Fig 1}%

\centering
\begin{minipage}{1.0\textwidth}
\hspace{1cm}
\includegraphics[width=0.9\textwidth]{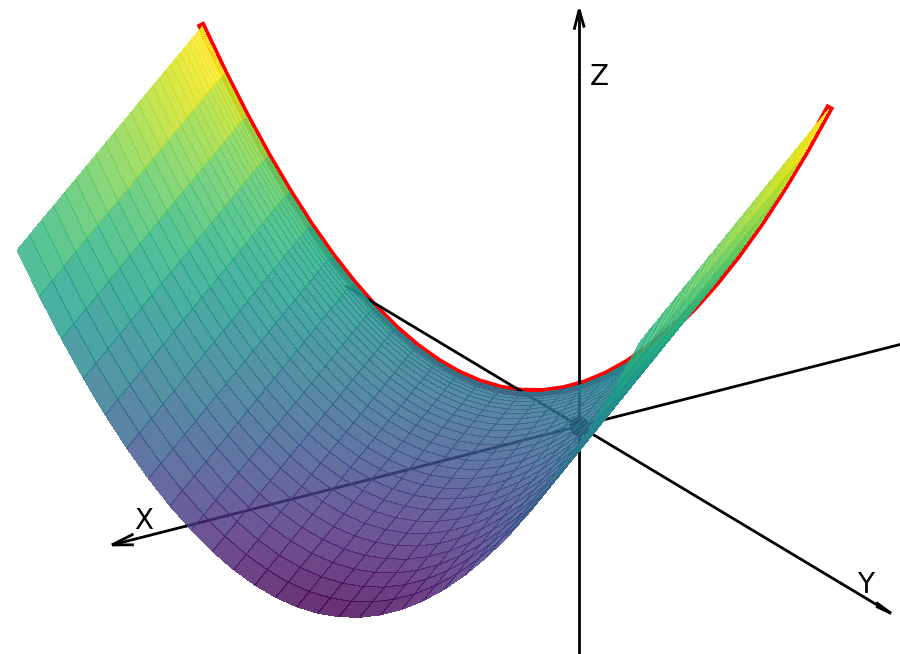}

\vspace{0cm} %
\end{minipage}
\caption{Portion of the cylinder $S(a)$ associated to the Rabi problem with parameters $g^{2}=a^{-1}=1$,  $\Delta\ge 0$, and  $E\in \mathbb{R}$.}%
\end{figure}

Since the four real coefficients $c_k$, $k=0,1,2,3$, of the
polynomial $P_0(z)$ depend on only three real parameters,
$\Delta$, $E$ and $g^2$, of the Rabi model, it is reasonable to
expect that some of the Stokes graphs and domain configurations of
$Q_0(z)\,dz^2$ described in Section~4 will not appear in the
framework of the Rabi problem. We start our discussion of possible
and impossible types of Stokes graphs with a simple result, which
excludes the possibility of graphs symmetric with respect to the
imaginary axis.

\begin{lemma} \label{Lemma 5.0} %
There are no Stokes graphs and domain configurations of
$Q_0(z)\,dz^2$ symmetric with respect to the imaginary axis, which
are associated with the Rabi problem having the parameters
$\Delta$, $E$, and $g$ such that $\Delta,E,g^2\in \mathbb{R}$,
$g\not=0$.
\end{lemma} %

\noindent %
 \emph{Proof.} %
Since the Stokes graphs appeared in this study already posses
symmetry with respect to the real axis, the symmetry with respect
to the imaginary axis occurs when the zeros of $P_0(z)$ are in
pairs symmetric with respect to the origin.
In this case, the polynomial $P_0(z)$
is
biquadratic; i.e. $ P_0(z)=z^4+c_2z^2+c_0$. In particular, $c_3=0$ in this case. %
Under our assumptions on the Rabi parameters, $c_3(g)=g^{-2}$
represents the boson-fermion coupling $g$. Thus, the case $c_3=0$
does not occur for finite values of $g$ and therefore Stokes graphs symmetric with respect to the imaginary axis
do not appear in the Rabi problem under our assumptions. %
 \hfill $\Box$ %

\smallskip


Although the symmetry of Stokes graphs with respect to the
imaginary axis does not occur for finite values of $g$, the
symmetric cases are possible as ``asymptotic cases'', which appear
in the Rabi model when the boson-fermion coupling tends to
infinity. Possible structures of Stokes graphs in these asymptotic
cases will be discussed in Section~6.



Next, we examine possibility of so-called ``breaks of symmetry''
in the Rabi model. Let
$\mathbf{\Omega}=\mathbf{\Omega}(\Delta,E,g)$ denote the domain
configuration of the quadratic differential $Q_0(z)\,dz^2$ defined
by~\eqref{s6aa} with the coefficients $c_k=c_k(\Delta,E,g)$,
$k=0,1,2,3$, given by equations~\eqref{eq2.9} and~\eqref{eq2.10}.
By ``break of symmetry'' in the Rabi model we understand a
situation, when a certain domain configuration
$\mathbf{\Omega}=\mathbf{\Omega}(\Delta,E,g)$ corresponds to some
values of the Rabi parameters $\Delta$, $E$, and $g$, such that
$\Delta,E,g^2\in \mathbb{R}$, $g\not=0$, but its mirror
configuration, call it $\widetilde{\mathbf{\Omega}}$, does not
correspond to any choice of such $\Delta$, $E$, and $g$. As the
following lemma shows, such breaks of symmetry never occur in the
settings of the Rabi problem with $\Delta,E,g^2\in
\mathbb{R}$, $g\not=0$. %

\begin{lemma} \label{Lemma 5.1} %
For any domain configuration
$\mathbf{\Omega}=\mathbf{\Omega}(\Delta,E,g)$ corresponding to the
Rabi problem with $\Delta,E,g^2\in \mathbb{R}$, $g\not=0$, there
exists a mirror domain configuration
$\widetilde{\mathbf{\Omega}}$, which corresponds to the Rabi
problem with the parameters $\Delta$, $-(E+1)$, and $ig$; i.e.
$\widetilde{\mathbf{\Omega}}=\mathbf{\Omega}(\Delta, -(E+1),ig)$.
\end{lemma} %

\noindent %
 \emph{Proof.} Let $\mathbf{\Omega}=\mathbf{\Omega}(\Delta,E,g)$
 be the domain configuration of $Q_0(z)\,dz^2$ with coefficients
 $c_k=c_k(\Delta,E,g)$ defined by formulas~\eqref{eq2.9},~\eqref{eq2.10} with $\Delta,E,g^2\in \mathbb{R}$, $g\not=0$. Let
 $e_k=e_k(\Delta,E,g)$, $k=1,2,3,4$, denote zeros of $Q_0(z)\,dz^2$. Since the
 coefficients $c_k$ are real, it follows that the mirror domain
 configuration $\widetilde{\mathbf{\Omega}}$ corresponds to the
 quadratic differential $\widetilde{Q}_0(z)\,dz^2$, which has
 zeros at the points $\widetilde{e}_k=-e_k(\Delta,E,g)$,
 $k=1,2,3,4$. From this, using Vieta's formulas
 (see, for instance,~\cite[Section 26, formula (1)]{van der Waerden}
 or~\cite[Remark 3.14]{Vinberg}), we conclude that
 the coefficients $\widetilde{c}_k$, $k=0,1,2,3$, of the numerator
 $\widetilde{P}_0(z)$ of $\widetilde{Q}_0(z)$ satisfy the
 following equations:
 $$ 
\widetilde{c}_3=-c_3(g), \ \widetilde{c}_2=c_2(\Delta,E,g), \
\widetilde{c}_1=-c_1(E,g), \ \widetilde{c}_0=c_0(\Delta,E,g).
$$ 

Matching these equations with appropriate equations for the
coefficients $c_k$ given by formulas~\eqref{eq2.9},~\eqref{eq2.10}, we obtain the following: %
\begin{eqnarray*} 
\widetilde{c}_3&=&-g^{-2}=(ig)^{-2},\\
\widetilde{c}_2&=&\frac{1}{4g^4}\left(8Eg^2+4\Delta^2+4g^2-1\right)
\\ %
&=&\frac{1}{4(ig)^4}\left(8(-(E+1))(ig)^2+4\Delta^2+4(ig)^2-1\right),\\ %
\end{eqnarray*} %
\begin{eqnarray*}
\widetilde{c}_1&=&\frac{1}{2g^4}\left(4g^2+2E+1\right)=-\frac{1}{2(ig)^4}\left(4(ig)^2+2(-(E+1))+1\right),\\
\widetilde{c}_0&=&-\frac{1}{4g^4}\left(4\Delta^2-4E^2-4E+1\right)
\\ %
&=&-\frac{1}{4(ig)^4}\left(4\Delta^2-4(-(E+1))^2-4(-(E+1))+1\right).
\end{eqnarray*} %

As the latter equations show, the domain configuration
$\widetilde{\mathbf{\Omega}}=\mathbf{\Omega}(\Delta,-(E+1),ig)$
corresponding to the Rabi parameters $\Delta$, $-(E+1)$, and $ig$
is the mirror domain configuration for the domain configuration
$\mathbf{\Omega}=\mathbf{\Omega}(\Delta,E,g)$ for a given set of
the Rabi parameters $\Delta$, $E$, and $g$. %
 \hfill $\Box$

 \smallskip




As Lemma~\ref{Lemma 5.0} shows, the Stokes graphs of
$Q_0(z)\,dz^2$ symmetric with respect to the imaginary axis do not
appear in the Rabi problem. Next, we discuss possibility of cases
when critical points of $Q_0(z)\,dz^2$ possess some ``partial
symmetries''. In our next  lemma, we study the case when all zeros
of $Q_0(z)\,dz^2$ lie on the same vertical line $\{z:\,\re
z=\alpha\}$. Since the mirror configuration always exists, in this
lemma we assume without loss of generality that $\alpha>0$.

\begin{lemma} \label{Lemma 5.2} %
For every $\alpha>0$ and $\beta_1\ge 0$, there is a unique
$\beta_2> \beta_1$ such that the quadratic differential
$Q_0(z)\,dz^2$ with zeros $e_1=\alpha+i\beta_1$,
$e_2=\alpha+i\beta_2$, $e_3=\alpha-i\beta_1$, and
$e_4=\alpha-i\beta_2$ is associated with the Rabi problem for some
$\Delta$, $E$, and $g$ such that $\Delta,E,g^2\in \mathbb{R}$,
$g\not=0$. Precisely, $\beta_2=\beta_2(\alpha,\beta_1)$ is given
by the following equation:
\begin{equation} \label{5.7.-1} %
\beta_2=\sqrt{\beta_1^2-8\alpha}.
\end{equation}  %

Furthermore, the Rabi parameters $g$, $E$, and $\Delta$
corresponding to the given values $\alpha$ and  $\beta_1$ are
defined by the following equations:
\begin{equation} \label{5.7.-2} 
g^{-2}=-4\alpha, \quad
E=\frac{1}{4\alpha}(\alpha^2-6\alpha+2+\beta_1^2), \quad
\Delta^2=\frac{1}{4\alpha^2}(3\alpha^2-4\alpha+1+\beta_1^2).
\end{equation}  %
\end{lemma} %

\noindent %
 \emph{Proof.} Let $e_1=\alpha+i\beta_1$, $e_2=\alpha+i\beta_2$,
 $e_3=\alpha-i\beta_1$, and $e_4=\alpha-i\beta_2$  with
 $\alpha>0$, $0<\beta_1<\beta_2$, be zeros of $P_0(z)$. Then %
 \begin{equation} \label{5.7} %
P_0(z)=z^4-4\alpha
z^3+(6\alpha^2+\beta_1^2+\beta_2^2)z^2-2\alpha(2\alpha^2+\beta_1^2+\beta_2^2)z+(\alpha^2+\beta_1^2)(\alpha^2+\beta_2^2).
 \end{equation} %
We define $a=g^{-2}-4\alpha$ and %
\begin{equation} \label{5.7.1} %
\delta_1=\beta_1^2+\beta_2^2, \quad \delta_2=\beta_1^2\beta_2^2. %
\end{equation} %
As in Theorems~\ref{Theorem 5.1} and~\ref{Theorem 5.2}, we will
use notations $X=c_2$, $Y=c_1$, and $Z=c_0$ for the appropriate
coefficients of $P_0(z)$. With these notations, the coefficients
of the polynomial~\eqref{5.7} are the following: %
 \begin{equation} \label{5.8} %
X=\frac{3}{8}a^2+\delta_1, \quad
Y=\frac{a^3}{16}+\frac{1}{2}a\delta_1, \quad
Z=\frac{a^4}{16^2}+\frac{1}{16}a^2\delta_1+\delta_2. %
 \end{equation} 

Equating the right-hand side of the second equation in~\eqref{5.8}
to the right-hand side of  equation~\eqref{5.6.1} and solving the resulting equation for $\delta_1$, we find that  %
\begin{equation} \label{5.8.1} %
\delta_1=-2aE-\frac{1}{8}a^2-a-4.
\end{equation} %

Substituting this expression for $\delta_1$ in the first equation
in~\eqref{5.8} and then equating its right-hand side to the
right-hand side of the first equation in~\eqref{5.6.2} and then
solving the resulting equation for $c$, we find that %
\begin{equation} \label{5.8.2} %
c=a\Delta^2=-4aE +\frac{1}{2}a^2-2a-4.
\end{equation} %

To find an expression for $\delta_2$, we replace $c$ in the second
equation in~\eqref{5.6.2} with the right-hand side of equation~\eqref{5.8.2} and we replace $\delta_1$ in the third equation in~\eqref{5.8} with the right-hand side of equation~\eqref{5.8.1}.
Then we equate the resulting expressions and solve this equation
for $\delta_2$ to get, after some algebra, the following:

\begin{equation} \label{5.8.3} %
\delta_2=\left(aE+\frac{1}{16}a^2-\frac{1}{2}a+2\right)\left(aE+\frac{1}{16}a^2+\frac{3}{2}a+2\right).
\end{equation} %

It follows from equations~\eqref{5.7.1} that $\beta_1^2$ and
$\beta_2^2$ are solutions of the quadratic equation %
\begin{equation} \label{5.8.4} %
\tau^2-\delta_1\tau+\delta_2=0
\end{equation} %
with $\delta_1$ and $\delta_2$ given in~\eqref{5.8.1}) and~\eqref{5.8.3}. Calculating the discriminant $\nabla$ of this equation, we find that%
$$ %
\nabla=\delta_1^2-4\delta_2=4a^2>0.
$$ %
Solving equation~\eqref{5.8.4} for $\tau=\beta_k^2$, $k=1,2$, we
find that %
\begin{equation} \label{5.8.5} %
\beta_1^2=-aE-\frac{1}{16}a^2-\frac{3}{2}a-2, \quad
\beta_2^2=\beta_1^2+2a=-aE-\frac{1}{16}a^2+\frac{1}{2}a-2.
\end{equation} %
The second of these equations implies~\eqref{5.7.-1}. Solving the
first of these equations for $E$, we obtain the second equation in~\eqref{5.7.-2}.

Next, we recall that $c=a^2\Delta^2$. Substituting this expression
for $c$ and the expression for $E$ given by the second equation in~\eqref{5.7.-2} in~\eqref{5.8.2} and then solving the resulting
equation for $\Delta^2$, we obtain the third equation in~\eqref{5.7.-2}). \hfill $\Box$ %

\begin{corollary} \label{Corollary 5.1} %
As equation~\eqref{5.7.-1} shows, degenerate configurations, when
$Q_0(z)\,dz^2$ has two conjugate double zeros or one real zero
$\not=\pm 1$ of order four, do not appear in the framework of the
Rabi problem with $\Delta,E,g^2\in \mathbb{R}$, $g\not=0$.
\end{corollary} %


\smallskip

Another case, when a ''partial symmetry'' may be important, is
when zeros of $Q_0(z)\,dz^2$ lie on two horizontal lines
$\{z:\,\im z=\pm \beta\}$, $\beta>0$. But, as the following lemma
shows, this case do not appear in solutions of the Rabi problem
with $\Delta,E,g^2\in \mathbb{R}$.

\begin{lemma} \label{Lemma 5.3} %
Suppose that the quadratic differential $Q_0(z)\,dz^2$ with
complex zeros $e_1=\alpha_1+i\beta_1$, $e_2=\alpha_2+i\beta_2$,
$e_3=\alpha_1-i\beta_1$, $e_4=\alpha_2-i\beta_2$, such that
$\beta_1>0$, $\beta_2>0$, is associated with the Rabi problem for
some values of the parameters $\Delta,E,g^2\in \mathbb{R}$. Then
$\beta_1\not=\beta_2$.
\end{lemma} %

\noindent %
 \emph{Proof.} Suppose, by contradictions, that $\beta_1=\beta_2$
 for some choice of $\Delta,E,g^2\in \mathbb{R}$.
 Using the ''mirror configuration'' argument once more, we may
assume without loss of generality that in this case  the zeros are
$e_1=(\alpha-\delta)+i\beta$, $e_2=(\alpha+\delta)+i\beta$,
$e_3=(\alpha-\delta)-i\beta$, and $e_4=(\alpha+\delta)-i\beta$
with $\alpha<0$, $\beta>0$, and $\delta>0$.
 With these zeros, the polynomial $P_0(z)$, which is the numerator of $Q_0(z)$, is the following:
\begin{eqnarray*} \label{5.9.3} %
P_0(z)&=&z^4-4\alpha
z^3+2(3\alpha^2-\delta^2+\beta^2)z^2-4\alpha(\alpha^2-\delta^2+\beta^2)z
\\
&+&(\alpha^2-\delta^2)^2
+\beta^2(2\alpha^2+2\delta^2+\beta^2).
\end{eqnarray*} %
We define $a=-4\alpha$ and, as before, we use $X$, $Y$, and $Z$ to denote the coefficients $c_2$, $c_1$, and $c_0$ of $P_0(z)$. Then%

$$ 
X=2((3/16)a^2+\beta^2-\delta^2), \quad
Y=a((1/16)a^2+\beta^2-\delta^2), \quad
 $$%
 $$
Z=((1/16)a^2-\delta^2)^2+\beta^2((1/8)a^2+2\delta^2+\beta^2). %
$$ 

Equating these expressions for $X$, $Y$, and $Z$ to the
corresponding expressions in the formulas (\ref{5.6.1}) and (\ref{5.6.2}),
and then solving the resulting equations for $E$, $c$, and $\delta$, we find that  %
$$ 
E=
-\frac{a^2\beta^2+16\beta^4+8a\beta^2+4a^2+32\beta^2}{16a\beta^2},
\quad c=\frac{16\beta^4+(3a^2+16)\beta^2+4a^2}{4\beta^2},
$$ %
and %
$$
\delta^2=-\frac{1}{4}\frac{a^2}{\beta^2}.
$$
Since $a^2>0$, $\beta^2>0$, the latter equation contradicts the
assumption that $\delta>0$, which proves the lemma.
 \hfill $\Box$ %

Next, we examine a possibility of two real zeros symmetric with
respect to the origin and two complex conjugate zeros.

\begin{lemma} \label{Lemma 5.4} %
The quadratic differential $Q_0(z)\,dz^2$ with zeros
$e_1=-\alpha$, $e_2=\alpha$,  $e_3=\delta+i\beta$,
$e_4=\delta-i\beta$, where $\alpha>0$, $\beta\ge 0$, is associated
with the Rabi problem for some $\Delta,E,g^2\in \mathbb{R}$ if and
only if one of the following conditions holds true: %
\begin{enumerate} %
\item[(a)] $\alpha\in (1,\sqrt{8/5})\cup (\sqrt{2},2)$ and
$\delta^2>\frac{\alpha^2(\alpha^2-1)}{4-\alpha^2}$, %
\item[(b)] $\sqrt{8/5})\le \alpha\le \sqrt{2}$ and
$\delta^2>\frac{4(\alpha^2-1)^2}{\alpha^2+2}$, %
\end{enumerate}
and if and only if %
\begin{equation}  \label{5.10.2}
\beta^2=\frac{(4-\alpha^2)\delta^2-\alpha^2(\alpha^2-1)}{\alpha^2-1}.
\end{equation}

Furthermore, the Rabi parameters $g$, $E$, and $\Delta$
corresponding to the given values $\alpha$ and  $\delta$,
satisfying conditions (a) and (b), are defined by the following
equations:
\begin{equation} \label{5.10.3} 
g^{-2}=-2\delta, \quad E=-\frac{\delta+\alpha^2-2}{2\delta}, \quad
\Delta^2=\frac{(\alpha^2+2)\delta^2-4(\alpha^2-1)^2}{4\delta^2(\alpha^2
- 1)}.
\end{equation}  %
\end{lemma} %

\noindent %
 \emph{Proof.} With zeros defined in this lemma, the numerator of the quadratic differential $Q_0(z)\,dz^2$ has the form
  \begin{equation} \label{5.10.4} %
P_0(z)=z^4-2\delta
z^3+(\delta^2-\alpha^2+\beta^2)z^2+2\alpha^2\delta
z-\alpha^2(\delta^2+\beta^2).
 \end{equation} %
Thus, in this case $c_3=g^{-2}=-2\delta$. As before, using
notations $X=c_2$, $Y=c_1$, and $Z=c_0$ for the appropriate
coefficients of $P_0(z)$, we find that %
 \begin{equation} \label{5.10.5} %
X=\delta^2-\alpha^2+\beta^2, \quad Y=2\alpha^2\delta, \quad Z=-\alpha^2(\delta^2+\beta^2). %
 \end{equation} 
Equating these expressions for $X$, $Y$, and $Z$ to the
corresponding expressions in formulas (\ref{5.6.1}) and
(\ref{5.6.2}), and then solving the resulting equations for $E$,
$\Delta^2=c/(4\delta^2)$, and $\beta^2$,
we obtain the second and the third equations in (\ref{5.10.3}) and equation (\ref{5.10.2}). %

Now, simple algebra shows that the right-hand side of equation
(\ref{5.10.2}) is  positive if and only if $1<\alpha<2$ and
$\delta^2>(\alpha^2(\alpha^2-1))/(4-\alpha^2)$. Furthermore, if
$1<\alpha<2$, then the right-hand side of the third equation in
(\ref{5.10.3}) is  positive if and only if
$\delta^2>(4(\alpha^2-1)^2)/(\alpha^2+2)$. Combining these cases,
we conclude that relation (\ref{5.10.2}) and  the inequalities %
\begin{equation}  \label{5.10.1} %
1<\alpha<2, \quad
\delta^2\ge\max\left\{\frac{\alpha^2(\alpha^2-1)}{4-\alpha^2},\frac{4(\alpha^2-1)^2}{\alpha^2+2}\right\}
\end{equation} %
 are necessary and
sufficient for the quadratic differential $Q_0(z)\,dz^2$ to be
associated with the Rabi problem and the corresponding physical
parameters are given by formulas
(\ref{5.10.3}). %

Finally, comparing functions
$\frac{\alpha^2(\alpha^2-1)}{4-\alpha^2}$ and
$\frac{4(\alpha^2-1)^2}{\alpha^2+2}$, one can easily find that
conditions (a) and (b)  of Lemma~5.10 are satisfied if and only if
inequalities (\ref{5.10.1}) are satisfied. %
 \hfill $\Box$ %

 \smallskip

A similar result, for the quadratic differential $Q_0(z)\,dz^2$
with four real zeros such that two of them are symmetric with
respect to the origin, is presented in the following lemma.

\begin{lemma} \label{Lemma 5.5} %
The quadratic differential $Q_0(z)\,dz^2$ with zeros
$e_1=-\alpha$, $e_2=\alpha$,  $e_3=\delta-\beta$,
$e_4=\delta+\beta$, where $\alpha>0$, $\beta> 0$, $\delta>0$, is
associated with the Rabi problem for some $\Delta,E,g^2\in
\mathbb{R}$ if and only if one of the following conditions holds
true: %
\begin{enumerate} %
\item[(a)] $0<\alpha<1$ and $\delta^2\le
\frac{4(\alpha^2-1)^2}{\alpha^2+2}$, %
\item[(b)] $\alpha\in (1,\sqrt{8/5}]\cup [\sqrt{2},2]$ and
$\frac{4(\alpha^2-1)^2}{\alpha^2+2}\le \delta^2< \frac{\alpha^2(\alpha^2-1)}{4-\alpha^2}$, %
\item[(c)] $\alpha>2$ and $\delta^2\ge
\frac{4(\alpha^2-1)^2}{\alpha^2+2}$,
\end{enumerate} %
and if and only if %
\begin{equation}  \label{5.10.20}
\beta^2=-\frac{(4-\alpha^2)\delta^2-\alpha^2(\alpha^2-1)}{\alpha^2-1}.
\end{equation}

Furthermore, the Rabi parameters $g$, $E$, and $\Delta$
corresponding to the given values $\alpha$ and  $\delta$,
satisfying conditions (a), (b), (c), are defined by equations
(\ref{5.10.3}).
\end{lemma} %

\noindent %
 \emph{Proof.} The proof is similar to the proof of the previous lemma.
 Under the assumptions, the numerator of the quadratic differential $Q_0(z)\,dz^2$ has the form
  \begin{equation} \label{5.10.40} %
P_0(z)=z^4-2\delta
z^3+(\delta^2-\alpha^2-\beta^2)z^2+2\alpha^2\delta
z-\alpha^2(\delta^2-\beta^2).
 \end{equation} %
Thus, in this case $c_3=g^{-2}=-2\delta$. With the notations
$X=c_2$, $Y=c_1$, and $Z=c_0$, we have the following: %
 \begin{equation} \label{5.10.50} %
X=\delta^2-\alpha^2-\beta^2, \quad Y=2\alpha^2\delta, \quad Z=-\alpha^2(\delta^2-\beta^2). %
 \end{equation} 
Equating these expressions for $X$, $Y$, and $Z$ to the
corresponding expressions in the formulas (\ref{5.6.1}) and
(\ref{5.6.2}), and then solving the resulting equations for $E$,
$\Delta^2=c/(4\delta^2)$, and $\beta^2$, we conclude that $g$,
$E$, and $\Delta$ are given by equations (\ref{5.10.3}) as in
Lemma~\ref{Lemma 5.4} and that $\beta$ is given by equation
(\ref{5.10.20}).

After simple algebra, left to the interested reader, we conclude
that, under the assumptions of Lemma~5.11, the right-hand side of
equation (\ref{5.10.20}) and the right-hand side of the third
equation in (\ref{5.10.3}) are non-negative if and only if
$\alpha$ and $\delta$ satisfy conditions (a), (b), (c) of the
lemma. %
 \hfill $\Box$ %

\smallskip

As we have shown earlier, the quadratic differentials
$Q_0(z)\,dz^2$ associated with the Rabi problem may have all zeros
on the same vertical line, but may not have all zeros on two
horizontal lines. Next, we show that a similar effect happens for
quadratic differentials with all zeros on the same circle centered
at the origin and all zeros on two rays issuing from the origin.

\begin{lemma} \label{Lemma 5.7} %
For every $r>0$ and $0\le \theta_1<\frac{\pi}{2}$, there is a
unique $\theta_2$, $0\le \theta_2\le \pi$, such that the quadratic
differential $Q_0(z)\,dz^2$ with zeros $e_1=re^{i\theta_1}$,
$e_2=re^{i\theta_2}$, $e_3=re^{-i\theta_1}$, and
$e_4=re^{-i\theta_2}$ is associated with the Rabi problem for some
$\Delta,E,g^2\in \mathbb{R}$. This unique $\theta_2$ does not
depend on $r$ and it is given by the following equation:
\begin{equation}  \label{5.10.200}
\theta_2=\pi-\arccos(\frac{1}{3}\cos \theta_1). 
\end{equation}

Furthermore, the Rabi parameters $g$, $E$, and $\Delta$
corresponding to the given values $r$ and $\theta_1$ are defined
by equations
\begin{equation} \label{5.10.30000} 
g^{-2}=-4r\alpha, \quad E=\frac{1}{4r\alpha}(r^2-2r\alpha+2),
\quad \Delta^2=\frac{r^2+1-2r^2\alpha^2}{4r^2\alpha^2},
\end{equation}  %
where $ \alpha=\frac{1}{3}\cos\theta_1$. %
\end{lemma} %

\noindent %
 \emph{Proof.}  Suppose that  $Q_0(z)\,dz^2$ has zeros $e_1=re^{i\theta_1}$,
$e_2=re^{i\theta_2}$, $e_3=re^{-i\theta_1}$, and
$e_4=re^{-i\theta_2}$. Then, the numerator of this quadratic
differential has the form
  \begin{equation} \label{5.10.400} %
P_0(z)=z^4-4r\delta z^3+2r^2(1+2(\delta^2-\beta^2))z^2-4r^3\delta
z+r^4,
 \end{equation} %
where
\begin{equation} \label{5.10.4000} 
\delta=(1/2)(\cos\theta_1+\cos\theta_2), \quad \quad
\beta=(1/2)(\cos\theta_1-\cos\theta_2).
 \end{equation} 

Thus, in this case $c_3=g^{-2}=-4r\delta$. Identifying the
coefficients $c_2$, $c_1$, and $c_0$ with the coordinates $X$,
$Y$, and $Z$ of $\mathbb{R}^3$, we obtain the following relations: %
 \begin{equation} \label{5.10.500} %
X=2r^2(1+2(\delta^2-\beta^2)), \quad Y=-4r^3\delta, \quad Z=r^4. %
 \end{equation} 

Equating these expressions for $X$, $Y$, and $Z$ to the
corresponding expressions in formulas (\ref{5.6.1}) and
(\ref{5.6.2}), and then solving the resulting equations for $E$,
$\Delta^2=c/(4\delta^2)$, and $\beta^2$, we obtain
\begin{equation} \label{5.10.301} 
E=\frac{1}{4r\delta}(r^2-2r\delta+2), \quad
\Delta^2=\frac{r^2+1-2r^2\delta^2}{4r^2\delta^2},
\end{equation}  %
and %
$$ 
\beta^2=4\delta^2.
$$ 
The latter equation together with (\ref{5.10.4000}) leads to the
following quadratic equation for the quotient
$y=\frac{\cos\theta_2}{\cos\theta_1}$:
$$ %
3y^2+10y+3=0,
$$
which solutions are $y=-\frac{1}{3}$ and $y=-3$. The solutions
correspond to two configurations, which are mirror configurations
to each other. Thus, without loss of generality, we assume that
$y=-\frac{1}{3}$. Then, $\cos\theta_2=-\frac{1}{3}\cos\theta_1$,
which gives (\ref{5.10.200}).

Furthermore, substituting $-\frac{1}{3}\cos\theta_1$ for
$\cos\theta_2$ into the first equation in (\ref{5.10.4000}), we
find that
$\delta=(1/2)(\cos\theta_1+\cos\theta_2)=\frac{1}{3}\cos\theta_1=\alpha$.
This, together with the relation $g^{-2}=-4r\delta$ and equations
(\ref{5.10.301}), gives equations (\ref{5.10.30000}).

It remains to verify that the right-hand side in the third
equation in (\ref{5.10.30000}) is non-negative. The later is
immediate from the following obvious inequality:
$\alpha^2=\frac{1}{9}\cos^2\theta_1<\frac{1}{2}+\frac{1}{2r^2}$. %
\hfill $\Box$

\smallskip

For the quadratic differential $Q_0(z)\,dz^2$ with zeros on two
rays issuing from the origin, which are symmetric to each other
with respect to the real axis, we have the following result.

\begin{lemma} \label{Lemma 5.8} %
Suppose that the quadratic differential $Q_0(z)\,dz^2$ with
complex zeros $e_1=r_1e^{i\theta_1}$, $e_2=r_2e^{i\theta_2}$,
$e_3=r_1e^{-i\theta_1}$, $e_4=r_2e^{-i\theta_2}$, such that
$r_k>0$, $0<\theta_k<\pi$, $k=1,2$,  is associated with the Rabi
problem for some values of the parameters $\Delta,E,g^2\in
\mathbb{R}$. Then $\theta_1\not=\theta_2$.
\end{lemma} %

\noindent %
 \emph{Proof.} Suppose, by contradictions, that $0<\theta_1=\theta_2=\theta<\pi$
 for some choice of $\Delta,E,g^2\in \mathbb{R}$.
 We may
assume without loss of generality that  the zeros are
$e_1=r_1e^{i\theta}$, $e_2=r_2e^{i\theta}$, $e_3=r_1e^{-i\theta}$,
$e_4=r_2e^{-i\theta}$ with $0<\theta<\pi/2$. Then, the polynomial
$P_0(z)$ has the form:
 $$ 
P_0(z)=z^4-4\delta t\,z^3+2(2\delta^2-\beta+2\beta
t^2)z^2-4\delta\beta t\,z+\beta^2,
$$ 
where %
\begin{equation} \label{eq1-Lemma5.8}%
\delta=(1/2)(r_1+r_2), \quad \beta=r_1r_2, \quad t=\cos\theta.
\end{equation}  %

As before, we use the coordinates $X$, $Y$, and $Z$ to denote the coefficients $c_2$, $c_1$, and $c_0$ of $P_0(z)$. Thus, %
$$ 
X=2(2\delta^2-\beta+2\beta t^2), \quad Y=-4\delta\beta, \quad
Z=\beta^2. %
$$ 

Equating these expressions for $X$, $Y$, and $Z$ to the
corresponding expressions for the coordinates $X$, $Y$, $Z$ in
formulas (\ref{5.6.1}) and (\ref{5.6.2}), one can solve the
resulting equations for $E$, $c$, and $\beta$. We only need the
following resulting expression for $\beta$:
$$ 
\beta=\frac{1+3t^2}{1-t^2}\,\delta^2.
$$ %
Using this equations and relations (\ref{eq1-Lemma5.8}), we obtain
the following quadratic equation for the ratio $y=r_1/r_2$:
$$ %
(1+3t^2)y^2+2(5t^2-1)y+(1+3t^2)=0.
$$ %
We recall that $t=\cos\theta$ and therefore the discriminant
$\bigtriangledown=16t^2(t^2-1)$ of the latter equation is
negative. Therefore, the ratio $r_1/r_2$ is not real contradicting
our assumption that $r_k>0$, $k=1,2$. This proves the lemma.
 \hfill $\Box$ %


\section{Asymptotic behavior for rescaled Rabi problem}\label{sec:Asymptotic}
In this section, we describe possible limit cases of the quadratic
differential $Q_0(z)dz^2$, when the boson-fermion coupling $g$
grows without bounds; i.e. when $|g|\to \infty$. To guarantee the
existence of the limit quadratic differential, we impose the
following 
conditions on the level
    of separation of the fermion mode $\Delta$ and on the eigenvalue $E$ of the
    Hamiltonian:

\begin{equation}\label{6.1} %
 E/g^2 \to E_a, \quad  \Delta^2/g^4\to \Delta_a^2 \quad \quad {\mbox{as $|g|\to
 \infty$.}}
\end{equation}
Here, $E_a\in \mathbb{R}$, $\Delta_a\ge 0$ and the subscript
``$a$'' stands for ``asymptotic''.

 Under these conditions, the polynomial $P_0(z)$
defined by equations (2.8) - (2.10) reduces to the biquadratic
polynomial
$P_a(z)=z^4+c_2z^2+c_0$ with the coefficients %
\begin{equation}\label{6.2} 
c_2 = 2E_a + \Delta^2_a, \quad c_0 = E_a^2 - \Delta^2_a,%
\end{equation} %
which zeros can be calculated as follows: %
\begin{alignat}{3} 
e_1 &= &\sqrt{-\frac{1}{2} \, \Delta_a^{2} - E_a - \frac{\Delta_a}{2} \, \sqrt{\Delta_a^2 + 4 \, E_a + 4}}, \nonumber\\
e_2 &= &\sqrt{-\frac{1}{2} \, \Delta_a^2 - E_a +
\frac{\Delta_a}{2} \, \sqrt{\Delta_a^2 + 4 \, E_a + 4}},\nonumber\\ %
e_3  &= -&\sqrt{-\frac{1}{2} \, \Delta_a^2  - E_a + \frac{\Delta_a}{2} \, \sqrt{\Delta_a^2 + 4 \, E_a + 4}},\nonumber\\
e_4 &= -&\sqrt{-\frac{1}{2} \, \Delta_a^2  - E_a-
\frac{\Delta_a}{2} \, \sqrt{\Delta_a^2 + 4 \, E_a + 4}}.\nonumber
\end{alignat} %
Accordingly, the quadratic differential $Q_0(z)\,dz^2$ reduces to
the quadratic differential %
\begin{equation}\label{6.3}
Q_a (z)dz^2 = -\frac{z^4 + c_2z^2 +
c_0}{(z-1)^2(z+1)^2}\,dz^2=-\frac{(z-e_1)(z-e_2)(z-e_3)(z-e_4)
}{(z-1)^2(z+1)^2}\,dz^2.
\end{equation}

We stress here that the expressions for the coefficients $c_2$,
$c_0$ in (\ref{6.2}) and the numeration of zeros $e_1$, $e_2$,
$e_3$, $e_4$ in this section may differ from those used in
Section~4. Notice that the coefficients of $Q_a(z)\,dz^2$ are real
and its set of zeros is symmetric with respect to both real and
imaginary axis and therefore its Stokes graph and domain
configuration are also symmetric with respect to both axis.
Since the zeros of $Q_a(z)\,dz^2$ are given explicitly, the type
of the associated Stokes graph can be easily identified in terms
of the asymptotic parameters $E_a$ and $\Delta_a$. Precisely, the
type of the Stokes graph of $Q_a(z)\,dz^2$ is determined by the
numbers of real and complex zeros of $Q_a(z)\,dz^2$ and therefore
it is determined by the combination of signs of the expressions
inside of the radicals in the formulas for $e_1$, $e_2$, $e_3$,
$e_4$ given
above; i.e. the type depends on the functions  %
$$ 
\Delta_a^2+4E_a+4 \quad {\mbox{and}} \quad
-\frac{1}{2}\Delta_a^2-E_a \pm \frac{\Delta_a}{2}
\sqrt{\Delta_a^2+4E_a+4}.
$$ 

 Next, we describe the
sets of the parameters $E_a$ and $\Delta_a$, which correspond to
the types of Stokes graphs and domain configurations
introduced in Section~4. Our designation 
of possible cases here is the same as in Section~4.

\smallskip

\textbf{I}-1.  The quadratic differential $Q_a(z)\,dz^2$ has four
distinct pure imaginary zeros if and only if the following
conditions hold: %
$$ %
\Delta_a>0, \quad \Delta_a^2+4E_a+4>0, \quad -\frac{1}{2}
\Delta_a^{2} - E_a +\frac{\Delta_a}{2} \sqrt{\Delta_a^2 + 4E_a +
4}<0. %
$$ %
Using these relations, we conclude, after simple algebra, that
this case occurs if and only if $(\Delta_a,E_a)\in \mathcal{I}_4$,
(here $\mathcal{I}_4$ stands for ``four distinct pure imaginary zeros),
where the set $\mathcal{I}_4\subset \mathbb{R}^2$ is defined as the following union: %
$$ %
\mathcal{I}_4=\{(X,Y):\, Y>X>0\}\cup \{(X,Y):\, X>2,\,
-1-\frac{1}{4}X^2<Y<-X\},
$$ %
see Figure~2, where the set $\mathcal{I}_4$ is shown in the yellow
color.

Under these conditions, $\im e_1>\im e_2>0>\im e_3>\im e_4$. Then
the intervals $(-i\infty,e_4)$, $(e_3,e_2)$, $(e_1,i\infty)$ of
the imaginary axis are critical trajectories of $Q_a(z)\,dz^2$ and
the intervals $(e_4,e_3)$ and $(e_2,e_1)$ are critical orthogonal
trajectories of $Q_a(z)\,dz^2$.  This implies that the domain
configuration of $Q_a(z)\,dz^2$ has a ring domain $\Omega_r$.
Therefore, in this case the Stokes graph and domain configuration
belong to the type described in part \textbf{I}-1 of Section~4,
see Figure~2.

\smallskip

\textbf{I}-2. The quadratic differential $Q_a(z)\,dz^2$ has four
distinct complex zeros, each  having non-zero real
and imaginary parts, if and only if %
$$ %
\Delta_a>0, \quad \Delta_a^2+4E_a+4<0
$$ %
or, equivalently, if and only if  $(\Delta_a,E_a)\in
\mathcal{C}_4$, where $\mathcal{C}_4$ stands for ``four complex
zeros'' and the set $\mathcal{C}_4\subset \mathbb{R}^2$ is defined
as $\mathcal{C}_4=\{(X,Y):\, X\not=0,\,Y<-\frac{1}{4}X^2-1\}$; see
Figure~2, where the set $\mathcal{C}_4$ is shown in the blue
color.

In this case, the imaginary axis is a trajectory of
$Q_a(z)\,dz^2$. This implies that the domain configuration has a
strip domain $\Omega_s(-i\infty,i\infty)$. Therefore, in this case
the Stokes graph and domain configuration belong to the type
described in part \textbf{I}-2 of Section~4, see Figure~2.

\smallskip

In the case \textbf{I}-3 discussed in Section~4, the Stokes graph
is not symmetric with respect to the imaginary axis and therefore
this type of graphs do not appear in the asymptotic cases
considered in this section. Also, the case \textbf{I}-4 of
Section~4 is a degenerate case when $Q_0(z)\,dz^2$ has a multiple
zero. These cases will be considered later in this section.

\smallskip

 \begin{figure}\label{asymptotic_plane.png}%
\centering
\begin{minipage}{1.0\textwidth}
\hspace{1cm}
\includegraphics[width=0.7\textwidth]{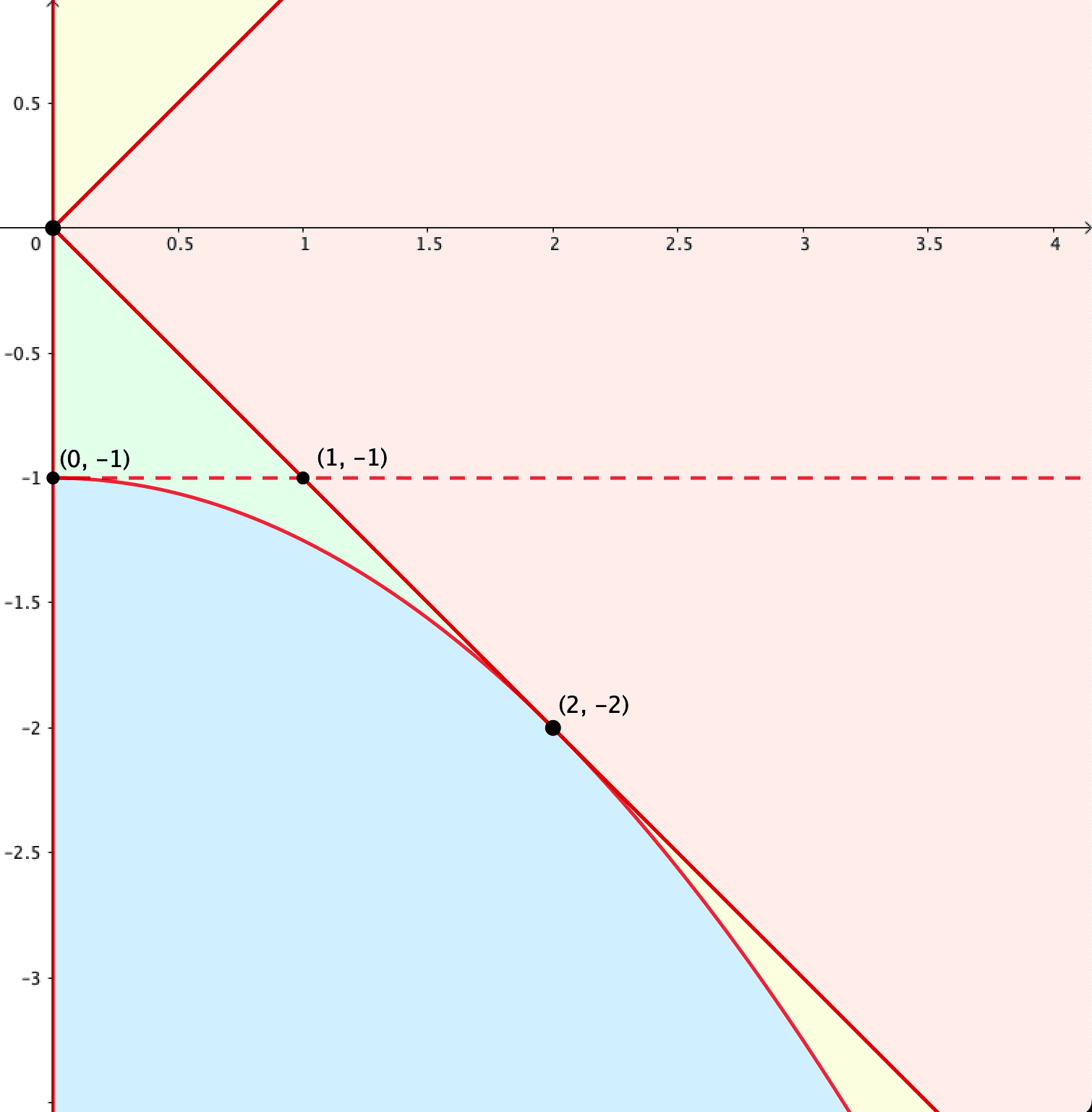}
\vspace{0cm} %
\end{minipage}
\caption{Sets $\mathcal{I}_4$, $\mathcal{C}_4$, $\mathcal{IR}$ and
$\mathcal{R}_4$
with different types of Stokes graphs of $Q_a(z)\,dz^2$.} 
\end{figure}

\textbf{II}. The polynomial $P_a(z)$ has four
distinct zeros, two real and two pure imaginary, if and only if %
\begin{equation} \label{6.4}
\Delta_a>0,  \quad \Delta_a^2+4E_a+4>0, \quad
\frac{\Delta_a}{2}\sqrt{\Delta_a^2+4E_a+4}>\left|\frac{1}{2}\Delta_a^2+E_a\right|.
\end{equation} 
Notice that these inequalities hold if and only if
$0<|E_a|<\Delta_a$ and therefore this case occurs  if and only if
$(\Delta_a,E_a)\in \mathcal{I}\mathcal{R}$, where the set
$\mathcal{I}\mathcal{R}\subset \mathbb{R}^2$ is defined as
$\mathcal{I}\mathcal{R}=\{(X,Y):\, 0<|Y|<X\}$; see Figure~2, where
the set $\mathcal{IR}$ is shown in the pink color. Furthermore,
using inequalities (\ref{6.4}) and elementary algebra one can show
that $e_2$ and $e_3$ defined earlier are real zeros such that %
$ -1\le e_3<0<e_2\le 1$. In the case when $e_2=-e_3=1$, which
occurs if and only if $E_a=-1$, $Q_a(z),dz^2$ reduces to the
depressed quadratics differential
$Q_s(z)\,dz^2=-\frac{z^2-c_0}{(z-1)(z+1)}\,dz^2$ with
$c_0==1-\Delta_a^2$, which has the points simple poles at $\pm 1$.
In the case under consideration,  the Stokes graphs as in the
subcases \textbf{II}-1, \textbf{II}-3 and \textbf{II}-4 of
Section~4 do not appear as an asymptotic case.

Also, the subcases \textbf{II}-1 and \textbf{II}-3 can be excluded
because the Stokes graphs in these cases are not symmetric with
respect to the imaginary axis and therefore these type of graphs
do not appear as asymptotic cases.

Thus, the only possible types of Stokes graphs and domain
configurations in the case under consideration are those discussed
in part \textbf{II}-2 of Section~4.

In this case, the intervals $(-i\infty,e_4)$ and $(e_1,i\infty)$
of the imaginary axis and the interval $(e_3,e_2)$ of the real
axis are critical trajectories of $Q_a(z)\,dz^2$ and the interval
$(e_4,e_1)$ is a critical orthogonal trajectory. This implies that
the domain configuration has a ring domain $\Omega_r$. Therefore,
in this case the Stokes graph and domain configuration belong to
the type described in part \textbf{II}-2-a of Section~4, as shown
in Figure~2, and the cases \textbf{II}-2-b and \textbf{II}-2-c do
not occur as asymptotic cases.

\smallskip

\textbf{III}. The polynomial $P_a(z)$ has four
distinct real zeros if and only if %
\begin{equation} \label{6.5}
\Delta_a>0, \quad \Delta_a^2+4E_a+4>0, \quad -\frac{1}{2}
\Delta_a^{2} - E_a -\frac{\Delta_a}{2} \sqrt{\Delta_a^2 + 4E_a +
4}>0. %
\end{equation} 
We perform algebraic operations to find that inequalities
(\ref{6.5}) hold if and only if $(\Delta_a,E_a)\in \mathcal{R}_4$,
where the set $\mathcal{R}_4\subset \mathbb{R}^2$ is defined as
$\mathcal{R}_4=\{(X,Y):\, 0<X<2,\,-\frac{1}{4}X^2-1<Y<-X\}$; see
Figure~2, where the set $\mathcal{R}_4$ is shown in the green
color. Furthermore, using inequalities (\ref{6.5}) and elementary
algebra one can show
that $e_1$, $e_2$, $e_3$, and $e_4$ defined above are real zeros such that %
$ -1\le e_3<e_4<0<e_1<e_2\le 1$. As in the previous case, if
$e_2=-e_3=1$, then
 $Q_a(z),dz^2$ reduces to the
depressed quadratics differential $Q_s(z)\,dz^2$ with  simple
poles at the points $\pm 1$. This implies that, if $e_3=-e_2<1$,
then only Stokes graphs described in the subcase \textbf{III}-2 of
Section~4 and shown in Figure~2 appear as an asymptotic case while
all other subcases described in part \textbf{III} of Section~4 do
not appear as asymptotic cases.

\medskip

Turning to the depressed and degenerate cases, we first mention
that if $(\Delta_a,E_a)\in \mathcal{L}(-1)$, where
$L(-1)=\{(X,Y):\, X\ge 0,\, Y=-1\}$ is the dash line shown in
Figure~2, then $P_a(z)$ has zeros at the points $\pm 1$ and
therefore $Q_a(z)\,dz^2$ reduces to the depressed quadratic
differential $Q_s(z)\,dz^2$ defined above. Thus, if $\Delta_a>1$,
then $Q_s(z)\,dz^2$ has two pure imaginary zeros and the Stokes
graph as in Figure~2. If $\Delta_a=1$, then $Q_s(z)\,dz^2$ has a
double zero at $z=0$ and the Stokes graph as in Figure~2.  If
$0<\Delta_a<1$, then $Q_s(z)\,dz^2$ has two real zeros and the
Stokes graph as in Figure~2. Finally,  if $\Delta_a=0$, then
$Q_s(z)\,dz^2=-dz^2$ and therefore its Stokes graph is empty and
the vertical lines are the trajectories of $Q_s(z)\,dz^2$ in this
case.

Next, we mention that if the point $(\Delta_2,E_a)$ lies on the
half-parabola $L_1=\{(X,Y):\, Y=-\frac{1}{4}X^2-1,\,X>0\}$ or on
one of the  half-lines $L_2=\{(X,Y):\, Y=X,\,X>0\}$,
$L_3=\{(X,Y):\, Y=-X,\,X>0\}$, $L_4\{(0,Y):\, Y>-1\}$,
$L_5=\{(0,Y):\, Y<-1\}$ (all of them are shown in red color,
except four black points,  in Figure~2), then $Q_a(z)\,dz^2$ has
two double zeros if $(\Delta_1,E_a)\not=(0,0)$ and
$(\Delta_1,E_a)\not=(2,-2)$ and it has zero of order $4$ at $z=0$
when $(\Delta_1,E_a)=(0,0)$ and when  $(\Delta_1,E_a)=(2,-2)$. If
$(\Delta_a,E_a)\in L_k$, $k=1,\ldots,5$, then the corresponding
quadratic differential $Q_a(z)\,dz^2$ has the Stokes graph of the
type shown in Figures~2--6, respectively.

        \appendix
\section{List of notation}

\begin{itemize} %
\item %
$\Delta$ - level of separation of the fermion mode in the Rabi
problem. %
\item %
 $g$ - boson-fermion coupling in the Rabi
problem. %
 \item %
 $E$ - eigenvalue of the Hamiltonian in the Rabi
problem. %
\item   %
$\mathbb C$ - complex plane. %
\item   %
 $\overline{\mathbb C}$ - Riemann sphere. %
\item  %
$\mathbb H_{+}$ - upper half-plane. %
\item    %
$\mathbb H_{-}$ - lower half-plane. %
\item    %
$(a,b)$ - open interval from $a$ to $b$. %
\item %
$[a,b]$ - closed interval from $a$ to $b$. %
\item  %
$Q(z)\,dz^2$ - general notation for a quadratic differential. 
\item  %
$G_Q$ - Stokes graph of $Q(z)\,dz^2$.  %
\item  %
$[a,b]_Q$ - integral $\int_a^b\sqrt{Q(z)}\,dz$ taken over the
interval $[a,b]$.
\item  %
$Q_0(z)\,dz^2$ - quadratic differential
$-\frac{z^4+c_3z^3+c_2z^2+c_1z+c_0}{(z-1)^2(z+1)^2}\,dz^2$. %
\item  %
$P_0(z)=z^4+c_3z^3+c_2z^2+c_1z+c_0$ - numerator of $Q_0(z)\,dz^2$. %
\item  %
$c_k$, $k=0,1,2,3$ - coefficients of $P_0(z)$. %
\item  %
$e_k$, $k=1,2,3,4$ - zeros of $P_0(z)$. %
\item  %
$e_{j,k}$ - double zero of $P_0(z)$ obtained by merging zeros $e_j$ and $e_k$. %
\item  %
$\mathbf{\Omega}=\mathbf{\Omega}(\Delta,E,g)$ - domain
configuration of $Q_0(z)\,dz^2$ associated with the Rabi
parameters $\Delta$, $E$, and $g$.  %
\item %
$\delta_k$ - $Q_0$-length of a trajectory/orthogonal trajectory
around $k=-1,1$. %
\item  %
$\gamma_{a,b}$ - closure of a critical trajectory of
$Q_0(z)\,dz^2$ oriented from $a$ to $b$.  %
\item  %
$\gamma_{a,b}^l$ - closure of a critical trajectory of
$Q_0(z)\,dz^2$ from $a$ to $b$ intersecting $(-\infty,-1)$.  %
\item  %
$\gamma_{a,b}^c$ - closure of a critical trajectory of
$Q_0(z)\,dz^2$ from $a$ to $b$ intersecting $(-1,1)$.  %
\item  %
$\gamma_{a,b}^r$ - closure of a critical trajectory of
$Q_0(z)\,dz^2$ from $a$ to $b$ intersecting $(1,\infty)$. %
\item  %
$\gamma_a^l$ - closure of a critical trajectory of
$Q_0(z)\,dz^2$ from $a$ to $a$ intersecting $(-\infty,-1)$, anticlockwise oriented. %
\item  %
$\gamma_a^c$ - closure of a critical trajectory of
$Q_0(z)\,dz^2$ from $a$ to $a$ intersecting $(-1,1)$, anticlockwise oriented. %
\item  %
$\gamma_a^r$ - closure of a critical trajectory of
$Q_0(z)\,dz^2$ from $a$ to $a$ intersecting $(1,\infty)$, anticlockwise oriented. %
\item  %
$\gamma_a^{l-}$ - closed curve $\gamma_a^l$ with reversed orientation.  %
\item  %
$\gamma_a^{c-}$ - closed curve $\gamma_a^c$ with reversed orientation.  %
\item  %
$\gamma_a^{r-}$ - closed curve $\gamma_a^r$ with reversed orientation.  %
\item  %
$\gamma_{a,i\infty}$ - closure of a critical trajectory of
$Q_0(z)\,dz^2$ starting at $a$ and approaching $\infty$ along
positive direction of some vertical line.  %
\item  %
$\gamma_{a,-i\infty}$ - closure of a critical trajectory of
$Q_0(z)\,dz^2$ starting at $a$ and approaching $\infty$ along
negative direction of some vertical line.  %
\item  %
$\gamma_{-i\infty,i\infty}$ - closure of a critical trajectory of
$Q_0(z)\,dz^2$ from $\infty$ to $\infty$, which approaches its
initial point along negative direction of some vertical line and
approached its terminal point  along positive direction of the
same vertical line.
\item  %
$\gamma_{i\infty,a}$ - arc $\gamma_{a,i\infty}$ with reversed orientation.  %
\item  %
$\gamma_{a,-i\infty}$ - arc $\gamma_{-i\infty,a}$ with reversed orientation.  %
\item  %
$\gamma_{i\infty,-i\infty}$ - closed curve $\gamma_{-i\infty,i\infty}$ with reversed orientation.  %
\item  %
$\gamma_{a,b}^+$ - closure of a critical trajectory of
$Q_0(z)\,dz^2$ from $a$ to $b$, $a,b\in \mathbb{R}\cup\{\infty\}$ lying in $\overline{\mathbb{H}}_+$.   %
\item  %
$\gamma_{a,b}^-$ - closure of a critical trajectory of
$Q_0(z)\,dz^2$ from $a$ to $b$, $a,b\in \mathbb{R}\cup\{\infty\}$ lying in $\overline{\mathbb{H}}_-$.   %
\item %
$\Omega_e^{l}$ - left end domain of $Q_0(z)\,dz^2$. %
\item %
$\Gamma_e^{l}$ - boundary of $\Omega_e^l$ positively oriented. %
\item %
$\Omega_e^{r}$ - right end domain of $Q_0(z)\,dz^2$. %
\item  %
$\Gamma_e^{r}$ - boundary of $\Omega_e^r$ positively oriented. %
\item %
$\Omega_c(k)$, $k=-1,1$ - circle domain of $Q_0(z)\,dz^2$ centered at $z=k$. %
\item  %
$\Gamma_c(k)$, $k=-1,1$ - boundary of $\Omega_c(k)$ positively oriented. %
\item %
$\Omega_r$ - ring domain of $Q_0(z)\,dz^2$. %
\item  %
$\Gamma_r^{(out)}$ - outer boundary component  of $\Omega_r$ oriented counterclockwise. %
\item  %
$\Gamma_r^{(inn)}$ - inner boundary component  of $\Omega_r$ oriented counterclockwise. %
\item  %
$\Omega_s(a,b)$ - strip domain of $Q_0(z)\,dz^2$ with vertices $a$
and $b$.
\item  %
$\Gamma_s^l(a,b)$ with $a=-i\infty$ and/or $b=i\infty$ - left side
of $\Omega_s(a,b)$. %
\item  %
$\Gamma_s^r(a,b)$ with $a=-i\infty$ and/or $b=i\infty$ - right
side of $\Omega_s(a,b)$. %
\item  %
$\Gamma_s^{(out)}(a,a)$ with $a=-1$ or $a=1$ - outer side of
$\Omega_s(a,a)$.  %
\item  %
$\Gamma_s^{(inn)}(a,a)$ with $a=-1$ or $a=1$ - inner side of
$\Omega_s(a,a)$.  %
\item  %
$\Gamma_s^+(-1,1)$ - side of $\Omega_s(-1,1)$ lying in the upper
half-plane. %
\item  %
$\Gamma_s^-(-1,1)$ - side of $\Omega_s(-1,1)$ lying in the lower
half-plane. %
\item  %
$S(a)$ - parabolic cylinder  $\{(X,Y,Z)\in \mathbb{R}^3:\,
(Y+a)^2-a^2X-a^2Z-(1/4)a^2(4+3a^2)=0\}$.%
\item  %
$Q_a(z)\,dz^2$ - asymptotic quadratic differential
$-\frac{z^4+c_2z^2+c_0}{(z-1)^2(z+1)^2}\,dz^2$. %
\end{itemize}

\clearpage
\section{Zoo of Stokes graphs.}\label{appendix}

\begin{figure}[h]
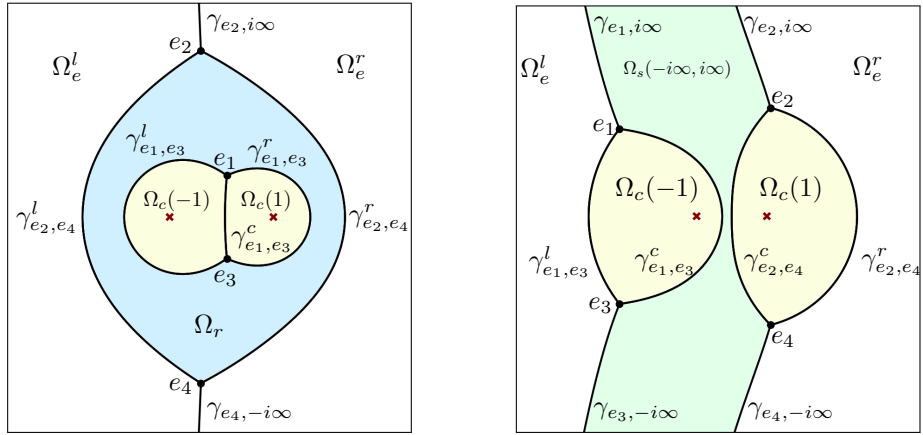

\centering
\begin{subfigure}{0.47\textwidth}

\end{subfigure}
\caption{Case I-1 and case I-2.}
\label{fig:I-1}
\end{figure}


\begin{figure}
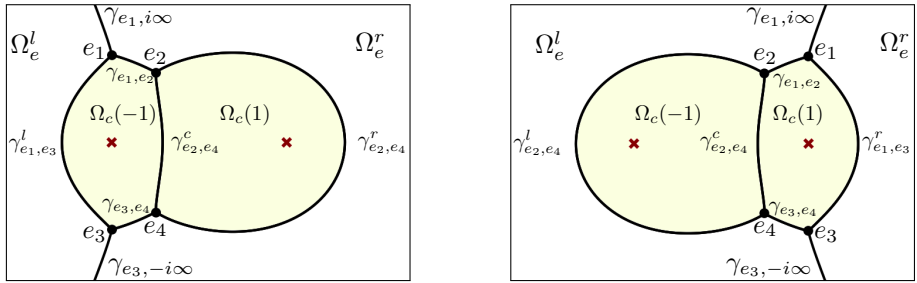

\centering
\begin{subfigure}{0.47\textwidth}
%
\end{subfigure}
\caption{Case I-3 and case I-3-m}
\label{fig:I-3}
\end{figure}

\begin{figure}
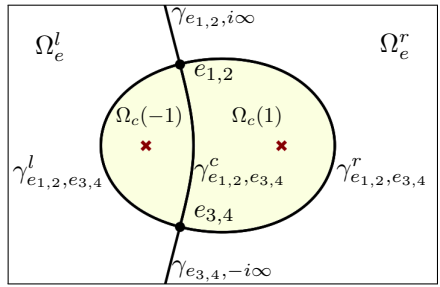

\centering
%
\caption{Case I-3-deg. }
\label{fig:I-3-deg}
\end{figure}

\begin{figure}
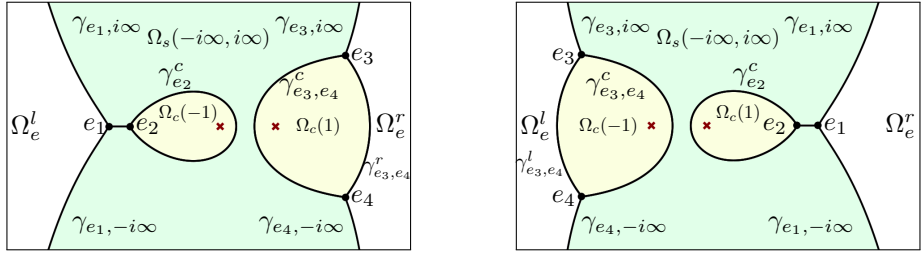
{}
\centering
\begin{subfigure}{0.47\textwidth}
%
\end{subfigure}
\caption{Case II-1-a and case II-1-a-m}
\label{fig:II-1-a}
\end{figure}

\begin{figure}
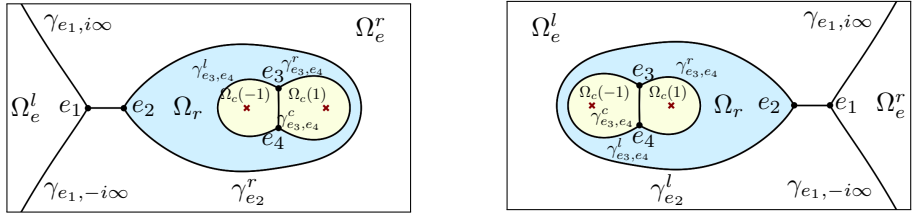

\centering
\begin{subfigure}{0.47\textwidth}
%
\end{subfigure}
\caption{Case II-1-b and case II-1-b-m}
\label{fig:II-1-b}
\end{figure}

\begin{figure}
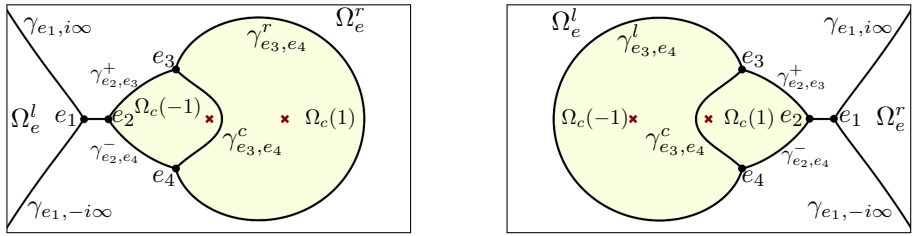

\centering
\begin{subfigure}{0.47\textwidth}
%
\end{subfigure}
\caption{Case II-1-c and case II-1-c-m}
\label{fig:II-1-c}
\end{figure}

\begin{figure}
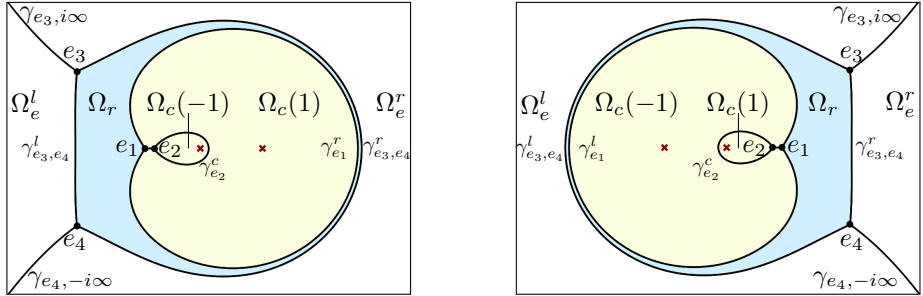

\centering
\begin{subfigure}{0.47\textwidth}
%
\end{subfigure}
\caption{Case II-1-d and case II-1-d-m}
\label{fig:II-1-d}
\end{figure}

\begin{figure}
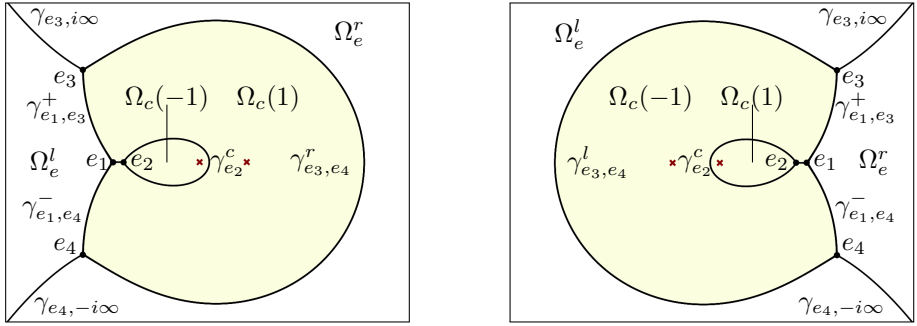

\centering
\begin{subfigure}{0.47\textwidth}
%
\end{subfigure}
\caption{Case II-1-e and case II-1-e-m}
\label{fig:II-1-e}
\end{figure}

\begin{figure}
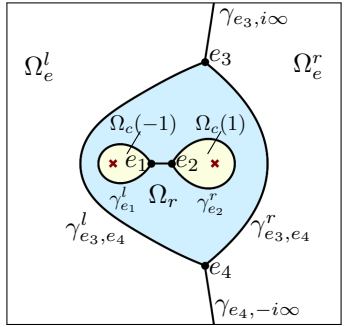

\centering
%
\caption{Case II-2-a.} 
\label{fig:II-2-a}
\end{figure}

\begin{figure}
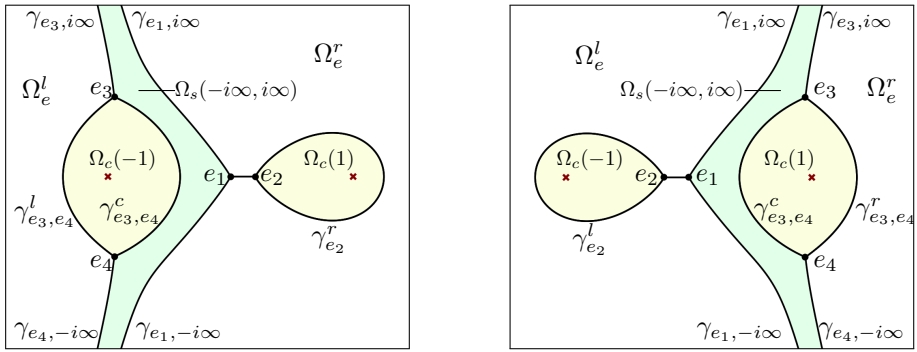

\centering
\begin{subfigure}{0.47\textwidth}
%
\end{subfigure}
\caption{Case II-2-b and case II-2-b-m}
\label{fig:II-2-b}
\end{figure}

\begin{figure}
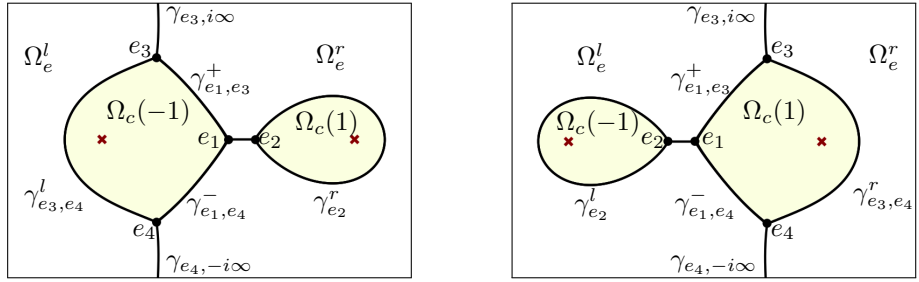

\centering
\begin{subfigure}{0.47\textwidth}
%
\end{subfigure}
\caption{Case II-2-c and case II-2-c-m}
\label{fig:II-2-c}
\end{figure}

\begin{figure}
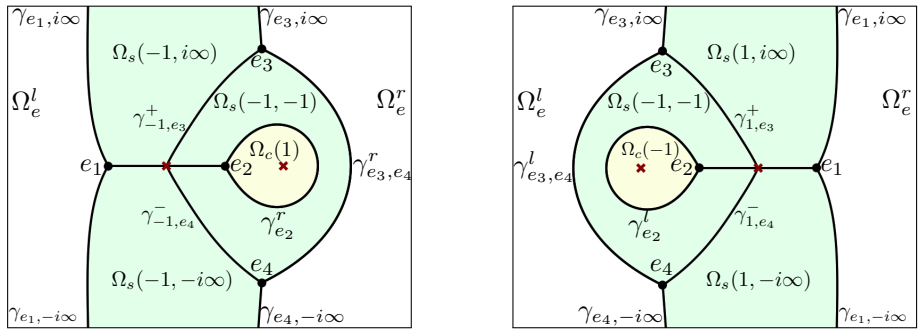

\centering
\begin{subfigure}{0.47\textwidth}
%
\end{subfigure}
\caption{Case II-3-a-$\alpha$ and case II-3-a-$\alpha$-m.}
\label{fig:II-3-a-alpha}
\end{figure}

\begin{figure}
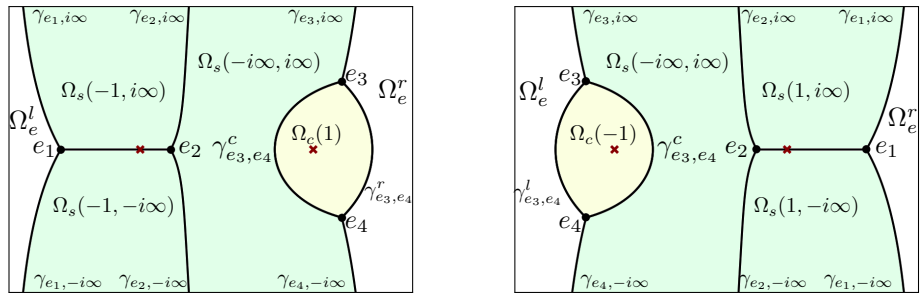

\centering
\begin{subfigure}{0.47\textwidth}
%
\end{subfigure}
\caption{Case II-3-a-$\beta$ and case II-3-a-$\beta$-m}
\label{fig:II-3-a-beta}
\end{figure}


\begin{figure}
\centering
\begin{subfigure}{0.47\textwidth}
%
\end{subfigure}
\caption{Case II-3-a-$\gamma$ and case II-3-a-$\gamma$-m}
\label{fig:II-3-a-gamma}
\end{figure}

\begin{figure}
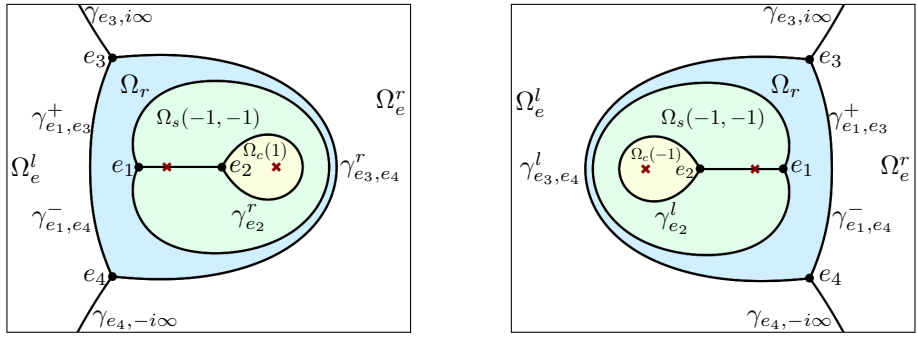

\centering
\begin{subfigure}{0.47\textwidth}
%
\end{subfigure}
\caption{Case II-3-b and case II-3-b-m}
\label{fig:II-3-b}
\end{figure}

\begin{figure}
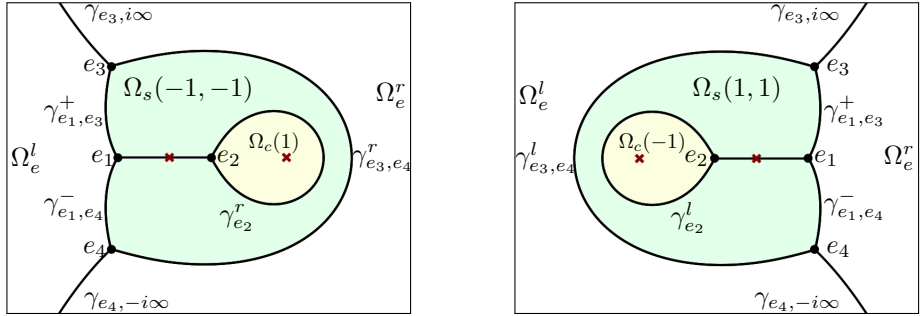

\centering
\begin{subfigure}{0.47\textwidth}
%
\end{subfigure}
\caption{Case II-3-c and case II-3-c-m}
\label{fig:II-3-c}
\end{figure}

\begin{figure}
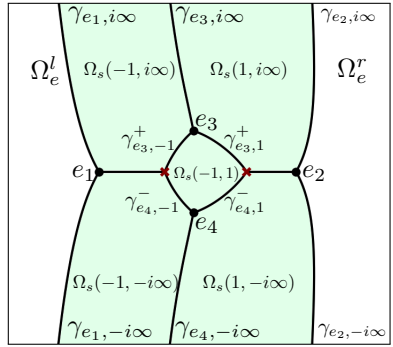

\centering
%
\caption{Case II-4-a-$\alpha$ } 
\label{fig:II-4-a-alpha}
\end{figure}

\begin{figure}
\centering
\begin{subfigure}{0.47\textwidth}
%
\end{subfigure}
\caption{Case II-4-a-$\beta$ and case II-4-a-$\beta$-m.}
\label{fig:II-4-a-beta}
\end{figure}

\begin{figure}
\centering
\begin{subfigure}{0.47\textwidth}
%
\end{subfigure}
\caption{Case II-4-a-$\gamma$ and case II-4-a-$\gamma$-m.}
\label{fig:II-4-a-gamma}
\end{figure}

\begin{figure}
\centering
\begin{subfigure}{0.47\textwidth}
%
\end{subfigure}
\caption{Case II-4-b-$\alpha$ and case II-4-b-$\alpha$-m.}
\label{fig:II-4-b-alpha}
\end{figure}

\begin{figure}
\centering
\begin{subfigure}{0.47\textwidth}
%
\end{subfigure}
\caption{Case II-4-c-$\alpha$ and case II-4-c-$\alpha$-m.}
\label{fig:II-4-c-alpha}
\end{figure}

\begin{figure}
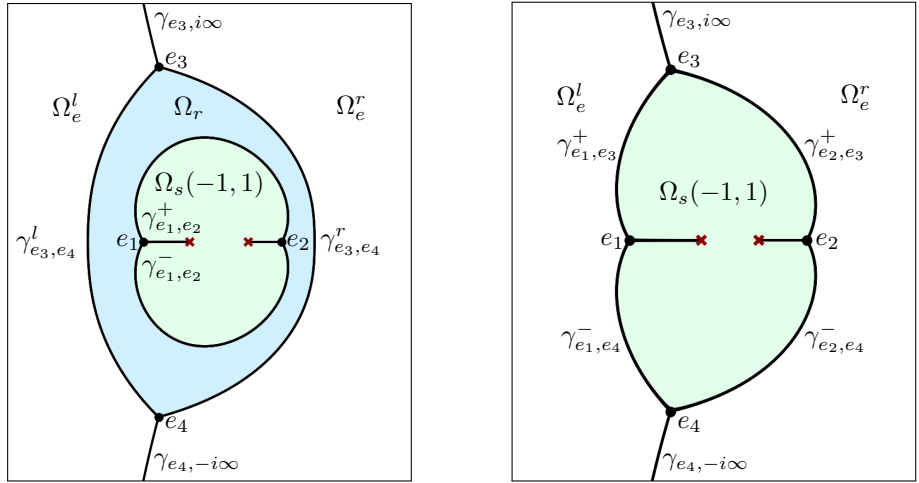

\centering
\begin{subfigure}{0.47\textwidth}
%
\end{subfigure}
\caption{Case II-4-b-$\beta$ and case II-4-c-$\beta$.}
\label{fig:II-4-b-beta}
\label{fig:II-4-c-beta}
\end{figure}

\begin{figure}
\centering
\begin{subfigure}{0.47\textwidth}
%
\end{subfigure}
\caption{Case III-1 and case III-1-m}
\label{fig:III-1}
\end{figure}

\begin{figure}
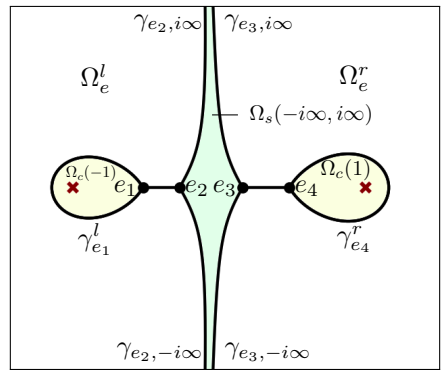

\centering
%
\caption{Case III-2}
\label{fig:III-2}
\end{figure}


\begin{figure}
\centering
\begin{subfigure}{0.47\textwidth}
%
\end{subfigure}
\caption{Case III-3 and case III-3-m}
\label{fig:III-3}
\end{figure}

\begin{figure}
\centering
\begin{subfigure}{0.47\textwidth}
%
\end{subfigure}
\caption{Case III-4-a-$\alpha$ and case III-4-a-$\alpha$-m}
\label{fig:III-4-a-alpha}
\end{figure}

\begin{figure}
\centering
\begin{subfigure}{0.47\textwidth}
%
\end{subfigure}
\caption{Case III-4-a-$\beta$ and case III-4-a-$\beta$-m}
\label{fig:III-4-a-beta}
\end{figure}

\begin{figure}
\centering
\begin{subfigure}{0.47\textwidth}
%
\end{subfigure}
\caption{Case III-4-a-$\gamma$ and case III-4-a-$\gamma$-m}
\label{fig:III-4-a-gamma}
\end{figure}

\begin{figure}
\centering
\begin{subfigure}{0.47\textwidth}
%
\end{subfigure}
\caption{Case III-4-b and case III-4-b-m}
\label{fig:III-4-b}
\end{figure}

\begin{figure}
\centering
\begin{subfigure}{0.47\textwidth}
%
\end{subfigure}
\caption{Case III-4-c and case III-4-c-m}
\label{fig:III-4-c}
\end{figure}

\begin{figure}
\centering
\begin{subfigure}{0.47\textwidth}
%
\end{subfigure}
\caption{Case III-5 and case III-5-m}
\label{fig:III-5}
\end{figure}

\begin{figure}
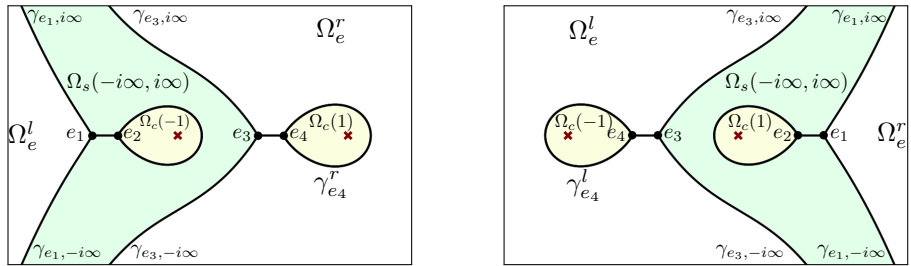
{}
\centering
\begin{subfigure}{0.47\textwidth}
%
\end{subfigure}
\caption{Case III-6-a and case III-6-a-m.}
\label{fig:III-6-a}
\end{figure}

\begin{figure}
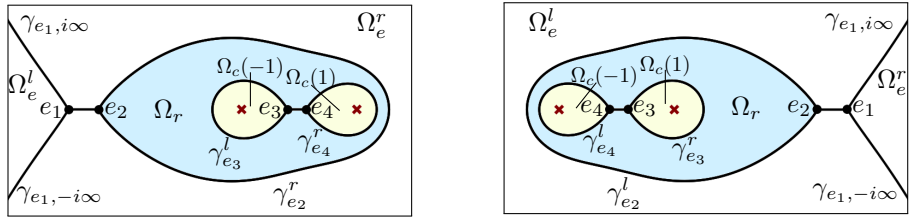
{}
\centering
\begin{subfigure}{0.47\textwidth}
%
\end{subfigure}
\caption{Case III-6-b and case III-6-b-m.}
\label{fig:III-6-b}
\end{figure}

\begin{figure}
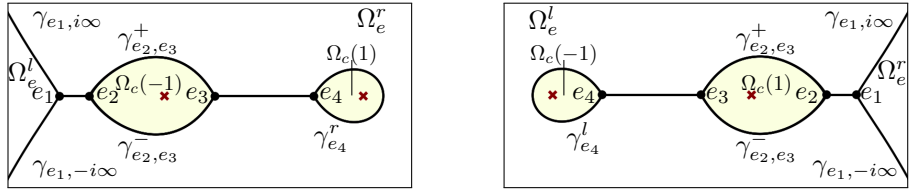

\centering
\begin{subfigure}{0.47\textwidth}
%
\end{subfigure}
\caption{Case III-6-c and case III-6-c-m}
\label{fig:III-6-c}
\end{figure}

\begin{figure}
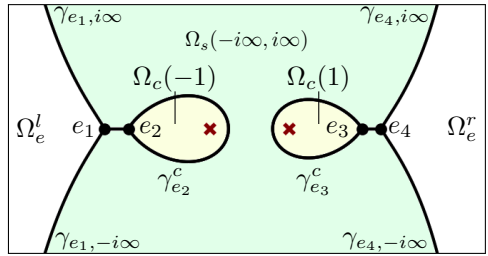

\centering
%
\caption{Case III-7-a.} 
\label{fig:III-7-a}
\end{figure}

\begin{figure}
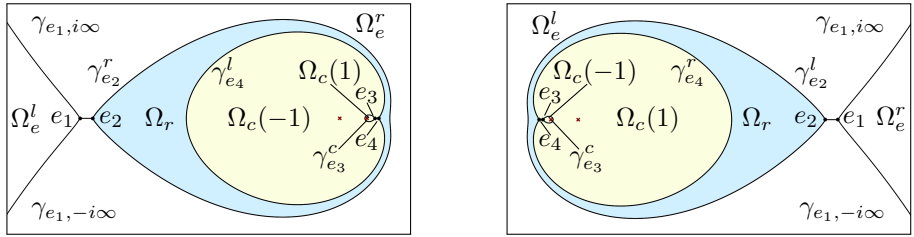

\centering
\begin{subfigure}{0.47\textwidth}
%
\end{subfigure}
\caption{Case III-7-b and case III-7-b-m.}
\label{fig:III-7-b}
\end{figure}

\begin{figure}
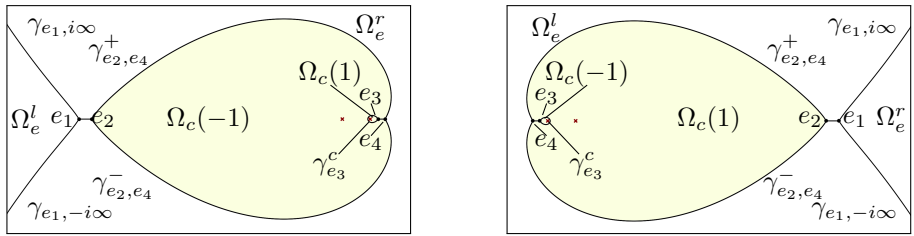

\centering
\begin{subfigure}{0.47\textwidth}
%
\end{subfigure}
\caption{Case III-7-c and case III-7-c-m}
\label{fig:III-7-c}
\end{figure}

\begin{figure}
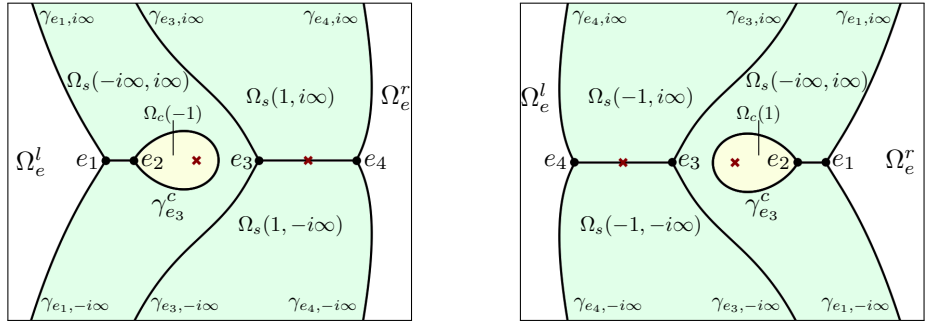

\centering
\begin{subfigure}{0.47\textwidth}
%
\end{subfigure}
\caption{Case III-8-a and case III-8-a-m.}
\label{fig:III-8-a}
\end{figure}

\begin{figure}
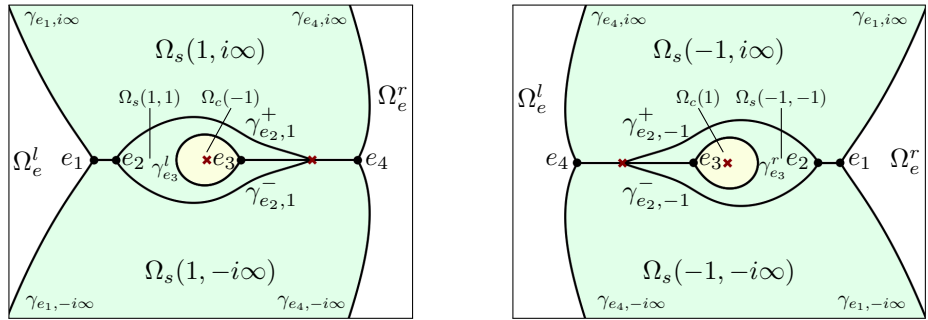

\centering
\begin{subfigure}{0.47\textwidth}
%
\end{subfigure}
\caption{Case III-8-b-$\alpha$ and case III-8-b-$\alpha$-m.}
\label{fig:III-8-b-alpha}
\end{figure}

\begin{figure}
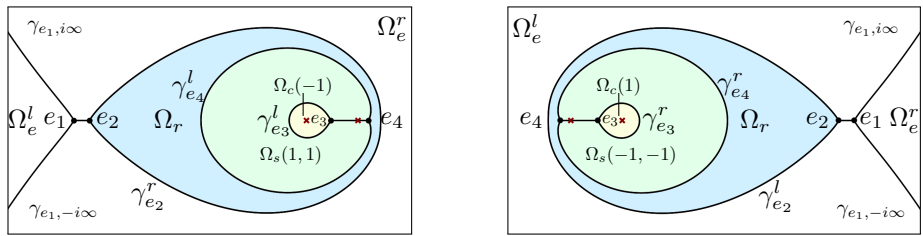

\centering
\begin{subfigure}{0.47\textwidth}
%
\end{subfigure}
\caption{Case III-8-b-$\beta$ and case III-8-b-$\beta$-m.}
\label{fig:III-8-b-beta}
\end{figure}

\begin{figure}
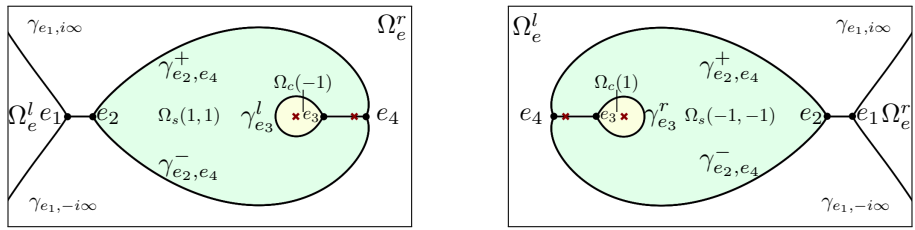

\centering
\begin{subfigure}{0.47\textwidth}
%
\end{subfigure}
\caption{Case III-8-b-$\gamma$ and case III-8-b-$\gamma$-m.}
\label{fig:III-8-b-gamma}
\end{figure}

\begin{figure}
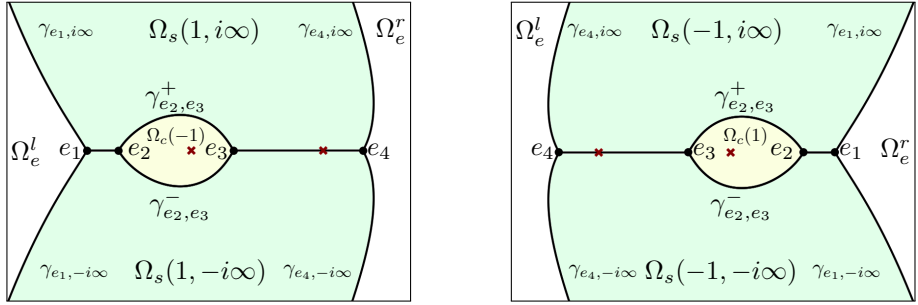

\centering
\begin{subfigure}{0.47\textwidth}
%
\end{subfigure}
\caption{Case III-8-c and III-8-c-m.}
\label{fig:III-8-c}
\end{figure}

\begin{figure}
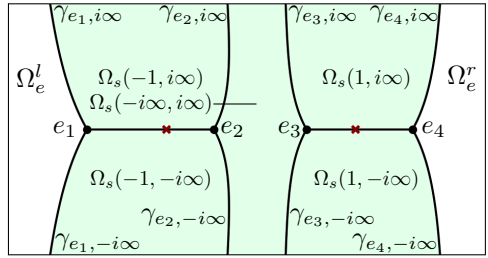

\centering
%
\caption{Case III-9.} 
\label{fig:III-9}
\end{figure}

\begin{figure}
\centering
\begin{subfigure}{0.47\textwidth}
%
\end{subfigure}
\caption{Case III-1-deg and case III-1-deg-m}
\label{fig:III-1-deg}
\end{figure}

\begin{figure}
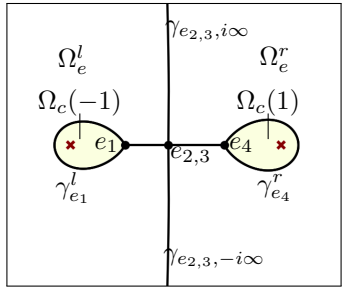

\centering
%
\caption{Case III-2-deg}
\label{fig:III-2-deg}
\end{figure}

\begin{figure}
\centering
\begin{subfigure}{0.47\textwidth}
%
\end{subfigure}
\caption{Case III-3-deg and case III-3-deg-m}
\label{fig:III-3-deg}
\end{figure}

\begin{figure}
\centering
\begin{subfigure}{0.47\textwidth}
%
\end{subfigure}
\caption{Case III-4-a-$\alpha$-deg and case III-4-a-$\alpha$-deg-m}
\label{fig:III-4-a-alpha-deg}
\end{figure}

\begin{figure}
\centering
\begin{subfigure}{0.47\textwidth}
%
\end{subfigure}
\caption{Case III-4-b-deg and case III-4-b-deg-m}
\label{fig:III-4-b-deg}
\end{figure}

\begin{figure}
\centering
\begin{subfigure}{0.47\textwidth}
%
\end{subfigure}
\caption{Case III-4-c-deg and case III-4-c-deg-m}
\label{fig:III-4-c-deg}
\end{figure}

\begin{figure}
\centering
\begin{subfigure}{0.47\textwidth}
%
\end{subfigure}
\caption{Case III-5-deg and case III-5-deg-m}
\label{fig:III-5-deg}
\end{figure}

\begin{figure}
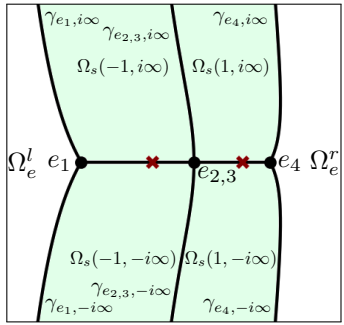

\centering
%
\caption{Case III-9-deg.} 
\label{fig:III-9-deg}
\end{figure}

\clearpage

\bibliographystyle{amsplain}

\end{document}